\documentclass[
11pt,
a4paper,
abstracton
]{scrartcl}

\usepackage[ngerman, english]{babel}
\usepackage[utf8]{inputenc}
\usepackage{graphicx}
\usepackage{url}
\usepackage{marvosym}
\usepackage{lmodern}
\usepackage{comment}
\usepackage{xcolor}
\usepackage[commandnameprefix=always]{changes}
\usepackage{changebar}
\usepackage{longtable}
\usepackage[doublespacing]{setspace}
\usepackage{fancyref}
\usepackage{enumitem}
\usepackage{tabularx}
\usepackage{longtable}
\usepackage{booktabs}
\usepackage{dcolumn}
\usepackage[para,online,flushleft]{threeparttable}
\usepackage{pdflscape}
\usepackage{amsmath, amssymb}
\usepackage{eurosym}
\usepackage{caption}
\usepackage{subcaption}

\usepackage[left=2.5cm, right=2.5cm, bottom=2.5cm, top=2.5cm]{geometry} 
\usepackage{graphicx}
\usepackage{mathtools}
\usepackage{mathpazo}
\usepackage{array}
\usepackage[title]{appendix}
\usepackage{titletoc}
\usepackage{afterpage}
\usepackage{pdflscape}

\newcolumntype{C}[1]{>{\centering\let\newline\\\arraybackslash\hspace{0pt}}m{#1}}
\newcolumntype{L}[1]{>{\raggedright\let\newline\\\arraybackslash\hspace{0pt}}m{#1}}

\newcommand{\sym}[1]{{#1}}

\usepackage[pdftex,bookmarks=true, colorlinks=false, pdfborder={0 0 0}]{hyperref} 	

\usepackage{apacite}

\begin{document}

\title{\LARGE The Devil is in the Details: \\Heterogeneous Effects of the German Minimum Wage \\ on Working Hours and Minijobs\vspace{0.5cm}}

\author{Mario Bossler%
\thanks{TH Nuremberg Georg Simon Ohm (TH Nuremberg), Institute for Employment Research (IAB), Institute of Labor Economics (IZA), and Labor and Socio-Economic Research Center of the University of Erlangen-Nuremberg (LASER). Address: TH Nuremberg, Kesslerplatz 12, 90489 Nuremberg, Germany, e-mail: \texttt{mario.bossler@th-nuernberg.de}.}\vspace{-0.3cm}
\\
\small{\emph {TH Nuremberg, IAB, IZA, LASER}}
\and
Ying Liang%
\thanks{Johannes Gutenberg-University Mainz (JGU Mainz). Address: JGU Mainz, Jakob-Welder-Weg 4, 55128 Mainz, e-mail: \texttt{liang.ying@uni-mainz.de}.}\vspace{-0.3cm}
\\
\small{\emph {JGU Mainz}}
\and
Thorsten Schank%
\thanks{Johannes Gutenberg-University Mainz (JGU Mainz),  Institute of Labor Economics (IZA), and Labor and Socio-Economic Research Center of the University of Erlangen-Nuremberg (LASER). Address: JGU Mainz, Jakob-Welder-Weg 4, 55128 Mainz, e-mail: \texttt{schank@uni-mainz.de}.
\newline
We gratefully acknowledge helpful comments from Matthias Umkehrer and those received at the 4th Scientific Workshop of the German Minimum Wage Commission in Berlin and from seminars and conferences at Goethe University Frankfurt, Leuphana University Lueneburg, IAB Nuremberg, and JGU Mainz.
We thank the Research Data Center of the German Statistical Office for their assistance in data provision.}\vspace{-0.3cm}
\\
\small{\emph {JGU Mainz, IZA, LASER}} \vspace{0.4cm}
}

\vspace{-0.3cm}
\date{April 2024}

\maketitle

\vspace{-0.9cm}
\begin{abstract}
\noindent \onehalfspacing In 2015, Germany introduced a national minimum wage. While the literature agrees on at most limited negative effects on the overall employment level, we go into detail and analyze the impact on the working hours dimension and on the subset of minijobs. Using data from the German Structure of Earnings Survey in 2010, 2014, and 2018, we find empirical evidence that the minimum wage significantly reduces inequality in hourly and monthly wages. While various theoretical mechanisms suggest a reduction in working hours, these remain unchanged on average. However, minijobbers experience a notable reduction in working hours which can be linked to the specific institutional framework. Regarding employment, the results show no effects for regular jobs, but there is a noteworthy decline in minijobs, driven by transitions to regular employment and non-employment. The transitions in non-employment imply a wage elasticity of employment of $-0.1$ for minijobs. Our findings highlight that the institutional setting leads to heterogeneous effects of the minimum wage. [160 words]

\vspace{0.3cm} 
\noindent \emph{JEL Classification}: J31, J38, J21 \\
\noindent \emph{Keywords}: minimum wage, working hours, monthly wages, hourly wages, minijobs
\end{abstract}

\thispagestyle{empty}
\clearpage
\setcounter{page}{1}
\doublespacing

\section{Introduction}
\label{sec:intro}

Minimum wages are a popular policy to improve the situation of the working poor, but also to address rising wage inequality and the declining bargaining power of workers. A critical and detailed evaluation of minimum wages is important because most economic theories predict a negative impact on employment, at least when minimum wages are sufficiently high. While there is considerable evidence at the extensive margin of employment adjustments in heads (see \citeA{Neumark2008} for a review), evidence of adjustments at the intensive margin of hours worked is scarce and not very conclusive. For the United Kingdom, for example, \citeA{Stewart2008} finds a reduction in hours. However, a recent study by \citeA{Datta2021} finds only a redistribution of hours, but no overall effect. In the USA the debate is even more controversial. For hours worked by teenage workers, who are strongly affected by the minimum wage, \citeA{Couch2001} find a negative effect in a state-level panel framework, while \citeA{Zavodny2000} obtains no adverse hours effects.

On 1 January 2015, Germany introduced a national minimum wage of \euro{8.50}, which was increased to \euro{8.84} on 1 January 2017. The minimum wage directly affected around 10-14 percent of the workforce, who were paid an average of 26 percent below the minimum wage threshold in 2014. Germany is an interesting case to study, as it was one of the few Western economies that did not have a nationwide minimum wage before 2015. We examine the impact of the introduction of the minimum wage by exploiting the variation in the \textit{bite} of the minimum wage (i.e., the share of workers affected) across regions in Germany. Thus, we identify minimum wage effects from difference-in-differences specifications, as first proposed in \citeA{Card1992}. While our focus is on the analysis of adjustments in working hours, we also provide a comprehensive analysis of how the minimum wage affects hourly and monthly wages as well as the number of employees. In particular, we are interested in whether the institutional setting leads to heterogeneous minimum wage effects between regular employees and minijobbers.

Minijobs denote jobs with a monthly salary up to \euro{450} and are largely exempted from social security contributions.\footnote{In 2022 and 2024, the minijob threshold increased in two steps to \euro{538}, which is outside our period of observation.} Hence, their net salary, by and large, corresponds with the gross salary, making them an attractive alternative compared to regular social security jobs, which start at a salary of \euro{451} per month. Minijobs are an interesting group to study, because these are, by definition, low-income workers with particular relevance for the analysis of wage inequality at the lower end of the monthly wage distribution. Minijobs made up as much as 21 percent of all jobs in 2014, with a fraction of 42 percent paid below the minimum wage. When analyzing minijobs, it is particularly relevant to analyze their interplay with regular jobs, as individuals move back and forth between both types of jobs and minijobs may also crowd out regular jobs, as documented in the literature \cite{Collischon2021}.

When hypothesizing hours adjustments in response to the introduction of the minimum wage, several theoretical mechanisms should be considered. (i) Employers may substitute away from labor to capital and may also produce less. Both effects imply that the demand for working hours would decrease. (ii) If productivity increases in the course of the minimum wage introduction \cite{Dustmann2022,Riley2017}, fewer hours are necessary to achieve a given output. However, rising productivity may also lead to an increase in the demand for hours worked. (iii) The introduction of the minimum wage increases the cost of an additional working hour while the fixed recruiting cost of an additional worker remains unchanged. Consequently, there may be a substitution from hours to workers. (iv) The labor supply response is ambiguous. The minimum wage may raise workers' incentive to work longer hours (dominating substitution effect). However, it is also possible that workers want to reduce their hours worked when receiving a higher hourly wage (dominating income effect). (v) There could be a rise in informal work. In consequence, unpaid hours may increase to avoid minimum wage payments. Hence, actual (reported plus unreported) hours may remain unchanged while contractual hours may decrease. (vi) The institutional setting can cause an hours reduction since subsidized minijobbers are not allowed to receive more than \euro{450} per month. Consequently, hours need to be reduced if employers and employees still want to take advantage of the subsidized social security contributions. Otherwise, minijobs have to be upgraded to regular jobs, which may be less attractive for employers, and possibly also for employees.

The empirical assessment of the German minimum wage introduction agrees on negligible aggregate employment effects, as pointed out in a literature review by \citeA{Bruttel2019}. At the same time, the literature documents a reallocation of minimum wage workers to more productive firms \cite{Dustmann2022}. Moreover, the effect on wages and income inequality was assessed to be positive \cite{Bossler2023,Burauel2020b}. However, the evidence concerning working hours adjustments is much more nuanced. Based on survey data, \citeA{Bossler2020} and \citeA{Ohlert2022} analyze establishment-level variation on hours from the IAB-Establishment Panel and the Earnings Survey respectively, and \citeA{Burauel2020} examine individual-level survey information from the German Socio-Economic Panel (SOEP). These studies obtain a reduction in working hours in the first year after the minimum wage introduction, but they cannot identify an effect in the second year. \citeA{Caliendo2022} and \citeA{Biewen2022} evaluate the impact of the minimum wage on working hours and combine them with the analysis of monthly and hourly wages. \citeA{Caliendo2022} use the SOEP and report a negative hours effect. By contrast, \citeA{Biewen2022} analyze the German Structure of Earning Survey and find no hours response to the minimum wage. 

We contribute to the literature in several ways. First, we examine the effects of the minimum wage introduction on hourly and monthly wages. In contrast to \citeA{Bossler2023}, whose analysis is limited to monthly wages, we are able to analyze whether their results regarding the reduction of inequality extend to hourly wages. Using the German Structure of Earnings Survey (SES), we can directly compare hourly and monthly wage adjustments using the same data and the same source of identifying variation. A comparison of hourly and monthly wage effects also allows us to infer potential hours adjustments. In addition, we can examine the extent of spillover effects on both hourly and monthly wages. The difference-in-differences wage regressions reveal effects of the minimum wage on monthly wages from the SES data that are virtually identical to those in \citeA{Bossler2023}, i.e., monthly wages increased by 5 percent on average. Our results also show that the effect is strikingly similar when looking at hourly wages. Further, while the hourly wage effect of the minimum wage is monotonously decreasing along the hours distribution, we observe significant spillovers up to the median. Finally, the effects show that the minimum wage significantly reduced the inequality of monthly and even more pronounced of hourly wages. 

Second, we directly analyze hours of work from the most comprehensive data source for working hours in Germany, which is the Structure of Earnings Survey. While \citeA{Biewen2022} demonstrate the absence of a working hours effect in the aggregate, we further disaggregate our analysis by looking at subgroups such as the severely affected minijobs. As in \citeA{Biewen2022}, we find no effect on either contractual or paid hours at the aggregate level. However, when we analyze heterogeneous effects, we find a more nuanced picture. There are negative effects along the hours distribution at percentiles where minijobs are located whose income after the minium wage introduction would increase beyond the minijob threshold if their working time was not reduced. We also find a negative effect on hours in small establishments and show that is not solely explained by a larger share of minijobbers in those establishments. 

Third, we investigate the minimum wage effect on employment. The absence of economically significant employment responses is a consensus in the literature (see e.g., \citeA{Bruttel2019}), which we can confirm by applying a region-level difference-in-differences specification determining aggregate employment effects. Furthermore, we distinguish between regular employment and minijobs. In their descriptive report to the German government, \citeA{VomBerge2018} document a reduction in the number of minijobs in January 2015, when the minimum wage was introduced. This reduction has been referred to as one of the most remarkable consequences of the minimum wage policy on the German labor market \cite{MLK2023}. The finding of a reduction of minijobs is corroborated in causal analyses by \citeA{Caliendo2018} and \citeA{Caliendo2023}. 

Consistent with the literature, we show that the employment effect of minijobs is significantly negative. We document that after the minimum wage introduction, minijobs disappeared at a certain point of the hours distribution. In absolute terms, 104,000 minijobs vanished due to the minimum wage. Moreover, we examine a hitherto unresolved question, namely whether the respective minijobbers were upgraded to regular jobs or went into non-employment. Using the Integrated Employment Biographies (IEB), which is the universe of administrative employment data, as an additional data source, we find that 54,000 minijobbers (which is about half of the vanished minijobs) entered non-employment, while the other half was upgraded to regular employment (mostly within the same establishment). The non-employment transition can be interpreted as a wage elasticity of employment for minijobbers of $-0.1$, while the employment elasticity for regular jobs is virtually zero. 

Fourth, we can quantify the extent of non-compliance with the minimum wage, which is highly controversial in the political debate \cite{MLK2023}. The SES data contain detailed and precise information on hours worked and monthly wages, thus defining the hourly wage that can be compared with the required minimum wage. Moreover, we can compare the initial and remaining wage gaps with the size of the wage effect to indirectly infer how much of the wage gap was closed by the minimum wage. Looking at the size of the observed wage effects, it is indeed the case that more than the initial wage gap has been closed by the minimum wage. This is only possible if there are significant spillovers, which is consistent with our observed wage effects along the wage distribution (up to the median). However, we still observe a significant descriptive number of non-compliant wages, pointing to an important policy area. 

The paper proceeds as follows: Section \ref{sec:data} presents the SES data, along with descriptions of the main variables of interest, and documents changes at the lower end of the wage distribution. Section \ref{sec:method} describes the empirical strategy, which is a difference-in-differences approach that identifies minimum wage effects from regional variation in the bite. Section~\ref{sec:results} reports and discusses the difference-in-differences effects on hourly wages, monthly wages, and hours worked. Section~\ref{sec:heterogeneities} presents heterogeneities of the hours effect across firm size, along the hours distribution, and between minijobs and regular jobs. Finally, we analyze the employment effects for different groups of workers, with a special focus on minijobbers. Since minijobs with a monthly working time of 50.9 hours or more would shift above the \euro{450} threshold due to the minimum wage of \euro{8.84} in 2018, we investigate what happened to these minijobbers. Section \ref{sec:conclusion} concludes.

\section{Data}\label{sec:data}

We use the years 2010, 2014, and 2018 of the German Structure of Earnings Survey (SES) which has been collected by the Federal Statistical Office of Germany every four years until 2018. It is a linked employer-employee data set consisting of repeated cross-sections. For each employee, it covers information on employment characteristics, gross monthly wages (which we deflated by the CPI to 2014 \euro{}), contractual working hours, and paid hours (which also include paid overtime hours). The sampled establishments are obliged to provide the earnings situation of a random sample of employees directly from the payroll accounting for April of the respective years.\footnote{Establishments with less than ten employees are obligated to report information on all employees. The proportion of sampled employees then falls with rising establishment size. For example, in 2014 and 2018 establishments with at least 100 and less than 250 employees are obliged to report information on every sixth employee.} Therefore, the information in the SES is highly accurate. We analyze the data of the 2010, 2014, and 2018 waves, covering two pre-periods and one post-period of the introduction of the minimum wage policy.

The sample size of the SES varies largely between 2010 and 2014/2018. Hence, a weighting factor is needed to ensure the representativeness of the sample. We calculate new weights with the target being the employment-population according to register data from the Establishment History Panel (BHP) of the IAB.\footnote{See Appendix~\ref{app:BHP} and \citeA{ganzer2020establishment} for more details on the BHP.} We impose the same sample restrictions in the population data as in our SES sample. Specifically, we exclude apprentices because they are exempted from the minimum wage legislation. To ensure time consistency of the sample, we drop establishments in agriculture as well as establishments with less than ten employees subject to social security (\textit{regular employees}).\footnote{We relax the restriction to establishments with at least 10 regular employees in the robustness checks reported at the end of Section~\ref{sec:results} and in the hours and employment regressions reported in Section~\ref{sec:heterogeneities}.} We also exclude the public sector due to the missing regional information. We calculate the weights based on stratification according to three establishment size categories, regular jobs and minijobs, manufacturing and services, and 401 counties where the latter is important for our regional-level identification of treatment effects.\footnote{The weights are based on establishment size categories of 1--9, 10--99, more than 100 employees. While the first category is not used in the main specifications, it is included in further analysis.}

Table \ref{tab:description} provides an overview of the employee-level data that we analyze. While the number of observations shrunk from 2010 to 2014 and again slightly in 2018, the weighted number of employees in plants with at least 10 employees grew from 23.27 million in 2010 to 25.24 million in 2014 and 27.40 million in 2018. These figures are consistent with relatively strong overall employment growth at that time. At the same time, the fraction of minijobs fell in exchange for growing shares of part-time (less than 30 weekly hours) and full-time jobs (at least 30 weekly hours). One year before the minimum wage was introduced (in 2014), 11.7 percent of the workers were still paid below the initial minimum wage of \euro{8.50}, demonstrating the relevance of the policy. In 2018, we observe that only 0.4 percent of the workers were paid below \euro{8.50}, while 2.5 percent were paid below the new minimum wage level which was increased to \euro{8.84} in 2017. While this shows a significant influence of the minimum wage on the reduction of low-wage employment, the remaining employees below the minimum wage also suggest some non-compliance with the policy.\footnote{According to the descriptive figures the faction of wages below the initial minimum wage decreased already between 2010 and 2014 (from 17.4 to 11.7 percent). However, in real terms, the fraction of workers below \euro{8.50} remained fairly constant (12.8 percent in 2010 compared with 11.7 percent in 2014). I.e., if we inflated the 2010 wages by the price increase between 2010 and 2014, the fraction of real wages below \euro{8.50} decreased. }

\begin{table}[ht!]
				\centering
				\caption{Weighted summary statistics }
\label{tab:description}
				\resizebox{1\textwidth}{!}{
				\begin{threeparttable}
				\begin{tabular}{p{5.7cm} rl rl rl}
			\hline\hline
&       \multicolumn{2}{c}{2010}     &         \multicolumn{2}{c}{2014}&         \multicolumn{2}{c}{2018}      \\
&     mean     & (std. dev.) &     mean     & (std. dev.)&     mean     & (std. dev.) \\
\hline
Gross hourly wage (2014 \euro{})   &      18.10\phantom{0}&      (12.57)&      17.60\phantom{0}&     (12.51) &      19.17\phantom{0} &      (14.26)\\
Gross monthly wage (2014 \euro{})  &    2,631.58\phantom{0}& (2,262.48)   & 2,593.35\phantom{0}  & (2,274.93) &    2,809.41\phantom{0} & (2,539.08)\\[1ex]
Log gross hourly wage (2014 \euro{})&       2.734& (0.561)  &   2.719& (0.522) &      2.818 &  (0.490) \\
Log gross monthly wage (2014 \euro{})&       7.479& (1.042)  &    7.481& (1.020)&      7.574 & (1.006)   \\[1ex]
Contractual monthly hours&     133.01\phantom{0}& (53.83)  &  135.26\phantom{0}& (53.59) &   135.08\phantom{0}   &  (53.00)\\
Paid monthly hours (incl. paid overtime)&     134.83\phantom{0}&  (55.29)  & 136.89\phantom{0}&(54.77) &    136.53\phantom{0}&  (53.97) \\[1ex]
Log contractual monthly hours&       4.734& (0.676)    &  4.751&(0.685)  &     4.746 &(0.701)\\
Log paid monthly hours&       4.745&   (0.681)   & 4.762& (0.688) &     4.756& (0.703)\\
 \hline
East Germany        &       0.165&  &     0.164&     &  0.164 &\\
Female              &       0.434&   &    0.445&    &   0.435&\\
Low education       &       0.189&    &   0.121&  &     0.110&\\
Medium education    &       0.693&    &   0.699&   &    0.684&\\
High education      &       0.118&    &   0.180&   &    0.207&\\
\hline
Regular job (weekly hours) & & & & &  & \\
\quad at least 30       &       0.689& &       0.701& &      0.708&\\
\quad between 18 and 30 &       0.099& &      0.100&  &     0.104&\\
\quad less than 18      &       0.042& &      0.036&  &     0.040&\\[1ex]
Minijob            &       0.171&    &   0.164&     &  0.148 &\\[1ex]
Nominal gross hourly wage below the 2015 min. wage (\euro{}8.50) &    0.174 &  &    0.117 &  & 0.004 & \\
... their relative wage gap to \euro{}8.50 
& 0.398 & (0.841)   & 0.256 & (0.477)    & 0.352 & (0.802)        \\[1ex]
Nominal gross hourly wage below the 2018 min. wage (\euro{}8.84) &0.195  &  &0.144 &  & 0.025 & \\
... their relative wage gap to \euro{}8.84 &0.408  &  (0.838)   & 0.254 & (0.461)  & 0.076  &  (0.374)   \\ 
\hline
Weighted No. of observations        &     23,270,552& &   25,238,142  &      &     27,395,012 & \\
Actual No. of observations           &  1,493,904  & &        605,352  & &  568,337  &  \\
\hline\hline
			\end{tabular}
			\begin{tablenotes}
            \begin{small} \textit{Notes:} Weighted sample averages and standard deviations by year of observation. No standard deviations reported for dummy variables. All wages are deflated to 2014 \euro{}. Low education denotes neither \textit{Abitur} nor vocational training; medium education denotes \textit{Abitur} and/or vocational training; high education denotes master craftsman/technician or university degree. The relative gap is calculated as (8.50 - nomimal gross hourly wage)/nominal gross hourly wage.
			\\\textit{Data:} SES, 2010, 2014, and 2018, weighted analysis sample, establishments with at least 10 regular employees. 
			\end{small}
		\end{tablenotes}
	\end{threeparttable}
}
\end{table}

Regarding the development of wages, Table \ref{tab:description} shows stagnating average real wages in the early 2010s which picked up after the minimum wage introduction. This trend holds for both, hourly wages and monthly wages. Looking at the standard deviation of the respective variables, we observe a continuous decrease in the employees' log wage dispersion. The decline in descriptive wage inequality concerning log monthly wages, which started to emerge already before the minimum wage introduction is in line with findings in \citeA{Bossler2023}. At the same time, the average number of working hours (with and without paid overtime) remained fairly constant in the period of analysis. 

\begin{figure}[ht!]
\captionabove{Changing mass of jobs along the wage distribution between 2014 and 2018}
\label{fig:bunching}
\centering
\includegraphics[width=0.9\textwidth]{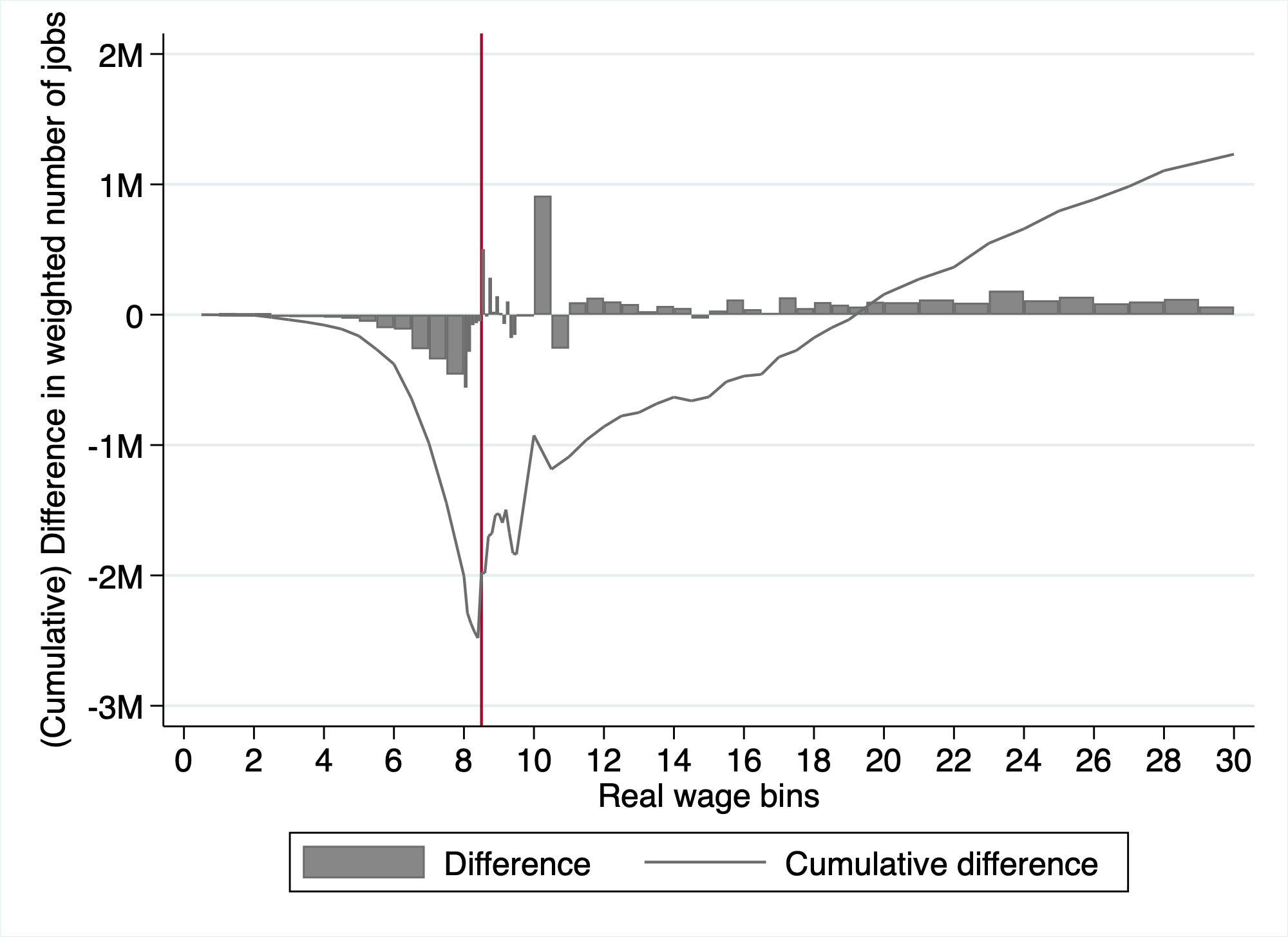}
\subcaption*{\textit{Notes:} Changes in the number of employees in millions (between 2018 and 2014) by real hourly wages (in 2014 \euro{}), as in \citeA{Cengiz2019}. Employees with real hourly wages above \euro{}30 not included. These amount to 11.2 percent (13.9 percent) of all employees in 2014 (2018). 
The vertical line is the initial minimum wage of \euro{8.50}, which corresponds with the real minimum wage of 2018 (see footnote \ref{fn:nominal}). 
\\\textit{Data:} SES, 2014 and 2018, weighted analysis sample, establishments with at least 10 regular employees.}
\end{figure}

Figure \ref{fig:bunching} provides a more precise description of changes along the wage distribution when the minimum wage was introduced. Inspired by the illustration from \citeA{Cengiz2019}, we examine the mass of employees along the real hourly wage distribution. The bars show the difference per wage bin in the weighted number of jobs between 2018 and 2014, i.e., after and before the minimum wage introduction in 2015. The figure clearly documents a reduction in the number of jobs up to the 2018 minimum wage level of \euro{8.84} (expressed in 2014 \euro), which is in line with the intention of the policy to reduce the number of jobs below the minimum wage. We also observe a spike to the right of the minimum wage, indicating an increased mass of jobs that are paid the minimum wage (or slightly above). These observations are consistent with workers receiving a wage increase and thereby moving on the x-axis to the right, but there is no panel data of this hourly wage precision available for Germany.

The solid line in Figure \ref{fig:bunching} illustrates the accumulative change in the mass of jobs along the wage distribution. It shows that not the entire mass of workers paid below the minimum wage in 2014 received a wage rise exactly to the level of the minimum wage. Rather, the line crosses the horizontal axis (indicating unchanged total employment) not before the \euro{18} mark. This may be because the minimum wage workers moved further up the wage distribution, possibly by changing employers, as suggested by \citeA{Dustmann2022}, or alternatively, due to respective workers entering non-employment. However, it becomes visible that overall employment paid below \euro{30} has grown, which may or may not be related to the minimum wage.

\section{Empirical strategy}\label{sec:method}

\begin{figure}[ht!]
\captionabove{Distribution of the bite across counties in Germany}
\label{fig:map}
\centering
\includegraphics[width=0.65\textwidth]{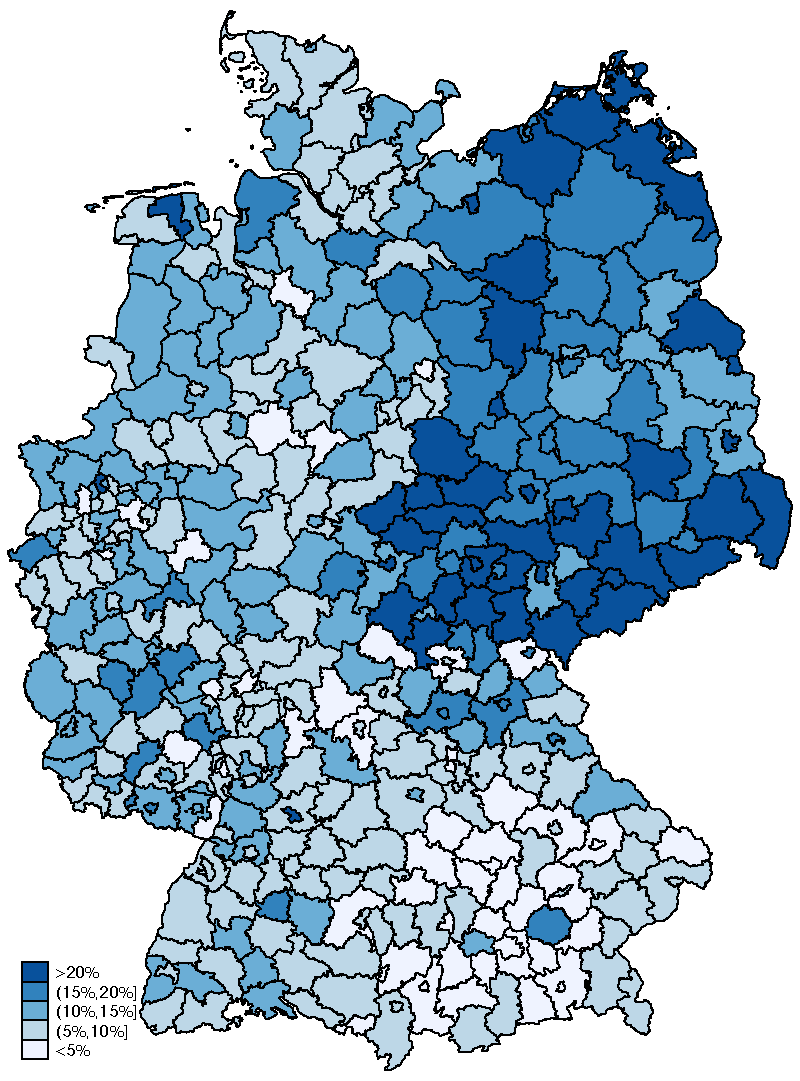}
\subcaption*{\textit{Notes:} The map displays the distribution of the bite across German counties, where the bite is defined as the share of workers paid in 2014 below the minimum wage of \euro{}8.50.\\ \textit{Data:} SES 2014, establishments with at least 10 regular employees.}
\end{figure}

We adopt a difference-in-differences framework that uses regional variation of the minimum wage bite to identify the effect of the minimum wage policy, as first proposed by \citeA{Card1992}. The bite is calculated as the share of workers in a county who are paid below the minimum wage threshold (\euro{8.5} per hour) in 2014, i.e., before the minimum wage came into force.\footnote{\label{fn:nominal}We base the bite on actual (i.e., nominal) wages paid in 2014 and on the nominal level of the minimum wage at its introduction in January 2015 which is 8.50. Although the nominal minimum wage was 8.84 at the beginning of 2018, which is our post-treatment year, in real terms it corresponds with the initial minimum wage in 2015. Hence, the bite based on the real minimum wage of 2018 is largely equivalent to the bite of the nominal minimum wage of 2015. We also checked that collectively bargained wages increased by more than the CPI between 2015 and 2018. Hence, the definition of our treatment variable assumes that workers paid between 8.50 and 8.84 in 2014 are not treated because they would have also received the same pay rise above 8.84 until 2018, even in the absence of the minimum wage.} Figure~\ref{fig:map} clearly demonstrates significant variation in the regional treatment intensity of the minimum wage policy across Germany. While the bite exceeds 20 percent in eastern German regions, is particularly low in southern Germany, falling below 5 percent in some of its counties. 

We specify the following regression equation for employee $i$ in year $t$, working in county $r$: 
\begin{multline}\label{eq:did_base}
	y_{it} = \gamma_{2014} + 
   \sum_{k=2010, \, 2018} \gamma_{k} * Year_{k,t} +
   \phi * Bite_{r} +
   \sum_{k=2010,\, 2018} \delta_{k}* Bite_{r} * Year_{k,t} +  \boldsymbol{X}_{i,t} \boldsymbol{\beta} +  \epsilon_{it} \\ t= 2010, 2014, 2018
\end{multline}
where $y_{it}$ denotes the logarithm of the respective dependent variable (hourly wage, monthly wage, monthly contractual hours, monthly paid hours). For hourly and monthly wages we do not only run conditional mean (and conditional variance) specifications, but also unconditional quantile regressions in which case $y_{it}$ is the RIF (re-centered influence function) of log hourly (monthly) wages at quantile $q$. In addition, when examining employment effects we also carry out regressions at the region-year level where the dependent variable measures employment in region $r$ in year $t$. $Bite_r$ is the share of minimum wage workers in region $r$ in the year $2014$. Since 2014 is the reference year, $\phi$ indicates whether the dependent variable correlates with the bite in 2014. The coefficient $\delta_{2018}$ of the interaction term $Bite_{r}*Year_{2018,t}$ captures the treatment effect of the minimum wage policy, while $\delta_{2010}$ serves as a placebo to test the plausibility of the parallel trends assumption. $\boldsymbol{X}_{it}$ denotes a set of control variables including age and dummies for gender, education, and industry. All regressions are weighted as described in Section~\ref{sec:data}; and standard errors are cluster-robust at the county-level.

\section{Results}\label{sec:results}

We start the regression analysis with the level and the variance of log hourly wages as the dependent variable. The estimates of the interaction effect are reported in Table~\ref{tab:baseline}.\footnote{The tables of the main text only report estimates of main interest, i.e., of the treatment and the placebo effect ($\widehat{\delta}_{2018}$ and $\widehat{\delta}_{2010}$). The extended tables including also the other estimated parameters (except $\widehat{\boldsymbol{\beta}}$) are presented in Appendix~\ref{app:full_tables}.} The first column shows a significant minimum wage-induced increase in hourly wages. A 10-percentage point increase in the regional bite causes wages to rise by 4.56 percent. Since the average bite in Germany is about 11.7 percent, we can therefore attribute an increase in average hourly wages by 5.3 percent to the minimum wage introduction. Note that the placebo interaction of the bite and the year 2010 is insignificant and small, supporting the internal validity of the reported minimum wage effect. As depicted in the second column, the minimum wage reduced the dispersion in hourly wages. A 10-percentage point increase in the regional bite implies a \textit{ceteris paribus} reduction in the hourly wage variance of 0.065. This is a meaningful effect size since the variance was 0.31 in 2010. Again, the respective placebo interaction effect is small and insignificant.

\begin{table}[ht!]\centering
\caption{Minimum wage effect on the mean and the variance of log hourly and monthly wages}
\label{tab:baseline}
\begin{threeparttable}
\begin{tabular}{lC{2.5cm}C{2.5cm}C{2.5cm}C{2.5cm}}
\hline\hline
	&\multicolumn{2}{c}{log hourly wage}&\multicolumn{2}{c}{log monthly wage}\\
\cmidrule(lr){2-3}\cmidrule(lr){4-5}
	&\multicolumn{1}{c}{mean}&\multicolumn{1}{c}{variance}&\multicolumn{1}{c}{mean}&\multicolumn{1}{c}{variance}\\[1ex]
\hline \\ [-1.4ex]
$Bite*Year_{2010}$ & -0.042 & -0.017 & -0.012 & -0.102  \\
	 & (0.044)  & (0.045) & (0.080)   &    (0.256)   \\[1ex]
$Bite*Year_{2014}$ & Reference & Reference & Reference & Reference \\[1ex]
$Bite*Year_{2018}$ & 0.456\sym{***}& -0.654\sym{***}& 0.499\sym{***}& -1.196\sym{***} \\
	 & (0.033) &    (0.034) &    (0.072) &    (0.242) \\[1ex] 
\hline 
Clusters            &         401         &         401         &         401         &      401          \\
Observations        &   2,667,593         &   2,667,593         &   2,667,593         &   2,667,593         \\[1ex]
\hline\hline
\end{tabular}
\begin{tablenotes}
\begin{small} \textit{Notes:} Treatment effect interactions from difference-in-differences estimations, as specified in equation (1). The dependent variable is either the log hourly or the log monthly wage or the RIF of the variance of log hourly or log monthly wages, as indicated by column titles. The $\boldsymbol{X}$-vector includes the following covariates: age, dummies for gender, education, and industry. All regressions are weighted. Standard errors in parentheses are clustered at the county level. Asterisks indicate significance levels: *~$p<0.05$, **~$p<0.01$, ***~$p<0.001$. \\\textit{Data:} SES, 2010, 2014, and 2018, weighted analysis sample, establishments with at least 10 regular employees.
\end{small}
\end{tablenotes}
\end{threeparttable}
\end{table}

A more comprehensive picture is provided by the results on monthly wages (third and fourth column). We find that monthly wages increase on average by 4.99 percent (Table \ref{tab:baseline}, third column) if the bite increases by 10 percentage points. This magnitude is close to the hourly wage effect, thereby providing the first evidence that on average working hours did not decrease by a significant margin. Also for log monthly wages, the variance decreases after the introduction of the minimum wage. A 10-percentage point increase in the regional bite leads to a reduction in the variance of log monthly wages of about 0.12. Since the variance of log monthly wages is at a higher level (1.09 in 2010), the reduction in relative terms is not as pronounced as for the variance in hourly wages. Again, the placebo interactions with the year 2010 are small and insignificant, supporting also for monthly wages the causal interpretation of the baseline estimates. 

It is interesting to note that the baseline hourly wage effect already points to significant spillover effects. If we extrapolate and compare the average hourly wage effect of 45.6 percent (for affected workers) with the average initial wage gap of affected workers of 26 percent, it demonstrates that the wage increase caused by the minimum wage is larger than the initial wage gap, thereby implying substantial spillovers.

In the presence of spillovers, positive wage effects can coexist with significant non-compliance. A descriptive assessment of the 2018 data shows that 2.5 percent of workers were still paid below minimum wage in 2018. Among those workers, the remaining wage gap is 7.6 percent, which is 30 percent of the initial wage gap in 2014 (see Table \ref{tab:description}). While the magnitude of non-compliance with the German minimum wage is still controversial in the literature \cite{Caliendo2019,MLK2023}, our share of non-compliant wages is meaningful given that it is self-reported employer-surveyed data. It points to the importance of institutions and inspections to monitor compliance \cite{Bossler2022,Goerke2023}. 
However, it is possible that effects on wages (but also adverse employment effects) could be stronger if non-compliance was reduced \cite{Garnero2022,Yaniv2001}. Note that it does not pose a threat to our identification because we estimate so-called unconditional reduced form effects which reflect ex-post realized outcomes after the minimum wage introduction.

Table~\ref{table:Rif_baseline} reports difference-in-differences effects of the minimum wage along the wage distribution from unconditional quantile regressions. As expected, the minimum wage effect is largest at the bottom of the distribution and then monotonously decreases. This matches the observed reduction in the variance discussed above. Interestingly, the treatment effect is statistically and economically significant up to the median which implies considerable spillover effects in the hourly wage. This is also consistent with Figure~\ref{fig:bunching}, which shows significant employment increases in wage bins that are far above the minimum wage threshold.

\afterpage{
\begin{landscape}

		\begin{table}[ht!]\centering
			\caption{RIF regressions along the hourly and the monthly wage distribution}
            \label{table:Rif_baseline}
				\begin{threeparttable}
				\begin{tabular}{L{2.7cm}C{1.8cm}C{1.8cm}C{1.8cm}C{1.8cm}C{1.8cm}C{1.8cm}C{1.8cm}C{1.8cm}C{1.8cm}}
				\hline\hline
			      &\multicolumn{1}{c}{q4}&\multicolumn{1}{c}{q8}&\multicolumn{1}{c}{q12}&\multicolumn{1}{c}{q16}&\multicolumn{1}{c}{q20}&\multicolumn{1}{c}{q30}&\multicolumn{1}{c}{q50}&\multicolumn{1}{c}{q70}&\multicolumn{1}{c}{q90}\\[1ex]
\hline \\[-1.4ex]
\multicolumn{10}{l}{A: Hourly wages} \\[1ex]
\hline \\[-1.4ex]
$Bite*Year_{2010}$  & -0.328\sym{*}   &  0.135\sym{**} &  0.144\sym{**} &  0.099\sym{*} &  0.084  &      0.008         &    -0.113\sym{*}  &      -0.155\sym{**} &   0.043 \\
			      &  (0.160)        &  (0.048)       &  (0.047)       &  (0.048)      &  (0.046)  &  (0.064)  &  (0.052)        &  (0.055)  &  (0.062)   \\[1ex]
$Bite*Year_{2014}$  & Reference & Reference & Reference & Reference & Reference & Reference & Reference & Reference & Reference \\[1ex]
$Bite*Year_{2018}$  &  2.717\sym{***} &  1.465\sym{***} & 1.089\sym{***} & 0.854\sym{***} & 0.775\sym{***}  &      0.715\sym{***}&      0.237\sym{***} &    0.015    &  -0.254\sym{***}\\
			      &  (0.092)  &    (0.037)  &  (0.040)  &  (0.042)  &    (0.046)  &    (0.061)         &    (0.049)         &    (0.040)         &   (0.067)          \\[1ex]
\hline \\[-1.4ex]
Clusters            &         401         &         401         &         401         &        401		&         401  &         401         &         401         &        401		&         401         \\		
Observations        &  2,667,593  &  2,667,593  &  2,667,593  &  2,667,593  &  2,667,593  &  2,667,593  &  2,667,593  &  2,667,593  &  2,667,593          \\[1ex]
\hline \\[-1.4ex]
\multicolumn{10}{l}{B: Monthly wages} \\[1ex]
\hline \\[-1.4ex]
$Bite*Year_{2010}$  & 0.015    &  0.084     &  -0.023    &   0.108    &   0.145 &   -0.030  &  -0.035  &  -0.105  &  0.032 \\
					& (0.424)  &  (0.177)   &  (0.054)   &   (0.083)  &  (0.275) &  (0.097) &  (0.058) &  (0.057) & (0.064)\\[1ex]
$Bite*Year_{2014}$  & Reference & Reference & Reference & Reference & Reference & Reference & Reference & Reference & Reference \\[1ex]
$Bite*Year_{2018}$  & 2.072\sym{***} & 0.705\sym{***} & 0.155\sym{**} & 0.249\sym{***} & 1.356\sym{***} & 1.217\sym{***}&      0.324\sym{***}&  0.023  &  -0.241\sym{***}\\
					&  (0.390) & (0.175) & (0.049) & (0.071) & (0.247) & (0.086) & (0.051) & (0.038) & (0.066) \\[1ex]
\hline \\[-1.4ex]
Clusters            &         401         &         401         &         401         &        401		&         401  &         401         &         401         &        401		&         401         \\		
Observations        &  2,667,593  &  2,667,593  &  2,667,593  &  2,667,593  &  2,667,593  &  2,667,593  &  2,667,593  &  2,667,593  &  2,667,593          \\[1ex]
			\hline\hline
					\end{tabular}
					\begin{tablenotes}
\begin{small} \textit{Notes:} Treatment effect interactions from difference-in-differences estimations, as specified in equation (1). The dependent variable is the RIF of the log hourly wage (upper panel) respectively of the log monthly wage (lower panel) defined for various percentiles, as indicated by column titles. The $\boldsymbol{X}$-vector includes the following covariates: age, dummies for gender, education, and industry. All regressions are weighted. Standard errors in parentheses are clustered at the county level. Asterisks indicate significance levels: 
 *~$p<0.05$, **~$p<0.01$, ***~$p<0.001$.
\\\textit{Data:} SES, 2010, 2014, and 2018, weighted analysis sample, establishments with at least 10 regular employees.
\end{small}
					\end{tablenotes}
				\end{threeparttable}
		\end{table}
  
\end{landscape}
}

When looking at the effects of the minimum wage introduction along the monthly wage distribution, we observe again spillover effects up to the median. Moreover, similar to the findings of \citeA{Bossler2023} there is an interesting hump-shaped pattern between the 4th and the 30th percentile, with the smallest treatment effect at the 12th percentile. This is exactly where subsidized minijobbers are located at the monthly wage distribution. The small treatment effect is consistent with many minijobbers not getting promoted to regular employees after the minimum wage increase (in which case the social security contributions would not be subsidized anymore). Hence, the results clearly demonstrate that the minijob institution dampened the minimum wage effect on monthly wages for the respective workers. 
Note that for both, hourly and monthly wages, there is a negative treatment effect at the 90th percentile.\footnote{More detailed RIF regressions for each percentile (untabulated) show already a negative (but smaller in size) coefficient at the 80th percentile ($-0.124$ for hourly wages, respectively $-0.106$ for monthly wages). Furthermore, the statistically significant negative effect is (in absolute terms) monotonously increasing until the top of the distribution. At the 97th percentile, the coefficient is $-0.371$  for hourly wages, respectively $-0.318$ for monthly wages.} This negative effect is in accordance with \citeA{gregory2022minimum}, who examine the minimum wage introduced since 1997 in parts of the German construction sector. While the authors also obtain positive spillovers above the minimum wage, they find that the minimum wage caused a reduction in wages for the highest quantiles.

Turning to the results of estimating the effects on working hours directly, we also cannot detect a consistent significant effect of the minimum wage (Table~\ref{tab:hours_baseline}). The effects on monthly contractual working hours and on paid working hours (both in levels and in logarithms) are statistically insignificant as well as small in economic terms. This confirms the finding above that the minimum wage affected average hourly and monthly wages similarly.

\begin{table}[ht!]
\centering
\caption{Minimum wage effect on monthly working hours}
\label{tab:hours_baseline}
\begin{threeparttable}
\begin{tabular}{lC{2.5cm}C{2.5cm}C{2.5cm}C{2.5cm}}
\hline\hline
	&\multicolumn{2}{c}{Contractual hours}&\multicolumn{2}{c}{Paid hours}\\\cmidrule(lr){2-3}\cmidrule(lr){4-5}
    & level & log & level & log \\[1ex]
\hline \\[-1.4ex]
$Bite*Year_{2010}$ & 0.088    & 0.020   & 1.987   & 0.031         \\
                   &    (3.592)         &    (0.059)         &    (3.566)         &    (0.059)         \\[1ex]
$Bite*Year_{2014}$ & Reference & Reference & Reference & Reference \\[1ex]
$Bite*Year_{2018}$ &     -4.029         &      0.049         &     -4.590         &      0.043         \\
				   &    (3.352)         &    (0.058)         &    (3.489)         &    (0.059)         \\[1ex]
\hline \\[-1.4ex]
Clusters           &  401         &         401         &         401         &        401        \\
Observations       &   2,667,593         &   2,667,593         &   2,667,593         &   2,667,593         \\[1ex]
\hline\hline
\end{tabular}
\begin{tablenotes}
\begin{small} \textit{Notes:} Treatment effect interactions from difference-in-differences estimations, as specified in equation (1). The dependent variable is either the (monthly) contractual or the (monthly) paid working hours in levels or in logarithms, as indicated by column titles. The $\boldsymbol{X}$-vector includes the following covariates: age, dummies for gender, education, and industry. All regressions are weighted. Standard errors in parentheses are clustered at the county level. Asterisks indicate significance levels: *~$p<0.05$, **~$p<0.01$, ***~$p<0.001$. \\\textit{Data:} SES, 2010, 2014, and 2018, weighted analysis sample, establishments with at least 10 regular employees.
\end{small}
\end{tablenotes}
\end{threeparttable}
\end{table}

We conduct the following robustness checks (all reported in Appendix~\ref{app:robustness}). 
(i) We add county fixed effects to account for the possibility that the unbalancedness of the sample across regions might influence the results, but these remain literally unaffected.
(ii) The inclusion of covariates in a difference-in-differences specification is often a contentious issue. Since they may be affected themselves by the treatment such that controlling for them may capture a genuine treatment effect. Therefore, we repeat the analysis above without covariates, but our findings do not change.
(iii) We censor hours worked above 250 hours per month to examine whether these 0.2 percent outlier observations in hours worked (and therefore also in hourly wages) influence the results, but this is not the case. 
(iv) In our baseline regressions, the treatment variable measures the share of workers paid below the minimum wage. In an alternative specification, we use instead the county-level average of the bite gap. For workers paid initially below the minimum wage, the bite gap is defined as the percentage rise in hourly wages required to comply with the minimum wage (while it is set to zero for those paid initially already above the minimum wage). However, the results are quantitatively similar when we use the bite gap.
(v) So far, the analysis has been based on establishments with at least 10 regular employees
since small establishments are not included in the 2010 wave.
To examine the severity of this restriction, we rerun the wage regressions using the waves 2014 and 2018 only, therefore allowing to include also those establishments with less than 10 regular employees. Hence, a placebo term for 2010 cannot be included. The minimum wage effect on hourly wages for all plants is very similar to the baseline results, while the coefficient of the treatment variable in the monthly wage specification has been reduced (0.350 for all plants versus 0.499 in the baseline). This indicates that the hours effect may differ by establishment size which we will examine in the next section.

\section{Heterogeneities}\label{sec:heterogeneities}

According to our baseline results, average hours worked remain unaffected after the minimum wage introduction. In this section, we examine whether there are heterogeneous effects by establishment size or along the hours distribution. Thereafter, we focus on minijobbers.

\subsection{Hours adjustment in small plants}\label{subsec:hoursmallplants}

Our baseline sample is restricted to establishments with at least 10 regular employees subject to social security. The exclusion of small establishments ensures a consistent sample over time because establishments with fewer than 10 regular employees were excluded from the SES wave of 2010. However, the wave of 2010 is crucial to test for the absence of a bite-specific trend between 2010 and 2014, i.e., before the minimum wage was introduced. Our difference-in-differences approach assumes that such a bite-specific trend is insignificant before the minimum wage came into force, thereby supporting the plausibility of the parallel trends assumption, i.e., in the absence of the minimum wage introduction we would not identify any treatment effect.

To analyze the effect of the minimum wage on working hours at small establishments we can only use data from the 2014 and 2018 waves. Consequently, we cannot check for a pre-treatment bite-specific trend in working hours. Since Table~\ref{tab:hours_baseline} shows no evidence of such a trend for establishments with at least 10 regular employees, we now have to assume that this is also the case for small establishments with less than 10 regular employees.

\begin{table}[ht!]\centering
	\caption{Minimum wage effect on monthly working hours; all versus small plants}
    \label{tab:hours_small_plants}
			\begin{threeparttable}
				\begin{tabular}{lC{2.5cm}C{2.5cm}C{2.5cm}C{2.5cm}}
					\hline\hline
					&\multicolumn{2}{c}{Contractual hours}
					&\multicolumn{2}{c}{Paid hours}\\
\cmidrule(lr){2-3}\cmidrule(lr){4-5}
					&\multicolumn{1}{c}{level}
					&\multicolumn{1}{c}{log}
					&\multicolumn{1}{c}{level}
					&\multicolumn{1}{c}{log}\\[1ex]
\hline \\ [-1.4ex]
\multicolumn{5}{l}{A: All plants} \\[1ex]
\hline \\ [-1.4ex]
$Bite*Year_{2014}$  & Reference & Reference & Reference & Reference \\[1ex]
$Bite*Year_{2018}$  &  -9.124\sym{***}&  -0.056  &  -9.455\sym{***}&  -0.060 \\
					&  (2.086)  &  (0.032)  &  (2.157)  &  (0.032) \\[1ex]
\hline \\ [-1.4ex]  
Clusters            &         401         &         401         &         401         &    401   \\
Observations        &   1,461,576         &   1,461,576         &   1,461,576         &   1,461,576       \\[1ex]
\hline \\ [-1.4ex]
\multicolumn{5}{l}{B: Small plants ($ < 10$ employees)} \\[1ex]
\hline \\ [-1.4ex]
$Bite*Year_{2014}$  & Reference & Reference & Reference & Reference \\[1ex]
$Bite*Year_{2018}$  &  -11.772\sym{***}&  -0.126\sym{***}&  -11.796\sym{***}&  -0.126\sym{***}     \\
					&  (2.439)  &  (0.035)  &  (2.468)  &  (0.035)  \\[1ex]
\hline \\ [-1.4ex]
Clusters            &         401         &         401         &         401         &    401              \\
Observations        &   277,791         &     277,791         &     277,791         &     277,791       \\[1ex]
					\hline\hline
				\end{tabular}
				\begin{tablenotes}
\begin{small} \textit{Notes:} Treatment effect interactions from difference-in-differences estimations, as specified in equation (1). The dependent variable is either the (monthly) contractual or the (monthly) paid working hours in levels or in logarithms, as indicated by column titles. The $\boldsymbol{X}$-vector includes the following covariates: age, dummies for gender, education, and industry. All regressions are weighted. Standard errors in parentheses are clustered at the county level. Asterisks indicate significance levels: *~$p<0.05$, **~$p<0.01$, ***~$p<0.001$.
      \\\textit{Data:} SES, 2014 and 2018, weighted analysis sample. 
\end{small}
			\end{tablenotes}
			\end{threeparttable}
\end{table}

Panel A of Table~\ref{tab:hours_small_plants} reports hours regressions for workers across all establishments. When the dependent variable denotes the level of contractual or paid hours worked, the treatment effect is now statistically significant and twice as big as in the baseline regressions. Correspondingly, the effect size increases (in absolute terms) when only including workers from small establishments (Panel B). Moreover, the effect on the logarithmic working hours also becomes significantly negative for the sample of small establishments in Panel B. 

These findings suggest that the working hours adjustments in the course of the minimum wage introduction are heterogeneous with respect to establishment size and tend to be significantly negative among small establishments. This finding is intuitive since the bite of the minimum wage was more significant at small establishments \cite{MLK2016}. Moreover, small establishments have less leeway for wage increases as compared to large establishments which are ascribed to pay significant wage premiums possibly due to higher productivity and perhaps also high market power \cite{Oi1999}. Finally, note that the share of minijobbers is considerably larger in establishments with less than 10 employees and amounted to 40.5 percent in 2014 compared to 16.4 percent in plants with at least 10 employees. In the following sections, we examine whether the hours response after the minimum wage introduction differs between minijobbers and regular employees and also whether the heterogenous effects across plant size are solely due to a different share of minijobbers.

One concern of a separate analysis of working hours by establishment size might be that the minimum wage led to a change in the establishment size composition, which would imply an endogenous sample split. In Appendix~\ref{app:size_selectivity}, we examine whether the minimum wage affects the firm size composition at the 10-employee threshold. We do not find any indication for such an effect, supporting the causal claim that the hours effect differs by establishment size. We also test whether minijobs moved to establishments with at least 10 employees in high-bite regions. However, between 2014 and 2018 we do not find any statistical difference (p-value of 0.529). 

\subsection{Hours adjustment along the unconditional hours distribution}\label{sec:hours_hoursdist}

\begin{figure}[ht!]
\captionabove{Minimum wage effects on log monthly working hours along the hours distribution, RIF regression estimates}
\label{fig:effect_along_hours_distribution}
\centering
\includegraphics[width=0.9\textwidth]{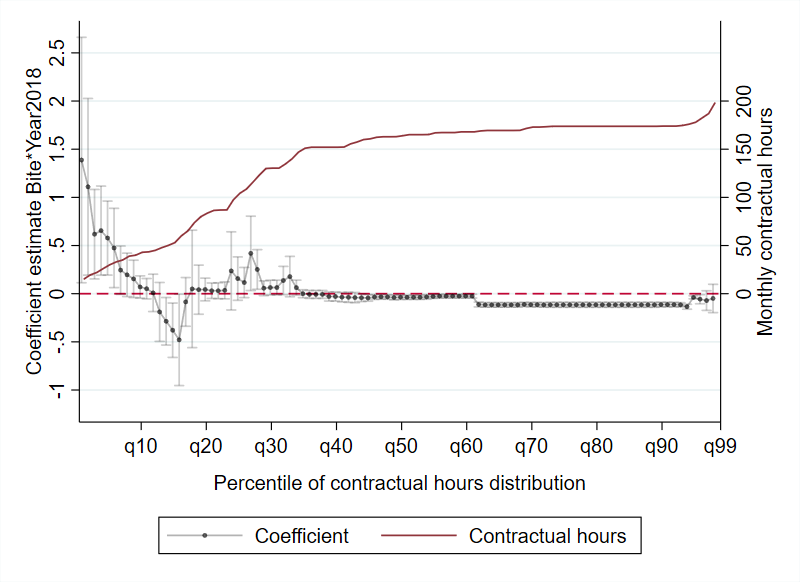}
\subcaption*{
\textit{Notes:} Black dots represent estimates of the treatment effect (the coefficient estimate of $Bite*Year_{2018}$) from difference-in-differences estimations, as specified in equation (1). Dependent variable is the RIF of log (monthly) contractual working hours, defined for the respective percentiles on the x-axis. The coefficient estimate of the first percentile (3.72 with a standard error of 1.94) is not displayed. The X-vector includes the following covariates: age, dummies for gender, education, and industry. All regressions are weighted. Confidence intervals are based on standard errors clustered at the county level. The upward-sloping line represents the contractual working hours at the respective percentile. \\\textit{Data:} SES, 2010, 2014, and 2018, weighted analysis sample, establishments with at least 10 regular employees. }
\end{figure}

To examine further potential heterogeneities of the impact of the minimum wage policy on hours worked, we estimate effects along the unconditional distribution of working hours. Although the differences are not statistically significant, from Table \ref{tab:hours_baseline} it is already prevalent that the hours effect slightly differs when looking at working hours in levels compared with working hours in logarithms. This finding suggests a closer look at effects along the (unconditional) hours distribution in more detail. For this purpose, we estimate RIF regressions of the difference-in-differences specification at each percentile. Again, we analyze the restricted sample of establishments with at least 10 regular employees, allowing us to assess the parallel trends assumption.
Figure~\ref{fig:effect_along_hours_distribution} visualizes the treatment effect at each percentile. The full regression results (including the placebo effects) are reported in Appendix Table~\ref{tab:hours_distribution}.

The results show negative treatment effects in the upper half of the hours distribution, indicating a negative hours effect among full-time workers. However, the effects are small in size, and moreover, some of these estimates come along with a significant placebo interaction for the year 2010 (see Appendix Table~\ref{tab:hours_distribution}). This casts some doubt that the coefficient estimates of the bite interacted with the year 2018 are genuine causal effects of the minimum wage on hours worked. Instead, they may reflect spurious trends in working hours over time. 

The hours effect is significantly positive and monotonously falls at the very bottom of the distribution, between the 1st and 7th percentile. By contrast, we observe a negative hours effect between the 15th and 17th percentile. For these effects, the causal claim is much stronger since the respective placebo effects remain statistically insignificant and are smaller in size relative to the treatment effect after the minimum wage introduction.

The negative hours effect between the 15th and the 17th percentile occurs where employees are working between 50 and 60 hours per month. Given a nominal minimum wage of \euro{8.84} per hour, the interval comprises the minijob threshold. Minijobs are a particularly interesting group of low-income jobs, which are largely subsidized in their personal income taxation as well as the required social security contributions. They are defined by a maximum gross monthly salary of \euro{450}. Hence, in the presence of the minimum wage, minijobbers can no longer work more than 50.9 hours ($\text{\euro{}}450/ \text{\euro}{8.84} \text{ per hour}$). Upgrading the respective minijob to a regular job which has to be paid above \euro{450} can only be avoided by reducing its working hours. According to the regression results, this seems to be the case for a significant number of minijobs. Hence, there is a very intuitive economic explanation for the negative hours effect at this point of the hours distribution which is to preserve the benefits of minijobs.\footnote{Note that minijobs are highly controversial in the public debate (see, for example, \textit{Frankfurter Allgemeine Zeitung}, 22.11.2021, \textit{Der Makel der Minijobs}). While they are characterized by subsidized social security contributions, they also offer little protection from the social security insurances, with little coverage by the unemployment insurance and no mandatory savings in the retirement pension plans. Hence, minijobs are criticized as being precarious. They are defined by a low monthly salary and also experience other disadvantages such as low tenure, low social recognition, and little scope to climb up the job ladder. For a discussion of the pros and cons of minijobs, see also \citeA{Walwei2019}.}

Regarding the positive hours effect at the very bottom of the hours distribution, where 96 percent of jobs are minijobs, we can only speculate about the reasons for the mechanism behind this observation. First, it may be less attractive to hire workers for only very few hours if they have to be paid the minimum wage, causing these jobs to move up the hours distribution.\footnote{If a new hire has to be paid the minimum wage, the hiring cost may increase because the employers want to make sure that the respective hire fulfills the required productivity. To compensate for the increased hiring costs, the firms may wish to extend the respective minijobbers' working hours.} Second, an increase of hours of other minijobs could compensate for the decreasing working hours between the 15th and the 17th percentile, i.e., hours-reallocation among minijobs may explain this effect. Third, jobs at the very bottom of the hours distribution may have been destructed. Thereby, the positive hours effect may have been caused by job destruction rather than a working hours increase within the same jobs.

\begin{figure}[ht!]
\captionabove{Minimum wage effects on log monthly working hours along the hours distribution for \textbf{small establishments}, RIF regression estimates}
\label{fig:effect_along_hours_distribution_small}
\centering
\includegraphics[width=0.9\textwidth]{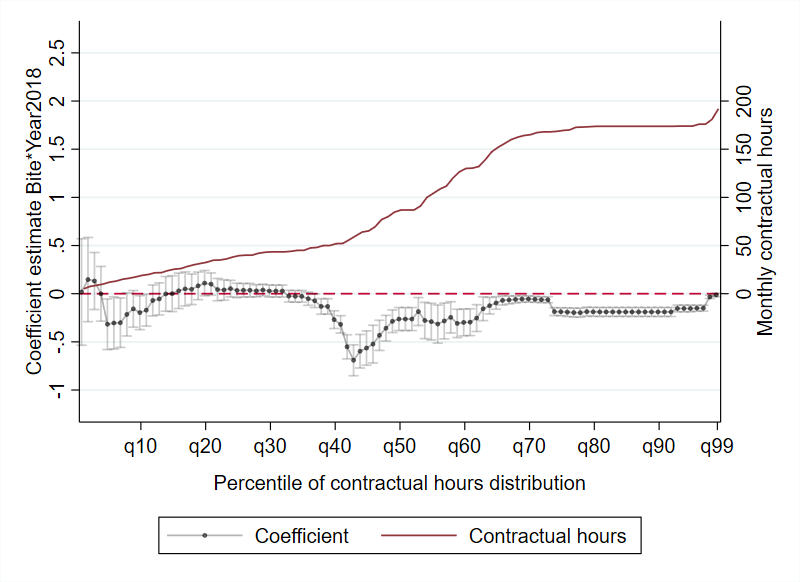}
\subcaption*{
\textit{Notes:} Black dots represent estimates of the treatment effect  (the coefficient estimate of $Bite*Year_{2018}$) from difference-in-differences estimations, as specified in equation (1). Dependent variable is the RIF
of log (monthly) contractual working hours, defined for the respective percentiles on the x-axis.  The X-vector includes the following covariates: age, dummies for gender, education, and industry. All regressions are weighted. Confidence intervals are based on standard errors clustered at the county level. The upward-sloping line represents the contractual working hours at the respective percentile. \\\textit{Data:} SES, 2014 and 2018, weighted analysis sample, establishments with less than 10 regular employees. }
\end{figure}

Given that the analysis in this section shows that the negative hours effect of the minimum wage introduction appears at the minijob threshold, one might speculate that the difference in the hours response across plant size documented in Section~\ref{subsec:hoursmallplants} is due to a different share in minijobbers. Therefore, we repeat the RIF regressions along the hours distribution of establishments with less than 10 regular employees. If a different share of minijobbers solely drives differences across plant size, then the RIF regressions for small establishments should again only show negative effects at those percentiles where (exact) minijobbers are located. The results are displayed in Figure~\ref{fig:effect_along_hours_distribution_small}.

While there are no significant positive effects at the bottom of the distribution anymore, we still observe large negative effects further up the distribution. However, they do not occur between the 15th and the 17th percentile (as in Figure~\ref{fig:effect_along_hours_distribution}) but beyond the 40th percentile. This is perfectly plausible since 50.9 hours (the threshold of minijobbers under the \text{\euro}{8.84} minimum wage) are for small establishments very close to the 40th percentile. Moreover, Figure~\ref{fig:effect_along_hours_distribution_small} displays a negative significant hours effect up to the 60th percentile, where working time amounts to 130 hours. While the share of minijobbers in small establishments is 66 percent at the 40th percentile, it has already fallen to 2.5 percent at the 49th percentile. This clearly suggests that the difference in the hours response of the minimum wage effect across plant size is \textit{not} solely driven by a different share of minijobbers. We rather observe a negative hours effect among the group of regular (part-time) employees in small establishments.

\subsection{Hours adjustment of minijobs}

The analysis along the unconditional hours distribution reported in Section~\ref{sec:hours_hoursdist} already suggests that working hours of minijobs were affected by the minimum wage introduction. However, the reported results refer to all employees including minijobs as well as regular social security jobs. Hence, the significant negative effect at the 16th percentile displayed in Figures~\ref{fig:effect_along_hours_distribution}, for example, could also result from changes in working hours of (well-paid) part-time jobs. To check more directly whether minijobs at the threshold of \euro{450} were affected by the working hours reduction discussed in Section~\ref{sec:hours_hoursdist}, we run our baseline difference-in-differences specification on this particular group of minijobs, i.e., we restrict the sample to jobs with monthly wages between 350 and 450\euro.

\begin{table}[ht!]\centering
\caption{Minimum wage effect on monthly working hours of exact minijobs (monthly wage between \euro{}350 and \euro{}450)}
\label{tab:hour_exact_mini}
	\begin{threeparttable}
	\begin{tabular}{lC{2.5cm}C{2.5cm}C{2.5cm}C{2.5cm}}
\hline\hline
				&\multicolumn{2}{c}{Contractual hours}
				&\multicolumn{2}{c}{Paid hours}\\
				\cmidrule(lr){2-3}\cmidrule(lr){4-5}
				&\multicolumn{1}{c}{hours}
				&\multicolumn{1}{c}{log hours}
				&\multicolumn{1}{c}{hours}
				&\multicolumn{1}{c}{log hours} \\[1ex]
\hline \\[-1.4ex]
				$Bite*Year_{2010}$        &     -3.573         &     -0.082         &     -3.120         &     -0.067         \\
				&    (4.195)         &    (0.094)         &    (4.231)         &    (0.095)         \\[1ex]
				$Bite*Year_{2014}$       & Reference & Reference & Reference & Reference \\[1ex]
				$Bite*Year_{2018}$        &    -37.995\sym{***}&     -0.653\sym{***}&    -38.250\sym{***}&     -0.677\sym{***}\\
				&    (4.068)         &    (0.089)         &    (4.094)         &    (0.086)         \\[1ex]
\hline \\ [-1.4ex]
				Clusters            &         401         &         401         &         401         &       401         \\
				Observations        &     117,684         &     117,684         &     117,684         &     117,684         \\[1ex]
				\hline\hline
			\end{tabular}
			\begin{tablenotes}
				\begin{small} \textit{Notes:} Treatment effect interactions from difference-in-differences estimations, as specified in equation (1). 
    The dependent variable is either the (monthly) contractual or the (monthly) paid working hours in levels or in logarithms, as indicated by column titles.
    The $\boldsymbol{X}$-vector includes the following covariates: age, dummies for gender, education, and industry. All regressions are weighted. Standard errors in parentheses are clustered at the county level. Asterisks indicate significance levels: 
      *~$p<0.05$, **~$p<0.01$, ***~$p<0.001$.
          \\\textit{Data:} SES, 2010, 2014, and 2018, only minijobs with monthly wage between \euro{}350 and \euro{}450, weighted analysis sample, establishments with at least 10 regular employees.
          \end{small}
        \end{tablenotes}
		\end{threeparttable}
\end{table}

\begin{figure}[ht!]
	\captionabove{Histograms of total monthly working hours, 5-hour bins}
\label{fig:histogram_hours_all_establishments}
	\centering
    \vspace{-0.3cm}
	\includegraphics[width=1\textwidth]{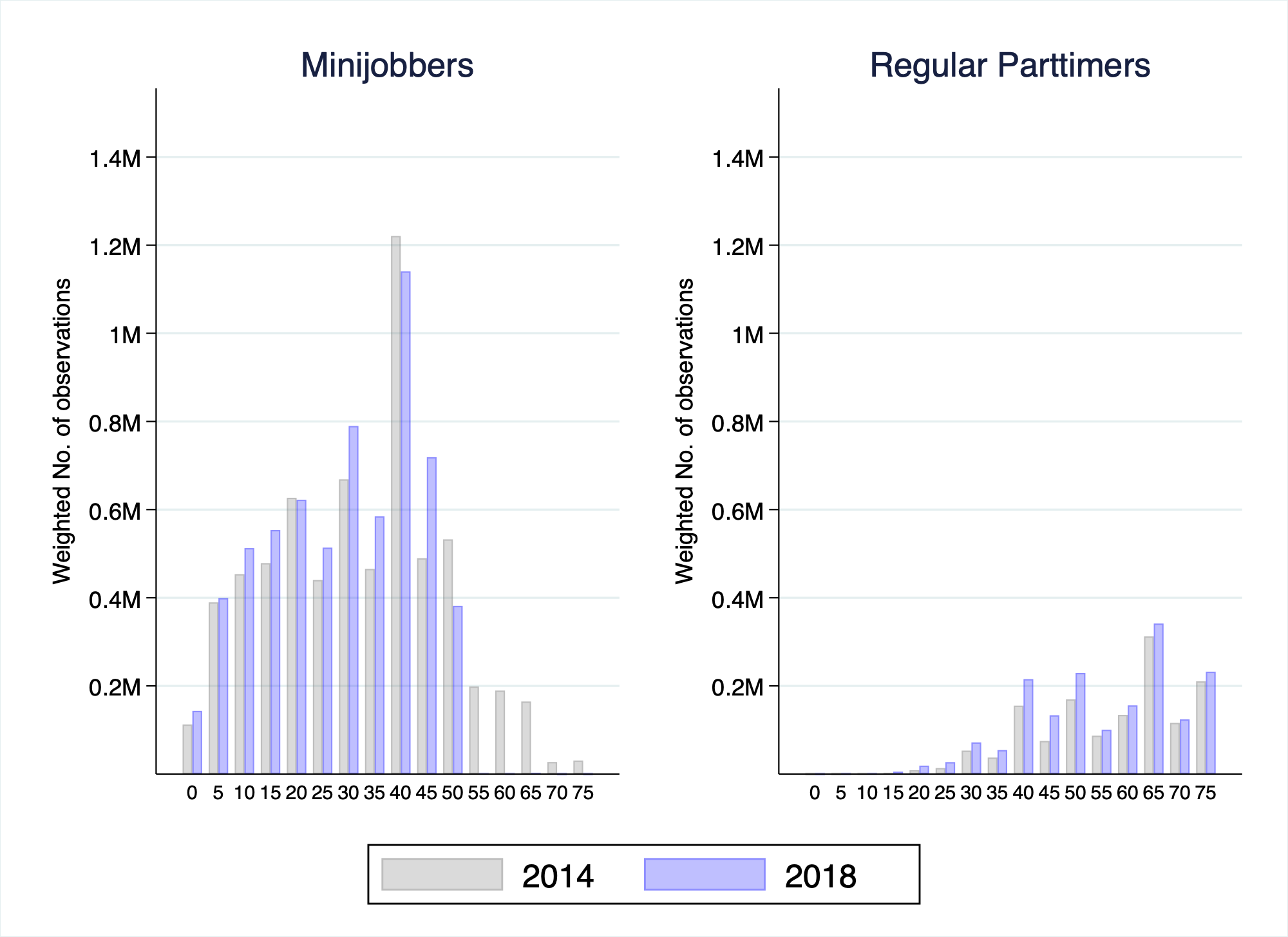}
\subcaption*{
\textit{Notes:} Histogram of monthly working hours of minijobs and part-time jobs, bin width is 5 hours. \\\textit{Data:} SES, 2014 and 2018, weighted analysis sample, all establishments.}
\end{figure}

Table~\ref{tab:hour_exact_mini} shows for minijobbers with a monthly wage between 350 and 450\euro\ a significant reduction in both, monthly contractual working hours and paid working hours. An increase in the average bite of 11.7 percentage points, for example, reduces contractual working time by four hours per month. While the estimates document a significant effect, it should be noted that they could include a bias due to selection in and out of this specific group of jobs. 

However, despite potential biases, the finding of reduced working hours of minijobbers is also corroborated by a descriptive inspection of the hours distribution of minijobbers and part-time workers. These are illustrated in Figure \ref{fig:histogram_hours_all_establishments} for 2014 and 2018. For minijobbers, it is evident that jobs above 55 working hours disappeared entirely. The change in the hours distribution implies that about 600,000 minijobbers with hours beyond 50 per month are no longer observed in 2018.\footnote{For ease of quantitative comparison with the individual-level analysis of minijobber transitions in Section~\ref{sec:employment}, Figure \ref{fig:histogram_hours_all_establishments} is based on all establishments. However, we obtain a very similar figure when restricting the sample to establishments with at least 10 employees, see Appendix Figure~\ref{fig:histogram_hours_10_plus}.}  Of these, 200,000 jobs may have been downgraded to 45-49 hours. Consequently, the other 400,000 jobs must have been either terminated or upgraded to regular (unsubsidized) jobs. The latter possibility, however, is not supported by the working hours distribution of part-time jobs which does not show a significant increase in the respective bins, i.e., between 50 and 70 hours there is no significant increase in regular part-time jobs. Of course, the descriptive distributions can only provide suggestive evidence rather than a decisive answer to the question of whether a significant number of minijobs have been destructed.

\subsection{Employment effects of minijobs}\label{sec:employment}

We finally inspect employment effects, where we look first at the aggregate and then stratify into the subgroups of regular jobs and minijobs to further examine the possibility of job destruction among minijobs. In addition, we investigate individual transitions of minijobbers to inspect whether effects are due to compositional changes in and out of employment, or due to individual transitions between labor market statuses. Hence, the analysis of employment effects may not solely help to provide an answer where all the minijobs went, but also an intuition about potential selectivities since changes in the employment composition can cause effects on the hours and wage distributions even if there are no effects on existing jobs.

We start by analyzing employment effects of the minimum wage policy in a difference-in-differences regression at the regional level, using the full population of administrative employment data, as provided in the Establishment History Panel (BHP) \cite{ganzer2020establishment}.\footnote{The use of administrative employment data from the BHP ensures that it yields exactly the same employment effect as the SES data since we calculate weights for our SES-analysis sample based on employment figures of the BHP (see Section \ref{sec:data}). A description of the BHP is provided in Appendix~\ref{app:BHP}.} As before, we use variation of the minimum wage bite across 401 counties $r$ in Germany interacted with the year 2018 (i.e., post minimum wage) to identify the treatment effect: 
\begin{multline}\label{eq:did_regional}
ln(employment_{rt}) = \tau_{2014}
+ \sum_{k=2014,\,2018} \tau_{k} * Year_{k,t}
+ \theta * Bite_{r} \\
+ \alpha_{DiD}* Bite_{r} * Year_{2018,t}    
 + \rho * Bite_r * Year_{t}
+  \mu_{rt}  \\ t= 2010, 2014, 2018
\end{multline}
Following the literature on employment effects of the German minimum wage, the model includes a bite-specific trend $Bite_r * Year_{t}$ which controls for employment trends that were present already before the minimum wage was introduced \cite{Ahlfeldt2018,Bossler2020,Bossler2023,Dustmann2022}.\footnote{Due to perfect multicollinearity the variable $Bite_{r} * Year_{2010,t}$ cannot be included.} The parallel trends assumption therefore assumes that such pre-determined trends would have continued in the absence of the minimum wage.\footnote{We inspect the necessity to include employment trends in Appendix~\ref{app:employment_trends}.}

\begin{table}[ht!]\centering
		\caption{Minimum wage effect on log employment}
        \label{tab:employment}
        \begin{threeparttable}
				\begin{tabular}{{lcccccc}}
					\hline\hline
					&\multicolumn{2}{c}{all jobs}               &\multicolumn{2}{c}{regular employment}     &\multicolumn{2}{c}{minijobs}               \\\cmidrule(lr){2-3}\cmidrule(lr){4-5}\cmidrule(lr){6-7}
					&\multicolumn{1}{c}{10+ plants}&\multicolumn{1}{c}{all plants}&\multicolumn{1}{c}{10+ plants}&\multicolumn{1}{c}{all plants}&\multicolumn{1}{c}{10+ plants}&\multicolumn{1}{c}{all plants}\\[1ex]\hline \\ [-1.4ex]
					$Bite*Year_{2014}$      & Reference & Reference & Reference & Reference & Reference & Reference \\[1ex]
					$Bite*Year_{2018}$     &      0.032     &      0.008         &     -0.011         &     -0.022         &     -0.243         &     -0.149         \\
					&    (0.054)         &    (0.032)         &    (0.042)         &    (0.029)         &    (0.300)         &    (0.121)         \\[1ex]
\hline \\[-1.4ex]
Clusters            &         401         &         401         &         401         &         401         &         401         &         401         \\
Observations        &        1203         &        1203         &        1203         &        1203         &        1203         &        1203         \\[1ex]
\hline\hline
\end{tabular}
\begin{tablenotes}
\begin{small} \textit{Notes:} Treatment effect interactions from difference-in-differences estimations, as specified in equation~(\ref{eq:did_regional}). The regressions are run at the county-level and weighted by county-level employment. The dependent variable is the logarithmic employment of either all jobs, regular employment, or minijobs, as indicated by column titles. Standard errors in parentheses are clustered at the county level. Asterisks indicate significance levels: *~$p<0.05$, **~$p<0.01$, ***~$p<0.001$.
\\ \textit{Data:} BHP of 2010, 2014, and 2018; full population aggregated to counties. 
\end{small}
\end{tablenotes}
\end{threeparttable}
\end{table}

Table \ref{tab:employment} presents the estimates of the treatment effect $\alpha_{DiD}$. The effect on total employment (all jobs) reported in the first two columns is insignificant irrespective of looking at establishments with at least 10 employees or at all establishments in the economy. This insignificant effect, which is virtually zero in magnitude, is in line with other minimum wage evaluations at the regional level \cite{Ahlfeldt2018, Bossler2023, Dustmann2022}. 

The third and fourth column report the effect of the minimum wage policy on regular employment (again differentiated by establishment size) which turns out to be unaffected, both in terms of statistical significance and regarding the size of the coefficients. According to the latter columns, the coefficients for minijobs are still statistically insignificant, but relatively large in size. It could either be that the effect is largely heterogeneous or the identification strategy has too little power to identify a potentially meaningful effect. 

To address the ambiguity concerning minijobs, we further investigate potential employment effects by examining individual-level labor market transitions of minijobbers. Since the SES does not allow to follow individuals over time, we rely on the German administrative social security data, namely the Integrated Employment Biographies (IEB) of the Institute for Employment Research (IAB). The IEB covers the full population of employment spells except those of civil servants and the self-employed which are by and large irrelevant to the minimum wage policy. We use all employment spells as of 30 June in an annual panel from 2012 to 2015, exempting only apprentices, interns, and individuals with multiple jobs.\footnote{A description of the data is presented in Appendix \ref{app:IEB} and in \citeA{Mueller2020}. The exclusion of multiple jobs restricts our analysis to the group of low-income minijobbers for which the minijob is the only source of labor income. The relevance of side jobs for individuals affected by the minimum wage is addressed in \citeA{VomBerge2023}.} While the IEB allows for a comprehensive analysis of labor market transitions, it does not include information on hours worked. Therefore, we merge the county-specific bite from our main data source, the SES.

We estimate the following difference-in-differences specification: 
\begin{multline}\label{eq:minijob_trans}
T(status_{t} \rightarrow status_{t+1})_{it} = 
\kappa_{2012} 
+ \sum_{k=2013,  \, 2014} \kappa_{k} * Year_{k,t} 
+ \psi * Bite_{r}
\\+ \sum_{ k=2013,  \, 2014   } \pi_{k}* Bite_{r} * Year_{k,t}  +  \nu_{rt} \\ t= 2012, 2013, 2014
\end{multline}
The dependent variable $T(status_{t} \rightarrow status_{t+1})_{it}$ indicates whether or not individual $i$ has transited between two exclusively defined labor market statuses between periods $t$ and $t+1$. If $T_{it}$ denotes transitions out of minijobs, it is defined for all minijobbers in period $t$ and takes the value $0$ if an individual remains in a minijob in $t+1$, while it takes the value $1$ if an individual is no longer in a minijob in $t+1$.\footnote{In our transition analysis, we look at yearly changes because tenure in minijobs is typically low. Hence, the transition between 2014 and 2015 should be relevant to capture the treatment effect. For the pre-treatment period, we restrict our sample to the years 2012--2014 due to a major break in the definitions of the administrative data.} In the latter case, the individual may have entered non-employment, a regular social security job, or in rare cases it may have entered one of the excluded labor market statuses.\footnote{The status non-employment captures individuals who are registered as unemployed or individuals who are non-employed but who have not registered as unemployed at the local employment agency either because they do not fulfill the eligibility criteria to receive benefits or because they do not want to register as unemployed for other reasons. In fact, the literature documents far from complete take-up \cite{Bruckmeier2021} of benefits, which is why we prefer to analyze transitions in non-employment rather than transitions in registered unemployment. However, in rare cases, transitions to non-employment could also imply transitions into retirement or to an education program.} Note that the coefficient $\pi_{2014}$ of the interaction term $Bite_r * Year_{2014}$ denotes the treatment effect, since observations of $T_{it}$ in 2014 capture transitions between 2014 and 2015 and therefore the date of the minimum wage introduction (1 January 2015). Standard errors are clustered at the county level because the bite again varies by county.

\begin{table}[ht!]\centering
\caption{Minimum wage effect on forward-looking transitions out of minijobs}
\label{tab:transition_mini}
\begin{threeparttable}
\begin{tabular}{lC{3.7cm}C{3.7cm}C{3.7cm}}
\hline\hline
&\multicolumn{1}{c}{out of minijob}&\multicolumn{1}{c}{minijob to regular job}&\multicolumn{1}{c}{minijob to non-empl.}\\
\hline \\ [-1.4ex]
$Bite * Year_{2012}$ & Reference & Reference & Reference \\[1ex]
$Bite * Year_{2013}$ & -0.018\sym{*}   & -0.003         &  -0.014\sym{*}  \\
	            & (0.007)            &    (0.004)     &    (0.006)      \\
$Bite * Year_{2014}$ & 0.169\sym{***}  &  0.081\sym{***}&   0.088\sym{**} \\
	          & (0.027)            &    (0.007)     &    (0.027)     \\[1ex]
\hline \\ [-1.4ex]
Clusters   &    401         &         401    &         401      \\
Observations  & 13,679,174  &    13,679,174  &    13,679,174    \\
Minijobs in 2014 & 4,542,156 &  4,542,156    &  4,542,156       \\[1ex]
\hline\hline
\end{tabular}
\begin{tablenotes}
\begin{small} 
\textit{Notes:} Treatment effect interactions from difference-in-differences estimations, as specified in equation~(\ref{eq:minijob_trans}). 
The dependent variable captures forward-looking transitions out of minijobs, as indicated by column titles. 
Standard errors in parentheses are clustered at the county level. Asterisks indicate significance levels: *~$p<0.05$, **~$p<0.01$, ***~$p<0.001$.\\ \textit{Data:} IEB, 2012-2015, population of all minijobbers, multiple jobs excluded.
\end{small}
\end{tablenotes}
\end{threeparttable}
\end{table}

The regression results are presented in Table \ref{tab:transition_mini}.\footnote{For our transition analyses, we look at the full population of all minijobs, allowing us to infer absolute numbers of transitions in regular employment and non-employment. However, the estimates remain fully robust when analyzing the transitions with the subsample of establishments with more than 10 employees, as reported in Appendix \ref{app:transitions_10+}.} According to the first column, the treatment effect presents a positive and significant increase in the probability of leaving a minijob.\footnote{Appendix \ref{app:in_minijobs} shows that the effects of transitions out of minijobs are not offset by excess transitions in minijobs.} In absolute size, the coefficient closely matches the coefficient of the last column of Table \ref{tab:employment}. While the standard errors are still clustered at the regional level (the variation of the treatment), the coefficient capturing labor market transitions is very precisely estimated. Also, note the placebo coefficient of the year before the treatment is close to zero. Given the average regional bite of 0.135 (weighted by the sample used in Table~\ref{tab:transition_mini}), the treatment effect of 0.169 implies that an excess of 2 percent of minijobs entered another labor market status between 2014 and 2015, corresponding to 104,000 minijobs that disappeared. This number of destructed minijobs falls short of the descriptively calculated 400,000 minijobs that disappeared from the hours distributions, as illustrated by Figure \ref{fig:histogram_hours_all_establishments}.\footnote{The discrepancy is even larger since we checked whether the 104,000 causal transitions out of minijobs are entirely driven by those who worked long hours (i.e., $\geq 55 \text{ hours}$, which is --by visual inspection-- the group that disappeared from the histogram in Figure \ref{fig:histogram_hours_all_establishments}). In fact, we also observe significant transitions from minijobs to non-employment even among minijobbers who worked fewer hours, suggesting that there is also a genuine labor demand effect, see Appendix \ref{app:additional}.} There are several possible explanations for this discrepancy. It could be that the descriptive inspection is misleading because it does not necessarily reflect a causal effect of the minimum wage. Moreover, the descriptive inspection is not conclusive about transitions. It is certainly possible that these jobs were not destroyed, but instead moved to various other parts of the distribution. There is also a possibility that minijobs were destroyed beyond individual transitions out of minijobs. This is because the usual tenure in a minijob is short, and terminated minijobs may simply not be renewed by hiring a replacement. However, we can show that the transitions out of minijobs are not accompanied by decreased transitions in minijobs, see Appendix \ref{app:in_minijobs}. 

In the remaining two columns of Table \ref{tab:transition_mini}, we disentangle the transitions out of minijobs into transitions in regular jobs (second column) and into transitions in non-employment (third column). It turns out that about half of the employees who moved out of minijobs entered regular jobs, i.e., 0.81 percent of all minijobbers if the bite increases by 10 percent. Hence, these employees experienced a promotion to socially insured jobs. Note, however, that the promotions from minijobs to regular jobs are partially offset by evidence of increased demotions from regular jobs to minijobs, as presented in Appendix Table \ref{tab:in_minijobs}. Most interestingly, as indicated in the third column, the treatment effect on transitions from minijobs to non-employment is 0.088. Given the average bite of 13.5 in the sample of minijobbers, about 1 percent of all minijobs entered non-employment as a result of the minimum wage introduction. In absolute terms, it implies that 54,000 minijobs were destructed.\footnote{These results remain fully robust when we add covariates to the regression specification, as presented in Appendix \ref{app:transitions_controls}. If anything, the treatment effect on transitions out of minijobs increases slightly in size.} To convert this number into an economic parameter, we calculate an employment elasticity with respect to wages, i.e., 
\[
\mbox{\Large $\eta$}_{\text{employment, wage}} = \frac{\partial \ln \text{employment}}{\partial \ln \text{wage}} = \frac{\partial \ln \text{employment} \big/ \partial \text{ minimum wage}}{\partial \ln \text{wage} \big/ \partial \text{ minimum wage} }
\]
where the right part of the equation can be inferred from our regression results. The numerator is the disemployment effect of minijobs of 0.088 and the denominator can be obtained from Panel~B of Table~\ref{table:Rif_baseline}, where the percentile estimates 4 to 16 correspond to the wage effect of minijobs, the average of which is 0.795. Together, these estimates result in an employment elasticity of $-0.11$. The corresponding elasticity for regular jobs is zero, given the zero employment effect reported in Table \ref{tab:employment}. 

For those employees who were upgraded from minijobs to regular jobs (second column), the question emerges whether they were promoted internally (within the same employer) or externally (by switching employers).\footnote{The German administrative employment data contain a unique identifier for establishments, which are local units, but no information on the employing enterprise. Hence, we can identify internal promotions to regular jobs in a narrow sense.} At first sight, it may be intuitive that minijobbers are promoted internally, simply because of a monthly wage increase due to the minimum wage the respective employees' then exceed the minijob limit. However, external promotions of minijobbers are certainly possible, because minijobs are typically characterized by low job tenure and frequent job changes.\footnote{Using the IEB data between 2012 and 2015, we observe that 48 percent of the minijobbers left their jobs each year, which is defined as either leaving the respective establishment or leaving the minijob within the establishment.} Due to turnover, it could be possible that some minijobs are destructed and the respective employees may find a regular job at another employer, especially since the level of unsatisfied labor demand, as measured by vacancies was growing in the years after the minimum wage introduction \cite{Bossler2023b}. 

The possibility of external upgrading of jobs is also closely linked to the literature. \citeA{Dustmann2022} show that employees affected by the German minimum wage introduction reallocated from lower-paying employers to higher-paying employers. It is interesting to see whether this upgrading of jobs through reallocation corresponds with the observed promotions of employees from minijobs to regular jobs.

\begin{table}[ht!]\centering
\caption{Minimum wage effect on minijobber promotions within and across establishments}
\label{tab:transition_mini_within}
\begin{threeparttable}
\begin{tabular}{lC{4.4cm}C{4.4cm}}
\hline\hline
 & \multicolumn{2}{c}{minijob to regular job}\\
                    &\multicolumn{1}{c}{within establishments}&\multicolumn{1}{c}{across establishments}\\[1ex]
\hline \\[-1.4ex]
$Bite * Year_{2012}$ & Reference  & Reference     \\[1ex]
$Bite * Year_{2013}$ &      -0.002         &      -0.002         \\
                    &     (0.003)         &     (0.003)         \\
$Bite * Year_{2014}$ &       0.078\sym{***}&       0.002         \\
                    &     (0.006)         &     (0.004)         \\[1ex]
\hline \\[-1.4ex]
Clusters   &    401         &         401       \\
Observations        &    13,679,174         &    13,679,174  \\
Minijobs in 2014 & 4,542,156 &  4,542,156       \\[1ex]
\hline\hline
\end{tabular}
\begin{tablenotes}
\begin{small} \textit{Notes:} Treatment effect interactions from difference-in-differences estimations, as specified in equation~(\ref{eq:minijob_trans}).
The dependent variable captures forward-looking transitions out of minijobs, as indicated by column titles. 
Standard errors in parentheses are clustered at the county level. Asterisks indicate significance levels: *~$p<0.05$, **~$p<0.01$, ***~$p<0.001$. \\ \textit{Data:} IEB, 2012-2015, population of all minijobbers, multiple jobs excluded.
\end{small}
\end{tablenotes}
\end{threeparttable}
\end{table}

Based on the reasoning above, we decompose upgrading to a regular job into the outcome variables (1) promotion to regular jobs within the establishment and (2) promotion to regular jobs across establishments. The results of these regressions are reported in Table \ref{tab:transition_mini_within}. Interestingly, all promotions to regular jobs due to the minimum wage introduction are observed within establishments.\footnote{Plausibly, these promotions to regular jobs are mostly observed among minijobbers who worked long hours initially (see Appendix \ref{app:additional}). This is in line with the institutional need to promote these individuals since long minijob hours can no longer exist after the minimum wage was introduced.} Hence, the upgrading of minijobs is unrelated to the reallocation effect detected in \citeA{Dustmann2022}. The internal promotions of minijobs are in line with results in \citeA{Bossler2023}, which show a significant positive wage effect for existing jobs. Note, however, that this monthly wage effect is much smaller (although still meaningful and statistically significant) at the minijob threshold, as implied by the estimate for 12th percentile reported in Table 3.

\begin{figure}[ht!]
\captionabove{Effect heterogeneities for transitions out of minijobs}\label{fig:heterogeneities}
\centering
\begin{subfigure}[a]{0.70\textwidth}	
\centering
\caption{Transitions in regular jobs}
\includegraphics[width=\textwidth]{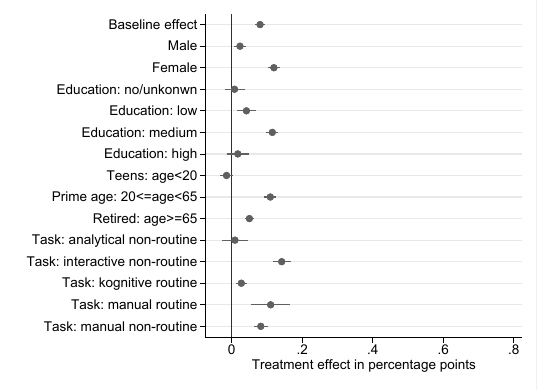}
\end{subfigure}
\hfill
\begin{subfigure}[b]{0.70\textwidth}	
\centering
\caption{Transitions in non-employment}
\includegraphics[width=\textwidth]{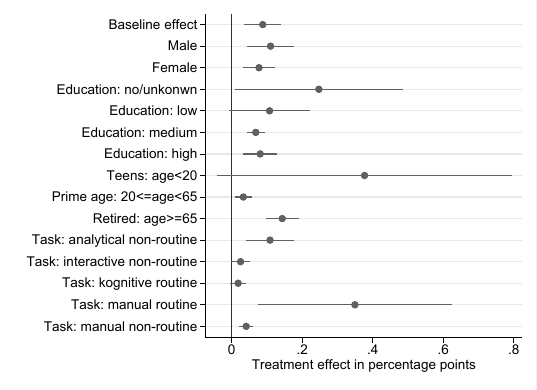}
\end{subfigure}
\subcaption*{
\textit{Notes:} Transition regressions in regular jobs and non-employment respectively, as specified in the second and third column of Table \ref{tab:transition_mini}. The heterogeneities are estimated from samples that are split by the respective worker and job characteristics. No/unknown education denotes no secondary or unknown education; low education denotes neither post-secondary education nor vocational training; medium education denotes vocational training; high education denotes master craftsman / technician or university degree. \\\textit{Data:} IEB, 2012-2015, population of all minijobbers, multiple jobs excluded.}
\end{figure}

To assess whether upgrading to a regular job respectively whether the transition of a minijobber to non-employment after the minimum wage introduction depends on socio-demographic characteristics and job task, we also run regressions by employee subgroups, as shown in Figure \ref{fig:heterogeneities}. This subgroup analysis helps us characterize the 54,000 minijobbers who lost their jobs due to the minimum wage.\footnote{Supplementary heterogeneities by establishment size, the establishments' AKM wage effect, and industry are presented in Appendix \ref{app:transtions_establ_heterogeneities}.} The individuals' gender might be important, as the literature has shown that the minimum wage disproportionately affected the wages of females to increase \cite{Caliendo2022b}. Interestingly, we do not observe a significant gender difference for transitions in non-employment, but females more likely received an upgrade to regular jobs, which is in line with the disproportionate female effect documented in the literature \cite{Caliendo2022b}. 

Concerning education, the results indicate that individuals without or with only a low levels of education are most likely to enter non-employment due to the minimum wage. At the same time, upgrades to regular jobs are more likely among the medium-skilled workers. This educational heterogeneity is very much in line with a simple argument that education is a proxy for productivity. Hence, the least productive minijobbers face the highest risk of job loss.

When looking at the age of minijobbers, the results show that prime-age workers are most likely promoted to regular jobs. By contrast, teenagers and elderly workers (above retirement age) face a much higher risk of transitioning to non-employment. However, we have to note that the age-specific heterogeneities regarding non-employment transitions are inaccurately estimated (with large point estimates). The age-specific heterogeneities show that individuals with other sources of income who use the minijob as supplementary income, such as teens or retired individuals, face the highest risk of job loss. The prime-age workers, who most likely have to rely on the minijob as their main income source, by contrast, have good chances of receiving a promotion due to the minimum wage. The age-specific non-employment transitions closely match the descriptive finding of \citeA{VomBerge2017}, who show that the termination of minijobs was relatively more likely among young and old workers compared with prime-age workers. 

Finally, we inspect heterogeneities by the major task, as indicated by the occupation of the minijobs. Manual routine minijobs are by far the most likely to enter non-employment. This finding is in line with the assumption that manual routine jobs face the highest risk of being replaced (potentially by other input factors).

\section{Conclusion}\label{sec:conclusion}

Using identifying variation from a regional bite variable in a difference-in-differences framework, we document compelling evidence that the 2015 introduction of the minimum wage in Germany induced significant wage increases in hourly and monthly wages. Thereby, the minimum wage contributed to a significant reduction in both hourly and monthly wage inequality. The wage effects are most pronounced at the bottom of the respective distributions but extend up to the median, implying significant spillover effects beyond the workers directly affected by the policy. Furthermore, the minimum wage effect on monthly (but not hourly) wages follows a hump-shaped pattern along the distribution, with a local minimum at the 12th percentile. This is where the minijob threshold of \euro{450} is located, demonstrating that this institutional threshold suppresses the effect of the minimum wage. All jobs below \euro{450} are minijobs, which are exempted from income taxation and only require reduced social security contributions. There are supply- and demand-side incentives to keep jobs below this threshold, which clearly hampers the effect of the minimum wage.

On average, contractual and paid working hours are hardly affected by the minimum wage introduction, corroborating the findings by \citeA{Biewen2022}. However, we document two important heterogeneities in the hours response. First, we show a negative hours effect among small establishments which are disproportionately affected by the minimum wage. Second, we observe a negative hours effect for minijobs. While small establishments have a larger share of minijobbers, we show that the negative hours response in small establishments goes beyond percentiles where minijobbers are located.

We observe clear indications that minijobs with more than 50 hours were eliminated. This observation is very intuitive since in the presence of an hourly minimum wage and the definition of minijobs (by a monthly wage up to \euro{450}). Hence, the number of working hours for minijobs was effectively limited by the minimum wage legislation. A central question remains whether the respective minijobs were terminated or whether they were upgraded to regular social security jobs. The empirical literature is not conclusive concerning employment effects of minijobs. While some studies show that minijobs reduced \cite{Caliendo2018}, indicating a negative employment effect, others show that regular jobs have increased in compensation \cite{Garloff2019}, suggesting an upgrade of minijobs. Finally, studies that look at employment as a whole do not identify an overall employment effect, indirectly confirming the latter argument \cite{Ahlfeldt2018, Bossler2023, Dustmann2022}. We are the first to causally examine individual-level transitions of minijobbers using social security data. Our results show that about half of the destructed minijobs were upgraded to regular social security jobs, while half of the individuals entered non-employment. This corresponds to 54,000 minijobbers entering non-employment, implying an employment elasticity with respect to wages of $-0.1$ for minijobs but a zero elasticity for regular jobs. 

Clearly, our results suggest that the institutional setting leads to heterogeneous effects of the minimum wage. Effects on working hours and transitions in non-employment are not revealed when analyzing the aggregate workforce. However, significant non-negligible effects are uncovered when looking at the particular subgroup of minijobs.

\clearpage
\begin{onehalfspacing}
\bibliographystyle{chicago}
\bibliography{_mw_hours.bib}
\end{onehalfspacing}

\clearpage
\begin{appendix}

\begin{center}

\Large
Online Appendix for \\
\vspace{0.6cm}

The Devil is in the Details: \\ Heterogeneous Effects of the German Minimum Wage \\ on Working Hours and Minijobs

\vspace{0.6cm}
\normalsize
by \\
Mario Bossler, Ying Liang, and Thorsten Schank

\vspace{1.0cm}
\Large 
\textbf{Content}
\normalsize

\singlespacing
\startcontents[sections] 
\printcontents[sections]{l}{1}{\setcounter{tocdepth}{2}} 

\end{center}

\clearpage


\section{Description of additional data}
\label{app:data_description}
\renewcommand*{\thefigure}{\thesection\arabic{figure}}
\renewcommand*{\thetable}{\thesection\arabic{table}}
\setcounter{figure}{0}
\setcounter{table}{0}

\subsection{Establishment History Panel (BHP)}
\label{app:BHP}

The Establishment History Panel is provided by the Institute for Employment Research (IAB). It is an annual establishment-level panel dataset derived from an aggregation of the mandatory reports of each employer on each employee to the German social insurance system. Thus, the data cover all establishments with at least one employee subject to social insurance. The panel information aggregates establishment-level snapshots of the social security data as of 30 June each year. As a result, the BHP covers the universe of German establishments and aggregates daily accurate individual social security information at the level of local establishments. A description of the data is provided by \citeA{ganzer2020establishment}. 

The data contain some structural information on each establishment, such as the industry code of its main economic activity and its location. The location is crucial for us because we identify treatment effects based on regional variation defined at the level of counties, of which there are 402 in Germany (so-called ``Kreise''). The data also include information on employment, i.e., the total number of employees on 30 June each year and the composition of employment. For our analysis of minijobs, we exploit the information on the number of minijobs and regular social security jobs (excluding apprentices). As the data come from the social security system, it excludes civil servants, the self-employed, and family workers who are exempt from social security contributions. However, for these groups of workers, the minimum wage is either not applicable (self-employed, family workers) or mostly irrelevant (civil servants).

We use the BHP in two ways. First, we use it to compute sampling weights for the SES survey data to make them fully consistent with the administrative population of employees in Germany, see Section \ref{sec:data}. Second, we use the BHP when estimating aggregate employment effects in Table \ref{tab:employment_full}. Since the BHP covers the full administrative employment-population (at the establishment level), we use it to construct county-level employment figures, which are then logged and used as dependent variables.

\subsection{Integrated Employment Biographies (IEB)}
\label{app:IEB}

The Integrated Employment Biographies (IEB) are administrative social security data hosted by the Institute for Employment Research (IAB). The IEB are compiled from several sources. These sources include employment records from mandatory employer reports on each employee to the German social security system and records on individuals receiving unemployment benefits or participating in labor market programs. Detailed information can be found in \citeA{Mueller2020}. 

We use the employment information of the IEB, which covers almost the entire working population in Germany, except for civil servants, the self-employed, and family workers.\footnote{In addition, we exclude apprentices who are exempted from the minimum wage.} Employers are required by law to provide employment information on all employees covered by the social security system at least once a year. The employment data include information on each employment spell with start and end dates, total earnings, part-time or full-time status, industry, and occupation. Additional information such as age, sex, and highest level of education is also available. In addition, each individual in the IEB has a unique personal identifier that allows us to track their employment transitions over time.

We construct an annual data panel from 2012 to 2015 that includes all employment spells on 30 June of each year. To define exclusive employment statuses, we exclude individuals who have jobs in multiple establishments on 30 June. We can then classify each individual into regular jobs and minijobs. After constructing these employment statuses, we can generate forward-looking transitions between $t$ and $t+1$: $T(status_{t} \rightarrow status_{t+1})_{it}$. These transitions allow us to analyze at the individual level whether minijobbers stayed in a minijob, were upgraded to a regular social security job, or moved into non-employment, which allows us to estimate equation~(\ref{eq:minijob_trans}).

Until 2014, the IEB additionally includes information on working hours from the compulsory injury insurance. For minimum wage analyses, this information has been used in \citeA{Bossler2023} and \citeA{Dustmann2022}. Unfortunately, employers were allowed to report either actual hours or contractual hours, and there is no information on which of the two is reported. Hence, we apply the heuristic outlined in \citeA{Berge2023} to harmonize the hours information to depict contractual hours plus overtime. We use the working hours information to estimate transitions of minijobs depending on their initial working time, presented in Appendix \ref{app:additional}.

\clearpage
\section{Extended regression tables}\label{app:full_tables}
\renewcommand*{\thefigure}{\thesection\arabic{figure}}
\renewcommand*{\thetable}{\thesection\arabic{table}}
\setcounter{figure}{0}
\setcounter{table}{0}

\begin{table}[ht!]\centering
	\caption{Minimum wage effect on the mean and the variance of log hourly and monthly wages, Table \ref{tab:baseline} amended by all DiD-coefficients}
	\label{tab:baseline_full}
		\begin{threeparttable}
			\begin{tabular}{lC{2.5cm}C{2.5cm}C{2.5cm}C{2.5cm}}
				\hline\hline
    			&\multicolumn{2}{c}{log hourly wage}
				&\multicolumn{2}{c}{log monthly wage}\\
				\cmidrule(lr){2-3}\cmidrule(lr){4-5}
				&\multicolumn{1}{c}{mean}&\multicolumn{1}{c}{variance}&\multicolumn{1}{c}{mean}&\multicolumn{1}{c}{variance}\\[1ex]
\hline \\ [-1.4ex]
 $Bite*Year_{2010}$ &     -0.042         &     -0.017           &     -0.012         &     -0.102         \\        
                    &    (0.044)         &    (0.045)           &    (0.080)         &    (0.256)         \\[1ex]   
$Bite*Year_{2014}$  & \multicolumn{4}{c}{Reference}                                                        \\[1ex]  
$Bite*Year_{2018}$  &      0.456\sym{***}&     -0.654\sym{***}  &      0.499\sym{***}&     -1.196\sym{***} \\       
                    &    (0.033)         &    (0.034)           &    (0.072)         &    (0.242)          \\[1ex]  
$Year_{2010}$       &      0.014\sym{*}  &      0.053\sym{***}  &     -0.004         &      0.044          \\       
                    &    (0.005)         &    (0.005)           &    (0.009)         &    (0.023)          \\[1ex]  
$Year_{2014}$       & \multicolumn{4}{c}{Reference}                                                        \\[1ex]  
$Year_{2018}$       &      0.018\sym{***}&      0.019\sym{***}  &      0.001         &      0.062\sym{**}  \\       
                    &    (0.004)         &    (0.004)           &    (0.007)         &    (0.021)          \\[1ex]  
$Bite$              &     -1.841\sym{***}&      0.215\sym{***}  &     -1.547\sym{***}&     -0.133          \\       
                    &    (0.087)         &    (0.050)           &    (0.156)         &    (0.277)          \\[1ex]  
\hline\\[-1.4ex]
Clusters            &         401         &         401         &         401         &      401          \\
Observations        &   2,667,593         &   2,667,593         &   2,667,593         &   2,667,593         \\[1ex]
\hline\hline
\end{tabular}
\begin{tablenotes}
\begin{small} \textit{Notes:} Difference-in-differences regression coefficients as specified in equation (1). 
The dependent variable is either the log hourly or the log monthly wage or the RIF of the variance of log hourly or log monthly wages, as indicated by column titles. Estimate of the constant not displayed.
The $\boldsymbol{X}$-vector includes the following covariates: age, dummies for gender, education, and industry. All regressions are weighted. Standard errors in parentheses are clustered at the county level. Asterisks indicate significance levels: 
   *~$p<0.05$, **~$p<0.01$, ***~$p<0.001$.
\\\textit{Data:} SES, 2010, 2014, and 2018, weighted analysis sample, establishments with at least 10 regular employees.
\end{small}
\end{tablenotes}
\end{threeparttable}
\end{table}

\begin{landscape}
		\begin{table}[ht!]\centering
			\caption{RIF regressions along the hourly wage distribution, Panel A of Table \ref{table:Rif_baseline}, amended by all DiD-coefficients}
           \label{table:Rif_baseline_hour_full}
				\begin{threeparttable}
				\begin{tabular}{L{2.7cm}C{1.8cm}C{1.8cm}C{1.8cm}C{1.8cm}C{1.8cm}C{1.8cm}C{1.8cm}C{1.8cm}C{1.8cm}}
				\hline\hline
			      &\multicolumn{1}{c}{q4}&\multicolumn{1}{c}{q8}&\multicolumn{1}{c}{q12}&\multicolumn{1}{c}{q16}&\multicolumn{1}{c}{q20}&\multicolumn{1}{c}{q30}&\multicolumn{1}{c}{q50}&\multicolumn{1}{c}{q70}&\multicolumn{1}{c}{q90}\\[1ex]
\hline \\ [-1.4ex]
$Bite*Year_{2010}$  & -0.328\sym{*}   &  0.135\sym{**} &  0.144\sym{**} &  0.099\sym{*} &  0.084  &      0.008         &    -0.113\sym{*}  &      -0.155\sym{**} &   0.043 \\
			      &  (0.160)        &  (0.048)       &  (0.047)       &  (0.048)      &  (0.046)  &  (0.064)  &  (0.052)        &  (0.055)  &  (0.062)   \\[1ex]
$Bite*Year_{2014}$  & \multicolumn{9}{c}{Reference} \\[1ex]
$Bite*Year_{2018}$  &  2.717\sym{***} &  1.465\sym{***} & 1.089\sym{***} & 0.854\sym{***} & 0.775\sym{***}  &      0.715\sym{***}&      0.237\sym{***} &    0.015    &  -0.254\sym{***}\\
			      &  (0.092)  &    (0.037)  &  (0.040)  &  (0.042)  &    (0.046)  &    (0.061)         &    (0.049)         &    (0.040)         &   (0.067)          \\[1ex]
	$Year_{2010}$     &     -0.056\sym{***} &     -0.037\sym{***}&     -0.033\sym{***} & -0.020\sym{***}& -0.022\sym{***}        &      0.002    &     0.035\sym{***}&       0.050\sym{***}&         0.048\sym{***}\\
 	               &     (0.016)    &    (0.005)     &    (0.005)    &    (0.005)     &    (0.005)     &    (0.007)   &    (0.006)       &    (0.007)       &    (0.008)         \\[1ex]
 $Year_{2014}$       & \multicolumn{9}{c}{Reference} \\[1ex]
 $Year_{2018}$       &     -0.008   &      0.001     &     -0.019\sym{***}&   0.014\sym{**} &     -0.008         &     -0.005         &      0.018\sym{***}&        0.033\sym{***}&         0.054\sym{***}\\
		          &    (0.009)    &    (0.004)  &    (0.004)   &    (0.004)      &    (0.005)  &    (0.007)          &    (0.006)             &    (0.005)           &    (0.009)         \\[1ex]
 	$Bite$    &      -2.824\sym{***}&        -1.526\sym{***}&     -1.528\sym{***}&       -1.574\sym{***}&     -1.580\sym{***}    &     -2.274\sym{***}&     -1.881\sym{***}&        -1.681\sym{***}&         -1.580\sym{***}\\
			     &    (0.096)   &    (0.038)  &    (0.043)  &    (0.056)     &    (0.063)     &    (0.105)         &    (0.101)         &    (0.104)       &    (0.134)   \\[1ex]
\hline \\ [-1.4ex]
Clusters            &         401         &         401         &         401         &        401		&         401  &         401         &         401         &        401		&         401         \\		
Observations        &  2,667,593  &  2,667,593  &  2,667,593  &  2,667,593  &  2,667,593  &  2,667,593  &  2,667,593  &  2,667,593  &  2,667,593          \\[1ex]
			\hline\hline
					\end{tabular}
					\begin{tablenotes}
\begin{small} \textit{Notes:} Difference-in-differences regression coefficients as specified in equation (1). The dependent variable is the RIF of log hourly wages defined for various percentiles, as indicated by column titles. 
Estimate of the constant not displayed.
The $\boldsymbol{X}$-vector includes the following covariates: age, dummies for gender, education, and industry. All regressions are weighted. Standard errors in parentheses are clustered at the county level. Asterisks indicate significance levels: *~$p<0.05$, **~$p<0.01$, ***~$p<0.001$.
\\\textit{Data:} SES, 2010, 2014, and 2018, weighted analysis sample, establishments with at least 10 regular employees.
\end{small}
					\end{tablenotes}
				\end{threeparttable}
		\end{table}

\end{landscape}

\begin{landscape}

		\begin{table}[ht!]\centering
			\caption{RIF regressions along the monthly wage distribution, Panel B of Table \ref{table:Rif_baseline}, amended by all DiD-coefficients}
            \label{table:Rif_baseline_month_full}
				\begin{threeparttable}
				\begin{tabular}{L{2.7cm}C{1.8cm}C{1.8cm}C{1.8cm}C{1.8cm}C{1.8cm}C{1.8cm}C{1.8cm}C{1.8cm}C{1.8cm}}
				\hline\hline
			      &\multicolumn{1}{c}{q4}&\multicolumn{1}{c}{q8}&\multicolumn{1}{c}{q12}&\multicolumn{1}{c}{q16}&\multicolumn{1}{c}{q20}&\multicolumn{1}{c}{q30}&\multicolumn{1}{c}{q50}&\multicolumn{1}{c}{q70}&\multicolumn{1}{c}{q90}\\[1ex]
\hline \\ [-1.4ex]

$Bite*Year_{2010}$  & 0.015    &  0.084     &  -0.023    &   0.108    &   0.145 &   -0.030  &  -0.035  &  -0.105  &  0.032 \\
					& (0.424)  &  (0.177)   &  (0.054)   &   (0.083)  &  (0.275) &  (0.097) &  (0.058) &  (0.057) & (0.064)\\[1ex]
$Bite*Year_{2014}$       & \multicolumn{9}{c}{Reference} \\[1ex]
$Bite*Year_{2018}$  & 2.072\sym{***} & 0.705\sym{***} & 0.155\sym{**} & 0.249\sym{***} & 1.356\sym{***} & 1.217\sym{***}&      0.324\sym{***}&  0.023  &  -0.241\sym{***}\\[1ex]
					&  (0.390) & (0.175) & (0.049) & (0.071) & (0.247) & (0.086) & (0.051) & (0.038) & (0.066) \\[1ex]
 	$Year_{2010}$            &      0.049           &     -0.047\sym{*}  &       -0.002         &         -0.132\sym{***}&      -0.157\sym{***}  &     -0.036\sym{**} &         0.006         &       0.030\sym{***}&          0.048\sym{***}\\
      &    (0.038)           &    (0.020)          &    (0.007)            &    (0.011)         &    (0.035) 	&    (0.011)    &    (0.007)     &    (0.007)       &    (0.008)             \\[1ex]
  	$Year_{2014}$       & \multicolumn{9}{c}{Reference} \\[1ex]
   	$Year_{2018}$      &     -0.077\sym{*}        &      0.006            &      0.021\sym{***}&      -0.062\sym{***}&        0.011    &     -0.045\sym{***}&         0.003         &   0.024\sym{***}&       0.049\sym{***}\\
    	   &    (0.036)              &    (0.018)          &    (0.006)           &    (0.008)           &    (0.028)   &    (0.009)       &    (0.005)       &    (0.005)        &    (0.009)         \\[1ex]
    	$Bite$     &    -2.504\sym{***}&        -0.760\sym{**} &        -0.058               &      0.024             &     -0.286       &     -1.787\sym{***}&      -1.965\sym{***}&       -1.602\sym{***}&      -1.606\sym{***}\\
     	   &    (0.413)             &    (0.230)          &    (0.086)    &    (0.148)     &    (0.572)   &    (0.166)         &    (0.110)           &    (0.100)    &    (0.136)         \\[1ex]
\hline \\ [-1.4ex]
Clusters            &         401         &         401         &         401         &        401		&         401  &         401         &         401         &        401		&         401         \\		
Observations        &  2,667,593  &  2,667,593  &  2,667,593  &  2,667,593  &  2,667,593  &  2,667,593  &  2,667,593  &  2,667,593  &  2,667,593          \\[1ex]
			\hline\hline
					\end{tabular}
					\begin{tablenotes}
\begin{small} \textit{Notes:} Difference-in-differences regression coefficients as specified in equation (1). The dependent variable is the RIF of log monthly wages defined for various percentiles, as indicated by column titles.
Estimate of the constant not displayed.
The $\boldsymbol{X}$-vector includes the following covariates: age, dummies for gender, education, and industry. All regressions are weighted. Standard errors in parentheses are clustered at the county level. Asterisks indicate significance levels: 
   *~$p<0.05$, **~$p<0.01$, ***~$p<0.001$.
 \\\textit{Data:} SES, 2010, 2014, and 2018, weighted analysis sample, establishments with at least 10 regular employees.
 \end{small}
					\end{tablenotes}
				\end{threeparttable}
		\end{table}

\end{landscape}

\begin{table}[ht!]\centering
\caption{Minimum wage effect on monthly working hours,  Table \ref{tab:hours_baseline} amended by all DiD-coefficients}
         \label{tab:hours_baseline_full}
			\begin{threeparttable}
				\begin{tabular}{lC{2.5cm}C{2.5cm}C{2.5cm}C{2.5cm}}
					\hline\hline
					&\multicolumn{2}{c}{Contractual hours}
					&\multicolumn{2}{c}{Paid hours}\\
                       \cmidrule(lr){2-3}\cmidrule(lr){4-5}
					&\multicolumn{1}{c}{level}
					&\multicolumn{1}{c}{log}
					&\multicolumn{1}{c}{level}
					&\multicolumn{1}{c}{log} \\[1ex]
\hline \\ [-1.4ex]
$Bite*Year_{2010}$  &      0.088         &      0.020         &      1.987         &      0.031         \\
					&    (3.592)         &    (0.059)         &    (3.566)         &    (0.059)         \\[1ex]
					$Bite*Year_{2014}$       & \multicolumn{4}{c}{Reference} \\[1ex]
					$Bite*Year_{2018}$         &     -4.029         &      0.049         &     -4.590         &      0.043         \\
					&    (3.352)         &    (0.058)         &    (3.489)         &    (0.059)         \\[1ex]
					$Year_{2010}$           &     -2.376\sym{***}&     -0.017\sym{**} &     -2.375\sym{***}&     -0.017\sym{**} \\
					&    (0.444)         &    (0.006)         &    (0.444)         &    (0.006)         \\[1ex]
					$Year_{2014}$       & \multicolumn{4}{c}{Reference} \\[1ex]
					$Year_{2018}$          &     -0.608         &     -0.017\sym{**} &     -0.656         &     -0.017\sym{**} \\
					&    (0.339)         &    (0.005)         &    (0.357)         &    (0.005)         \\[1ex]
					$Bite$                &     36.554\sym{***}&      0.298\sym{**} &     36.195\sym{***}&      0.295\sym{**} \\
					&    (8.076)         &    (0.107)         &    (7.901)         &    (0.106)                \\[1ex]
\hline \\ [-1.4ex]
					Clusters            &         401         &         401         &         401         &        401        \\
					Observations        &   2,667,593         &   2,667,593         &   2,667,593         &   2,667,593         \\[1ex]
					\hline\hline
				\end{tabular}
				\begin{tablenotes}
					\begin{small} \textit{Notes:} Difference-in-differences regression coefficients as specified in equation (1). 
     The dependent variable is either the (monthly) contractual or the (monthly) paid working hours in levels or in logarithms, as indicated by column titles.
     Estimate of the constant not displayed.
     The $\boldsymbol{X}$-vector includes the following covariates: age, dummies for gender, education, and industry. All regressions are weighted. Standard errors in parentheses are clustered at the county level. Asterisks indicate significance levels: 
        *~$p<0.05$, **~$p<0.01$, ***~$p<0.001$.
     \\\textit{Data:} SES, 2010, 2014, and 2018, weighted analysis sample, establishments with at least 10 regular employees.
\end{small}
				\end{tablenotes}
			\end{threeparttable}
\end{table}

\begin{table}[ht!]\centering
\caption{Minimum wage effect on monthly working hours, all versus small plants, Table \ref{tab:hours_small_plants} amended by all DiD-coefficients}
\label{tab:hours_small_plants_full}
\begin{threeparttable}
\begin{tabular}{lC{2.5cm}C{2.5cm}C{2.5cm}C{2.5cm}}
\hline\hline
	&\multicolumn{2}{c}{Contractual hours}&\multicolumn{2}{c}{Paid hours}\\
\cmidrule(lr){2-3}\cmidrule(lr){4-5}
	&\multicolumn{1}{c}{level}&\multicolumn{1}{c}{log}&\multicolumn{1}{c}{level}&\multicolumn{1}{c}{log}\\[1ex]
\hline \\[-1.4ex]
\multicolumn{5}{l}{A: All plants} \\[1ex]
\hline \\[-1.4ex]
$Bite*Year_{2014}$  & \multicolumn{4}{c}{Reference} \\[1ex]
$Bite*Year_{2018}$  &  -9.124\sym{***}&  -0.056  &  -9.455\sym{***}&  -0.060 \\
					&  (2.086)  &  (0.032)  &  (2.157)  &  (0.032) \\[1ex]
$Year_{2014}$       & \multicolumn{4}{c}{Reference} \\[1ex]
$Year_{2018}$            &      0.458         &     -0.000        &      0.421        &     -0.000         \\
 					&    (0.325)         &    (0.005)         &    (0.338)         &    (0.005)         \\[1ex]
 $Bite$                 &     36.790\sym{***}&      0.345\sym{**} &     36.691\sym{***}&      0.346\sym{**} \\
 					&    (8.648)         &    (0.112)         &    (8.478)         &    (0.111)         \\[1ex]
\hline \\ [-1.4ex]  
Clusters            &         401         &         401         &         401         &    401   \\
Observations        &   1,461,576         &   1,461,576         &   1,461,576         &   1,461,576       \\[1ex]
\hline \\ [-1.4ex]
\multicolumn{5}{l}{B: Small plants ($ < 10$ employees)} \\[1ex]
\hline \\ [-1.4ex]
$Bite*Year_{2014}$  & \multicolumn{4}{c}{Reference} \\[1ex]
$Bite*Year_{2018}$  &  -11.772\sym{***}&  -0.126\sym{***}&  -11.796\sym{***}&  -0.126\sym{***}     \\
					&  (2.439)  &  (0.035)  &  (2.468)  &  (0.035)  \\[1ex]
$Year_{2014}$       & \multicolumn{4}{c}{Reference} \\[1ex]
$Year_{2018}$        &    1.164        &      0.005         &      1.226         &      0.005         \\
				&   (0.845)         &    (0.012)         &    (0.856)         &    (0.012)        \\[1ex]
$Bite$               & 50.593\sym{***}&      0.631\sym{***}&     50.912\sym{***}&      0.634\sym{***}    \\
				 &   (5.121)         &    (0.075)         &    (5.081)         &    (0.074)              \\[1ex]
\hline \\[-1.4ex]
Clusters            &         401         &         401         &         401         &    401              \\
Observations        &   277,791         &     277,791         &     277,791         &     277,791       \\[1ex]
\hline\hline
\end{tabular}
\begin{tablenotes}
\begin{small} \textit{Notes:} Difference-in-differences regression coefficients as specified in equation (1). 
The dependent variable is either the (monthly) contractual or the (monthly) paid working hours in levels or in logarithms, as indicated by column titles.
Estimate of the constant not displayed.
The $\boldsymbol{X}$-vector includes  the following covariates: age, dummies for gender, education, and industry. All regressions are weighted. Standard errors in parentheses are clustered at the county level. Asterisks indicate significance levels: 
   *~$p<0.05$, **~$p<0.01$, ***~$p<0.001$.
 \\\textit{Data:} SES, 2014 and 2018,  weighted analysis sample.
 \end{small}
\end{tablenotes}
\end{threeparttable}
\end{table}

\begin{landscape}

\vspace*{\fill}

\begin{table}[h!]\centering
\caption{RIF regressions on log monthly contractual working hours along the hours distribution, Table \ref{tab:hours_distribution} amended by all DiD-coefficients}
\label{tab:hours_distribution_full}
\begin{threeparttable}
\begin{tabular}{L{2.7cm}C{1.8cm}C{1.8cm}C{1.8cm}C{1.8cm}C{1.8cm}C{1.8cm}C{1.8cm}C{1.8cm}C{1.8cm}}
\hline\hline
	&\multicolumn{1}{c}{q4}&\multicolumn{1}{c}{q8}&\multicolumn{1}{c}{q12}&\multicolumn{1}{c}{q16}&\multicolumn{1}{c}{q20}&\multicolumn{1}{c}{q30}&\multicolumn{1}{c}{q50}&\multicolumn{1}{c}{q70}&\multicolumn{1}{c}{q90}\\[1ex]
\hline \\ [-1.4ex]
$Bite*Year_{2010}$ &  0.207  &  -0.054  &  -0.003  &  0.115  &  0.016  &  0.036  &  0.003  &  -0.058\sym{***}&  -0.065\sym{***} \\
				&  (0.271)  &  (0.162)  &  (0.075)  &  (0.180)  &  (0.146)  &  (0.045)  &  (0.015)  &  (0.012)  &  (0.014)  \\[1ex]
$Bite*Year_{2014}$  & \multicolumn{9}{c}{Reference} \\[1ex]
$Bite*Year_{2018}$  &  0.618\sym{**} &  0.246  &  0.052  &  -0.380\sym{**} &  0.042  & 0.057  &  -0.039\sym{**} &  -0.111\sym{***}&  -0.115\sym{***}\\
			&  (0.237)  &  (0.128)  &  (0.055)  &  (0.144)  &  (0.130)  &  (0.039)  &  (0.014)  &  (0.012)  &  (0.013)  \\[1ex]
	$Year_{2010}$        &      0.022              &      0.008          &     -0.013        &   -0.075\sym{***}&       -0.085\sym{***}    &     -0.036\sym{***}&        -0.007\sym{***}&        -0.004\sym{***}&      -0.003\sym{*}  \\
					&    (0.024)   &    (0.017)   &    (0.009)     &    (0.022)           &    (0.018)    &    (0.006)            &    (0.002)             &    (0.001)           &    (0.001)         \\[1ex]
					$Year_{2014}$       & \multicolumn{9}{c}{Reference} \\[1ex]
					$Year_{2018}$      &     -0.085\sym{***}&        -0.068\sym{***}&     -0.034\sym{***}&        -0.047\sym{**} &     0.023     &     -0.005        &     -0.002        &      0.003\sym{*}    &      0.003         \\
					&    (0.021)            &    (0.014)             &    (0.006)            &    (0.015)              &    (0.015)       	&    (0.005)         &    (0.002)     &    (0.001)        &    (0.001)         \\[1ex]
					$Bite$               &     -0.545\sym{*}     &      0.189          &      0.291\sym{***}&        1.261\sym{***}&         1.114\sym{***}      &      0.536\sym{***}&        0.194\sym{***}&        0.151\sym{***}&     0.150\sym{***} \\
					&    (0.249)          &    (0.181)     &    (0.087)        &    (0.313)         &    (0.325)      &
					(0.109)            &    (0.027)              &    (0.015)            &    (0.016)      \\[1ex]
\hline \\ [-1.4ex]
Clusters            &         401         &         401         &         401         &       401      &         401         &         401         &         401         &       401     &         401              \\
Observations       &   2,667,593         &   2,667,593         &   2,667,593         &   2,667,593         &   2,667,593         &   2,667,593         &   2,667,593         &   2,667,593         &   2,667,593           \\[1ex]
\hline\hline
\end{tabular}
\begin{tablenotes}
\begin{small} \textit{Notes:} Difference-in-differences regression coefficients as specified in equation (1). 
The dependent variable is the RIF of log (monthly) contractual working hours for various percentiles, as indicated by column titles.
Estimate of the constant not displayed.
The $\boldsymbol{X}$-vector includes  the following covariates: age, dummies for gender, education, and industry. All regressions are weighted. Standard errors in parentheses are clustered at the county level. Asterisks indicate significance levels: 
   *~$p<0.05$, **~$p<0.01$, ***~$p<0.001$.
 \\\textit{Data:} SES, 2010, 2014, and 2018, weighted analysis sample, establishments with at least 10 regular employees.	
 \end{small}		
\end{tablenotes}
\end{threeparttable}
\end{table}

\vspace*{\fill}

\end{landscape}

\begin{table}[ht!]\centering
\caption{Minimum wage effect on monthly working hours of exact minijobs (monthly wage between \euro{}350 and \euro{}450), Table \ref{tab:hour_exact_mini} amended by all DiD-coefficients}
\label{tab:hour_exact_mini_full}
	\begin{threeparttable}
	\begin{tabular}{lC{2.5cm}C{2.5cm}C{2.5cm}C{2.5cm}}
\hline\hline
				&\multicolumn{2}{c}{Contractual hours}
				&\multicolumn{2}{c}{Paid hours}\\
				\cmidrule(lr){2-3}\cmidrule(lr){4-5}
				&\multicolumn{1}{c}{hours}
				&\multicolumn{1}{c}{log hours}
				&\multicolumn{1}{c}{hours}
				&\multicolumn{1}{c}{log hours} \\[1ex]
\hline \\[-1.4ex]
$Bite*Year_{2010}$ &     -3.573         &     -0.082         &     -3.120         &     -0.067         \\
				&    (4.195)         &    (0.094)         &    (4.231)         &    (0.095)         \\[1ex]
$Bite*Year_{2014}$   & \multicolumn{4}{c}{Reference} \\[1ex]
$Bite*Year_{2018}$   & -37.995\sym{***} & -0.653\sym{***} & -38.250\sym{***} & -0.677\sym{***} \\
				&    (4.068)         &    (0.089)         &    (4.094)         &    (0.086)         \\[1ex]
$Year_{2010}$            &     -0.124        &     -0.014         &     -0.214         &     -0.017         \\
				&    (0.416)         &    (0.010)         &    (0.420)         &    (0.010)         \\[1ex]
$Year_{2014}$       & \multicolumn{4}{c}{Reference} \\[1ex]
$Year_{2018}$   &     -1.659\sym{***}&     -0.046\sym{***}&     -1.619\sym{***}&     -0.042\sym{***}\\
			&    (0.394)         &    (0.010)         &    (0.394)         &    (0.009)         \\[1ex]
$Bite$                 &     56.064\sym{***}&      1.143\sym{***}&     55.980\sym{***}&      1.137\sym{***}\\
             &    (3.806)         &    (0.078)         &    (3.831)         &    (0.078)          \\[1ex]
\hline \\[-1.4ex]
Clusters            &         401         &         401         &         401         &       401         \\
Observations        &     117,684         &     117,684         &     117,684         &     117,684         \\[1ex]
\hline\hline
\end{tabular}
\begin{tablenotes}
\begin{small} \textit{Notes:} Difference-in-differences regression coefficients as specified in equation (1). 
The dependent variable is either the (monthly) contractual or the (monthly) paid working hours in levels or in logarithms, as indicated by column titles.
Estimate of the constant not displayed.
The $\boldsymbol{X}$-vector includes  the following covariates: age, dummies for gender, education, and industry. All regressions are weighted. Standard errors in parentheses are clustered at the county level. Asterisks indicate significance levels:
   *~$p<0.05$, **~$p<0.01$, ***~$p<0.001$.
 \\\textit{Data:} SES, 2010, 2014, and 2018, only minijobs with monthly wage above \euro{350}, weighted analysis sample, establishments with at least 10 regular employees.
 \end{small}
\end{tablenotes}
\end{threeparttable}
\end{table}

\begin{table}[ht!]\centering
\caption{Minimum wage effect on log employment, Table \ref{tab:employment} amended by all DiD-coefficients}
\label{tab:employment_full}
\begin{threeparttable}
\begin{tabular}{{lcccccc}}
\hline\hline
	&\multicolumn{2}{c}{all jobs}&\multicolumn{2}{c}{regular employment}&\multicolumn{2}{c}{minijobs}\\
\cmidrule(lr){2-3}\cmidrule(lr){4-5}\cmidrule(lr){6-7}
	&\multicolumn{1}{c}{10+ plants}&\multicolumn{1}{c}{all plants}&\multicolumn{1}{c}{10+ plants}&\multicolumn{1}{c}{all plants}&\multicolumn{1}{c}{10+ plants}&\multicolumn{1}{c}{all plants}\\[1ex]\hline \\[-1.4ex]
$Bite*Year_{2014}$       & \multicolumn{6}{c}{Reference} \\[1ex]
$Bite*Year_{2018}$     &      0.032     &      0.008         &     -0.011         &     -0.022         &     -0.243         &     -0.149         \\
	&    (0.054)         &    (0.032)         &    (0.042)         &    (0.029)         &    (0.300)         &    (0.121)         \\[1ex]
	$Year_{2010}$  & -0.099\sym{***}& -0.091\sym{***}& -0.102\sym{***}&     -0.101\sym{***}&     -0.080\sym{***}&     -0.058\sym{***}\\
				&    (0.005)         &    (0.005)         &    (0.006)         &    (0.006)         &    (0.010)         &    (0.007)            \\[1.000ex]
				$Year_{2014}$       & \multicolumn{6}{c}{Reference} \\[1ex]
				$Year_{2018}$ & 0.095\sym{***}& 0.087\sym{***}& 0.111\sym{***}&      0.112\sym{***}&      0.058\sym{**} &      0.028\sym{**} \\
				&    (0.005)         &    (0.004)         &    (0.005)         &    (0.004)         &    (0.018)         &    (0.010)         \\[1ex]
				$Bite$                &     -4.859\sym{***}&     -4.714\sym{***}&     -5.076\sym{***}&     -4.752\sym{***}&     -4.015\sym{***}&     -4.694\sym{***}\\
				&    (1.376)         &    (1.230)         &    (1.445)         &    (1.295)         &    (1.176)         &    (1.043)     \\[1ex]
				$Bite*Trend$ &     -0.056\sym{***}& -0.051\sym{***}&     -0.031\sym{*}  &  -0.037\sym{***}   &     -0.128\sym{***}&    -0.073\sym{***} \\
				&    (0.012)         &   (0.008)          &    (0.012)         &        (0.009)     &    (0.027)         &   (0.012)  \\[1ex]
				$Constant$            &     12.014\sym{***}&     12.358\sym{***}&     11.827\sym{***}&     12.109\sym{***}&     10.065\sym{***}&     10.717\sym{***}\\
				&    (0.222)         &    (0.230)         &    (0.236)         &    (0.247)         &    (0.170)         &    (0.180)         \\[1ex]
\hline \\[-1.4ex]
Clusters            &         401         &         401         &         401         &         401         &         401         &         401 \\
Observations &        1203         &        1203         &        1203         &        1203         &        1203         &        1203 \\[1ex]
\hline\hline
\end{tabular}
\begin{tablenotes}
\begin{small} \textit{Notes:} Difference-in-differences regression coefficients as specified in equation~(\ref{eq:did_regional}). The regressions are run at the county-level and weighted by county-level employment.
The dependent variable is logarithmic employment of either all jobs, regular employment, or minijobs, as indicated by column titles. 
Standard errors in parentheses are clustered at the county level.
Asterisks indicate significance levels: 
   *~$p<0.05$, **~$p<0.01$, ***~$p<0.001$.
 \\ \textit{Data:} BHP of 2010, 2014, and 2018; full population aggregated to counties. 
\end{small}
\end{tablenotes}
\end{threeparttable}
\end{table}

\begin{table}[ht!]\centering
\caption{Minimum wage effect on forward-looking transitions out of minijobs, Table \ref{tab:transition_mini} amended by all DiD-coefficients}
\label{tab:transition_mini_full}
\begin{threeparttable}
\begin{tabular}{lC{3.7cm}C{3.7cm}C{3.7cm}}
\hline\hline
&\multicolumn{1}{c}{out of minijob}&\multicolumn{1}{c}{minijob to regular job}&\multicolumn{1}{c}{minijob to non-empl.}\\
\hline \\ [-1.4ex]
$Bite*Year_{2012}$ & Reference & Reference & Reference \\[1ex]
$Bite*Year_{2013}$ & -0.018\sym{*}   & -0.003         &  -0.014\sym{*}  \\
	            & (0.007)            &    (0.004)     &    (0.006)      \\
$Bite*Year_{2014}$ & 0.169\sym{***}  &  0.081\sym{***}&   0.088\sym{**} \\
	          & (0.027)            &    (0.007)     &    (0.027)     \\[1ex]
$ Year_{2012} $ & Reference & Reference & Reference \\[1ex]
$ Year_{2013} $   & 0.003\sym{**}   & 0.004\sym{***}& -0.001         \\
                   &    (0.001)      &    (0.001)    &    (0.001)     \\
$ Year_{2014} $   & 0.010\sym{***}  & 0.011\sym{***}& -0.001         \\
                   &    (0.003)      &    (0.001)    &    (0.003)     \\[1ex]
$Bite$               &      0.030      & 0.042\sym{*}  & -0.003         \\
                   &    (0.049)      &    (0.019)    &    (0.033)     \\[1ex]
$Constant$           &   0.334\sym{***}& 0.075\sym{***}&  0.241\sym{***}\\
                    &    (0.008)     &    (0.003)    &    (0.005)     \\[1ex]
\hline \\ [-1.4ex]
Clusters      &    401         &         401    &         401      \\
Observations  & 13,679,174  &    13,679,174  &    13,679,174    \\
Minijobs in 2014 & 4,542,156 &  4,542,156    &  4,542,156       \\[1ex]
\hline\hline
\end{tabular}
\begin{tablenotes}
\begin{small} \textit{Notes:} 
Difference-in-differences regression coefficients as specified in equation~(\ref{eq:minijob_trans}).
Dependent variable captures forward-looking transitions out of minijobs, as indicated by column titles. 
Standard errors in parentheses are clustered at the county level.
Asterisks indicate significance levels: 
   *~$p<0.05$, **~$p<0.01$, ***~$p<0.001$.
 \\\textit{Data:} IEB 2012-2015, population of all minijobbers, multiple jobs excluded.
 \end{small}
\end{tablenotes}
\end{threeparttable}
\end{table}

\begin{table}[ht!]\centering
\caption{Minimum wage effect on minijobber promotions within and across establishments, Table \ref{tab:transition_mini_within} amended by all DiD-coefficients}
\label{tab:transition_mini_within_full}
\begin{threeparttable}
\begin{tabular}{lC{4.4cm}C{4.4cm}}
\hline\hline
 & \multicolumn{2}{c}{minijob to regular job}\\
                    &\multicolumn{1}{c}{within establishments}&\multicolumn{1}{c}{across establishments}\\[1ex]
\hline \\[-1.4ex]
$Bite * Year_{2012}$ & Reference  & Reference     \\[1ex]
$Bite * Year_{2013}$ & -0.002     &      -0.002         \\
                              & (0.003)    &     (0.003)         \\
$Bite * Year_{2014}$ &  0.078\sym{***} &  0.002         \\
    &  (0.006)   &     (0.004)         \\[1ex]
$ Year_{2012} $    & Reference  & Reference     \\[1ex]
$ Year_{2013} $     &       0.001\sym{***}&       0.002\sym{***}\\
                    &     (0.000)         &     (0.000)         \\
$ Year_{2014} $     &       0.005\sym{***}&       0.006\sym{***}\\
                    &     (0.001)         &     (0.001)         \\[1ex]
$Bite$                &       0.031\sym{***}&       0.011         \\
                    &     (0.007)         &     (0.014)         \\[1ex]
$Constant$            &       0.032\sym{***}&       0.043\sym{***}\\
                    &     (0.001)         &     (0.002)         \\[1ex]
\hline \\[-1.4ex]
Clusters   &    401         &         401       \\
Observations        &    13,679,174         &    13,679,174  \\
Minijobs in 2014 & 4,542,156 &  4,542,156       \\[1ex]
\hline\hline
\end{tabular}
\begin{tablenotes}
\begin{small} \textit{Notes:} 
Difference-in-differences regression coefficients as specified in equation~(\ref{eq:minijob_trans}).
The dependent variable captures forward-looking transitions out of minijobs, as indicated by column titles. 
Standard errors in parentheses are clustered at the county level. Asterisks indicate significance levels: 
   *~$p<0.05$, **~$p<0.01$, ***~$p<0.001$.
 \\ \textit{Data:} IEB, 2012-2015, population of all minijobbers, multiple jobs excluded.
 \end{small}
\end{tablenotes}
\end{threeparttable}
\end{table}


\clearpage
\section{Robustness checks}\label{app:robustness}
\renewcommand*{\thefigure}{\thesection\arabic{figure}}
\renewcommand*{\thetable}{\thesection\arabic{table}}
\setcounter{figure}{0}
\setcounter{table}{0}

\begin{table}[ht!]\centering
\caption{Minimum wage effect on the mean and the variance of log hourly and monthly wages,
adding county fixed effects}
\label{tab:robust_county_fe_wage}
\begin{threeparttable}
\begin{tabular}{lC{2.5cm}C{2.5cm}C{2.500cm}C{2.5cm}}	
  \hline\hline
 			&\multicolumn{2}{c}{log hourly wage}
			&\multicolumn{2}{c}{log monthly wage}\\
			\cmidrule(lr){2-3}\cmidrule(lr){4-5}
			&\multicolumn{1}{c}{mean}&\multicolumn{1}{c}{variance}&\multicolumn{1}{c}{mean}&\multicolumn{1}{c}{variance}\\[1ex]
			\hline \\ [-1.4ex]
 					$Bite*Year_{2010}$      &     -0.014        &     -0.099         &     -0.038         &     -0.023         \\
 					&    (0.079)         &    (0.256)         &    (0.044)         &    (0.045)         \\[1ex]
 					$Bite*Year_{2014}$       & \multicolumn{4}{c}{Reference} \\[1ex]
 					$Bite*Year_{2018}$        &      0.486\sym{***}&     -1.185\sym{***}&      0.446\sym{***}&     -0.654\sym{***}\\
 					&    (0.073)         &    (0.231)         &    (0.034)         &    (0.033)         \\[1ex]
 					$Year_{2010}$                &     -0.004         &      0.047\sym{*}  &      0.014\sym{**} &      0.054\sym{***}\\
 					&    (0.009)         &    (0.023)         &    (0.005)         &    (0.005)         \\	[1ex]
 					$Year_{2014}$       & \multicolumn{4}{c}{Reference} \\[1ex]
 					$Year_{2018}$                &      0.006         &      0.060\sym{**} &      0.021\sym{***}&      0.020\sym{***}\\
 					&    (0.007)         &    (0.021)         &    (0.004)         &    (0.004)          \\[3ex]
County Fixed Effects  & \checkmark   & \checkmark   & \checkmark& \checkmark                                          \\[3ex]
\hline \\[-1.4ex]
 					Cluster             &         401      &         401         &         401      &         401       \\
 					Observations        &   2,667,593         &   2,667,593      &   2,667,593        &   2,667,593        \\[1ex]
\hline  
 				\end{tabular}
 				\begin{tablenotes}
\begin{small}  \textit{Notes:} Difference-in-differences regression coefficients as specified in equation (1). 
      The dependent variable is either the log hourly or the log monthly wage or the RIF of the variance of log hourly or log monthly wages, as indicated by column titles.  
      The $\boldsymbol{X}$-vector includes  the following covariates: age, dummies for gender, education, and industry. All regressions are weighted. Standard errors in parentheses are clustered at the county level. Asterisks indicate significance levels: 
         *~$p<0.05$, **~$p<0.01$, ***~$p<0.001$.
       \\\textit{Data:} SES, 2010, 2014, and 2018, weighted analysis sample, establishments with at least 10 regular employees.
\end{small}
\end{tablenotes}
\end{threeparttable}
 
\end{table}

\begin{table}[ht!]\centering
\caption{Minimum wage effect on monthly working hours, adding county fixed effects}
\label{tab:robust_county_fe_hour}
\begin{threeparttable}
\begin{tabular}{lC{2.5cm}C{2.5cm}C{2.5cm}C{2.5cm}}
			\hline\hline
			&\multicolumn{2}{c}{Contractual hours}
			&\multicolumn{2}{c}{Paid hours}\\
			\cmidrule(lr){2-3}\cmidrule(lr){4-5}
			&\multicolumn{1}{c}{hours}
			&\multicolumn{1}{c}{log hours}
			&\multicolumn{1}{c}{hours}
			&\multicolumn{1}{c}{log hours} \\[1ex]
			\hline \\[-1.4ex]
 					$Bite*Year_{2010}$     &     -0.601         &      0.013         &      1.291         &      0.024         \\
 					&    (3.496)         &    (0.059)         &    (3.471)         &    (0.058)         \\[1.000ex]
 					$Bite*Year_{2014}$       & \multicolumn{4}{c}{Reference} \\[1ex]
 					$Bite*Year_{2018}$      &     -4.028         &      0.046         &     -4.628         &      0.040         \\
 					&    (3.190)         &    (0.055)         &    (3.316)         &    (0.056)         \\[1ex]
 					$Year_{2010}$             &     -2.380\sym{***}&     -0.017\sym{**} &     -2.381\sym{***}&     -0.018\sym{**} \\
 					&    (0.433)         &    (0.006)         &    (0.433)         &    (0.006)         \\[1ex]
 					$Year_{2014}$       & \multicolumn{4}{c}{Reference} \\[1ex]
 					$Year_{2018}$         &     -0.494         &     -0.015\sym{**} &     -0.541         &     -0.015\sym{**} \\
 					&    (0.332)         &    (0.005)         &    (0.349)         &    (0.005)        \\[3ex]
County Fixed Effects  & \checkmark   & \checkmark   & \checkmark& \checkmark                                          \\[3ex]
\hline \\[-1.4ex]
 					Cluster             &         401      &         401       &         401         &         401      \\
 					Observations        &   2,667,593      &   2,667,593       &   2,667,593        &   2,667,593        \\[1ex]
 					\hline\hline
 				\end{tabular}
 				\begin{tablenotes}
\begin{small}  \textit{Notes:} Difference-in-differences regression coefficients as specified in equation (1). 
      The dependent variable is either the (monthly) contractual or the (monthly) paid working hours in levels or in logarithms, as indicated by column titles.
      The $\boldsymbol{X}$-vector includes  the following covariates: age, dummies for gender, education, and industry. All regressions are weighted. Standard errors in parentheses are clustered at the county level. Asterisks indicate significance levels: 
         *~$p<0.05$, **~$p<0.01$, ***~$p<0.001$.
       \\\textit{Data:} SES, 2010, 2014 and 2018, weighted analysis sample, establishments with at least 10 regular employees.
       \end{small}
\end{tablenotes}
\end{threeparttable}
\end{table}

\begin{table}[ht!]\centering
\caption{Minimum wage effect on the mean and the variance of log hourly and monthly wages, without covariates}
\label{tab:robust_no_cov_wage}
\begin{threeparttable}
\begin{tabular}{lC{2.5cm}C{2.5cm}C{2.5cm}C{2.5cm}}	
	\hline\hline
 			&\multicolumn{2}{c}{log hourly wage}
			&\multicolumn{2}{c}{log monthly wage}\\
			\cmidrule(lr){2-3}\cmidrule(lr){4-5}
			&\multicolumn{1}{c}{mean}&\multicolumn{1}{c}{variance}&\multicolumn{1}{c}{mean}&\multicolumn{1}{c}{variance}\\[1ex]
			\hline \\ [-1.4ex]
 					$Bite*Year_{2010}$        &     -0.081         &     -0.129         &     -0.083         &     -0.046         \\
 					&    (0.091)         &    (0.302)         &    (0.052)         &    (0.046)          \\[1ex]
 					$Bite*Year_{2014}$       & \multicolumn{4}{c}{Reference} \\[1ex]
 					$Bite*Year_{2018}$        &      0.501\sym{***}&     -1.367\sym{***}&      0.446\sym{***}&     -0.730\sym{***}\\
 					&    (0.108)         &    (0.287)         &    (0.044)         &    (0.038)         \\[1ex]
 					$Year_{2010}$            &      0.007         &      0.060\sym{*}  &      0.024\sym{***}&      0.046\sym{***}\\
 					&    (0.010)         &    (0.026)         &    (0.007)         &    (0.005)          \\[1ex]
 					$Year_{2014}$       & \multicolumn{4}{c}{Reference} \\[1ex]
 					$Year_{2018}$           &      0.043\sym{***}&      0.107\sym{***}&      0.054\sym{***}&      0.040\sym{***}\\
 					&    (0.010)         &    (0.025)         &    (0.005)         &    (0.005)         \\[1ex]
 					$Bite$                &     -1.728\sym{***}&     -0.617         &     -2.041\sym{***}&      0.025         \\
 					&    (0.208)         &    (0.326)         &    (0.124)         &    (0.087)         \\[1ex]
 					\hline\\[-1.4ex] 
 					Cluster            &         401         &         401         &         401         &         401         \\
 					Observations        &   2,667,593         &   2,667,593         &   2,667,593         &   2,667,593         \\[1ex]
 					\hline\hline
 				\end{tabular}
 				\begin{tablenotes}
\begin{small}  \textit{Notes:} Difference-in-differences regression coefficients as specified in equation (1). 
      The dependent variable is either the log hourly or the log monthly wage or the RIF of the variance of log hourly or log monthly wages, as indicated by column titles. Estimate of the constant not displayed.
      Standard errors in parentheses are clustered at the county level. Asterisks indicate significance levels: 
         *~$p<0.05$, **~$p<0.01$, ***~$p<0.001$.
       \\\textit{Data:} SES, 2010, 2014, and 2018, weighted analysis sample, establishments with at least 10 regular employees.
       \end{small}
\end{tablenotes}
\end{threeparttable}
\end{table}

\begin{table}[ht!]\centering
\caption{Minimum wage effect on monthly working hours, without covariates}
\label{tab:robust_no_cov_hour}
\begin{threeparttable}
\begin{tabular}{lC{2.5cm}C{2.5cm}C{2.5cm}C{2.5cm}}
 			    	\hline\hline
			&\multicolumn{2}{c}{Contractual hours}
			&\multicolumn{2}{c}{Paid hours}\\
			\cmidrule(lr){2-3}\cmidrule(lr){4-5}
			&\multicolumn{1}{c}{hours}
			&\multicolumn{1}{c}{log hours}
			&\multicolumn{1}{c}{hours}
			&\multicolumn{1}{c}{log hours} \\[1ex]
			\hline \\[-1.4ex]
 						$Bite*Year_{2010}$        &     -2.833         &     -0.009         &     -0.947         &      0.003         \\
 					&    (3.874)         &    (0.068)         &    (3.889)         &    (0.068)          \\[1ex]
 					$Bite*Year_{2014}$       & \multicolumn{4}{c}{Reference} \\[1ex]
 					$Bite*Year_{2018}$        &     -3.867         &      0.059         &     -4.114         &      0.055         \\
 					&    (4.904)         &    (0.080)         &    (5.064)         &    (0.081)         \\[1ex]
 					$Year_{2010}$            &     -1.987\sym{***}&     -0.017\sym{**} &     -1.976\sym{***}&     -0.017\sym{**} \\
 					&    (0.428)         &    (0.006)         &    (0.438)         &    (0.006)         \\	[1ex]
 					$Year_{2014}$       & \multicolumn{4}{c}{Reference} \\[1ex]
 					$Year_{2018}$           &      0.217         &     -0.011         &      0.061         &     -0.011         \\
 					&    (0.432)         &    (0.007)         &    (0.451)         &    (0.007)         \\[1ex]
 					$Bite$                &     36.522\sym{***}&      0.314\sym{*}  &     36.493\sym{***}&      0.313\sym{*}  \\
 					&    (9.625)         &    (0.125)         &    (9.332)         &    (0.124)         \\[1ex]
 					\hline \\[-1.4ex]
 					Cluster             &         401         &         401         &         401         &         401         \\
 					Observations        &   2,667,593         &   2,667,593         &   2,667,593         &   2,667,593         \\[1ex]
 					\hline\hline
 				\end{tabular}
 				\begin{tablenotes}
\begin{small}  \textit{Notes:} Difference-in-differences regression coefficients as specified in equation (1). 
      The dependent variable is either the (monthly) contractual or the (monthly) paid working hours in levels or in logarithms, as indicated by column titles. Estimate of the constant not displayed.
      All regressions are weighted. Standard errors in parentheses are clustered at the county level. Asterisks indicate significance levels: 
         *~$p<0.05$, **~$p<0.01$, ***~$p<0.001$.
      \\\textit{Data:} SES, 2010, 2014, and 2018, weighted analysis sample, establishments with at least 10 regular employees.
      \end{small}
\end{tablenotes}
\end{threeparttable}
\end{table}

 \begin{table}[ht!]\centering
\caption{Minimum wage effect on the mean and the variance of log hourly and monthly wages, monthly working hours are censored from above at 250}
\label{tab:robust_censor_hour_wage}
\begin{threeparttable}
\begin{tabular}{lC{2.5cm}C{2.5cm}C{2.5cm}C{2.5cm}}	
	\hline\hline
 			&\multicolumn{2}{c}{log hourly wage}
			&\multicolumn{2}{c}{log monthly wage}\\
			\cmidrule(lr){2-3}\cmidrule(lr){4-5}
			&\multicolumn{1}{c}{mean}&\multicolumn{1}{c}{variance}&\multicolumn{1}{c}{mean}&\multicolumn{1}{c}{variance}\\[1ex]
			\hline \\ [-1.4ex]
 					$Bite*Year_{2010}$       &     -0.011         &     -0.105         &     -0.042         &     -0.017         \\
 					&    (0.080)         &    (0.256)         &    (0.044)         &    (0.045)         \\[1ex]
 					$Bite*Year_{2014}$       & \multicolumn{4}{c}{Reference} \\[1ex]
 					$Bite*Year_{2018}$      &      0.500\sym{***}&     -1.198\sym{***}&      0.457\sym{***}&     -0.654\sym{***}\\
 					&    (0.072)         &    (0.242)         &    (0.033)         &    (0.034)         \\[1ex]
 					$Year_{2010}$           &     -0.004         &      0.045         &      0.014\sym{*}  &      0.053\sym{***}\\
 					&    (0.009)         &    (0.023)         &    (0.005)         &    (0.005)         \\	[1ex]
 					$Year_{2014}$       & \multicolumn{4}{c}{Reference} \\[1ex]
 					$Year_{2018}$       &      0.001         &      0.062\sym{**} &      0.018\sym{***}&      0.019\sym{***}\\
 					&    (0.007)         &    (0.021)         &    (0.004)         &    (0.004)         \\[1ex]
 					$Bite$                 &     -1.547\sym{***}&     -0.131         &     -1.842\sym{***}&      0.215\sym{***}\\
 					&    (0.157)         &    (0.278)         &    (0.087)         &    (0.050)         \\[1ex]
 					\hline \\[-1.4ex]
 					Cluster              &         401         &         401         &         401         &         401         \\
 						Observations        &   2,667,593         &   2,667,593         &   2,667,593         &   2,667,593         \\[1ex]
 					\hline\hline
 				\end{tabular}
 				\begin{tablenotes}
\begin{small}  \textit{Notes:} Difference-in-differences regression coefficients as specified in equation (1). 
      The dependent variable is either the log hourly or the log monthly wage or the RIF of the variance of log hourly or log monthly wages, as indicated by column titles.
      Monthly contractual and paid working hours are censored from above at 250 hours. 
      Estimate of the constant not displayed.
      The $\boldsymbol{X}$-vector includes  the following covariates: age, dummies for gender, education, and industry. All regressions are weighted. Standard errors in parentheses are clustered at the county level. Asterisks indicate significance levels: 
         *~$p<0.05$, **~$p<0.01$, ***~$p<0.001$.
       \\\textit{Data:} SES, 2010, 2014 and 2018, weighted analysis sample, establishments with at least 10 regular employees.
       \end{small}
\end{tablenotes}
\end{threeparttable}
\end{table}

\begin{table}[ht!]\centering
\caption{Minimum wage effect on monthly working hours, working hours are censored from above at 250}
\label{tab:robust_censor_hour_hour}
\begin{threeparttable}
\begin{tabular}{lC{2.5cm}C{2.5cm}C{2.5cm}C{2.5cm}}
 			    	\hline\hline
			&\multicolumn{2}{c}{Contractual hours}
			&\multicolumn{2}{c}{Paid hours}\\
			\cmidrule(lr){2-3}\cmidrule(lr){4-5}
			&\multicolumn{1}{c}{hours}
			&\multicolumn{1}{c}{log hours}
			&\multicolumn{1}{c}{hours}
			&\multicolumn{1}{c}{log hours} \\[1ex]
			\hline \\[-1.4ex]
 					$Bite*Year_{2010}$        &      0.152         &      0.020         &      1.921         &      0.031         \\
 					&    (3.574)         &    (0.059)         &    (3.541)         &    (0.059)         \\[1ex]
 					$Bite*Year_{2014}$       & \multicolumn{4}{c}{Reference} \\[1ex]
 					$Bite*Year_{2018}$         &     -3.992         &      0.049         &     -4.568         &      0.044         \\
 					&    (3.341)         &    (0.058)         &    (3.476)         &    (0.059)         \\[1ex]
 					$Year_{2010}$             &     -2.395\sym{***}&     -0.017\sym{**} &     -2.380\sym{***}&     -0.017\sym{**} \\
 					&    (0.442)         &    (0.006)         &    (0.441)         &    (0.006)         \\	[1ex]
 					$Year_{2014}$       & \multicolumn{4}{c}{Reference} \\[1ex]
 					$Year_{2018}$            &     -0.609         &     -0.017\sym{**} &     -0.639         &     -0.017\sym{**} \\
 					&    (0.338)         &    (0.005)         &    (0.356)         &    (0.005)         \\[1ex]
 					$Bite$                &     36.520\sym{***}&      0.297\sym{**} &     36.230\sym{***}&      0.295\sym{**} \\
 					&    (8.081)         &    (0.107)         &    (7.910)         &    (0.106)         \\[1ex]
 					\hline \\[-1.4ex]
 					Cluster             &         401         &         401         &         401         &         401         \\
 						Observations        &   2,667,593         &   2,667,593         &   2,667,593         &   2,667,593         \\[1ex]
 					\hline\hline
 				\end{tabular}
 				\begin{tablenotes}
\begin{small}  \textit{Notes:} Difference-in-differences regression coefficients as specified in equation (1). 
       The dependent variable is either the (monthly) contractual or the (monthly) paid working hours in levels or in logarithms, as indicated by column titles.
       Monthly contractual and paid working hours are censored from above at 250 hours. 
       Estimate of the constant not displayed.
       The $\boldsymbol{X}$-vector includes  the following covariates: age, dummies for gender, education, and industry. All regressions are weighted. Standard errors in parentheses are clustered at the county level. Asterisks indicate significance levels: 
          *~$p<0.05$, **~$p<0.01$, ***~$p<0.001$. 
       \\\textit{Data:} SES, 2010, 2014, and 2018, weighted analysis sample, establishments with at least 10 regular employees.
       \end{small}
\end{tablenotes}
\end{threeparttable}
\end{table}


\begin{table}[ht!]\centering
\caption{Minimum wage effect on the mean and the variance of log hourly and monthly wages,
using the bite gap as the treatment variable}
\label{tab:robust_bite_gap_wage}
\begin{threeparttable}
\begin{tabular}{lC{2.5cm}C{2.5cm}C{2.5cm}C{2.5cm}}	
	\hline\hline
 			&\multicolumn{2}{c}{log hourly wage}
			&\multicolumn{2}{c}{log monthly wage}\\
			\cmidrule(lr){2-3}\cmidrule(lr){4-5}
			&\multicolumn{1}{c}{mean}&\multicolumn{1}{c}{variance}&\multicolumn{1}{c}{mean}&\multicolumn{1}{c}{variance}\\[1ex]
			\hline \\ [-1.4ex]
 					$Bite\ gap*Year_{2010}$ &      0.341         &     -1.999         &     -0.026         &     -0.192         \\
 					&    (0.284)         &    (1.129)         &    (0.139)         &    (0.143)         \\[1ex]
					$Bite\ gap*Year_{2014}$   &    \multicolumn{4}{c}{Reference}       \\[1ex]
 					$Bite\ gap*Year_{2018}$  &      1.618\sym{***}&     -4.281\sym{***}&      1.234\sym{***}&     -1.900\sym{***}\\
 					&    (0.331)         &    (1.251)         &    (0.146)         &    (0.158)         \\[1ex]
 					$Year_{2010}$               &     -0.014         &      0.083\sym{***}&      0.009         &      0.056\sym{***}\\
 					&    (0.008)         &    (0.024)         &    (0.005)         &    (0.004)         \\[1.000ex]
 					$Year_{2014}$       & \multicolumn{4}{c}{Reference} \\[1ex]
 					$Year_{2018}$             &      0.009         &      0.049         &      0.031\sym{***}&      0.001         \\
 					&    (0.008)         &    (0.028)         &    (0.004)         &    (0.004)         \\[1ex]
 					$Bite\ gap$             &     -3.785\sym{***}&      1.664         &     -4.233\sym{***}&      1.014\sym{***}\\
 					&    (0.481)         &    (1.518)         &    (0.505)         &    (0.139)         \\[1ex]
 					\hline \\[-1.4ex]
 					Cluster             &         401         &         401         &         401         &         401         \\
 					Observations        &   2,667,593         &   2,667,593         &   2,667,593         &   2,667,593         \\[1ex]
 					\hline\hline
 				\end{tabular}
 				\begin{tablenotes}
\begin{small}  \textit{Notes:} Difference-in-differences regression coefficients as specified in equation (1). 
      The dependent variable is either the log hourly or the log monthly wage or the RIF of the variance of log hourly or log monthly wages, as indicated by column titles.
      Estimate of the constant not displayed.
      The $\boldsymbol{X}$-vector includes  the following covariates: age, dummies for gender, education, and industry. Bite gap is calculated based on wages in 2014 and equals to (real hourly wage - 8.5)/real hourly wage. Real hourly wage is censored from below  at 1 \euro. Standard errors in parentheses are clustered at the county level. Asterisks indicate significance levels: 
         *~$p<0.05$, **~$p<0.01$, ***~$p<0.001$.
      \\\textit{Data:} SES, 2010, 2014, and 2018, weighted analysis sample, establishments with at least 10 regular employees.
\end{small}
\end{tablenotes}
\end{threeparttable}
\end{table}

\begin{table}[ht!]\centering
\caption{Minimum wage effect on monthly working hours, using the bite gap as the treatment variable}
\label{tab:robust_bite_gap_hour}
\begin{threeparttable}
\begin{tabular}{lC{2.5cm}C{2.5cm}C{2.5cm}C{2.5cm}}
 			    	\hline\hline
			&\multicolumn{2}{c}{Contractual hours}
			&\multicolumn{2}{c}{Paid hours}\\
			\cmidrule(lr){2-3}\cmidrule(lr){4-5}
			&\multicolumn{1}{c}{hours}
			&\multicolumn{1}{c}{log hours}
			&\multicolumn{1}{c}{hours}
			&\multicolumn{1}{c}{log hours} \\[1ex]
			\hline \\[-1.4ex]
 					$Bite\ gap*Year_{2010}$    &     11.274         &      0.354         &     13.896         &      0.368         \\
 					&   (10.225)         &    (0.221)         &   (10.380)         &    (0.220)         \\[1ex]
					$Bite\ gap*Year_{2014}$   &    \multicolumn{4}{c}{Reference}       \\[1ex]
 			    		$Bite\ gap*Year_{2018}$   &      7.289         &      0.386         &      8.633         &      0.384         \\
 			    		&   (17.417)         &    (0.337)         &   (18.056)         &    (0.344)         \\[1ex]
 					$Year_{2010}$            &     -2.612\sym{***}&     -0.023\sym{***}&     -2.488\sym{***}&     -0.023\sym{***}\\
 					&    (0.359)         &    (0.006)         &    (0.367)         &    (0.005)         \\	[1ex]
 					$Year_{2014}$       & \multicolumn{4}{c}{Reference} \\[1ex]
 					$Year_{2018}$           &     -1.161\sym{**} &     -0.022\sym{**} &     -1.296\sym{**} &     -0.022\sym{**} \\
 					&    (0.409)         &    (0.008)         &    (0.428)         &    (0.008)         \\[1ex]
 					$Bite\ gap$             &     76.157\sym{*}  &      0.450         &     75.311\sym{*}  &      0.449         \\
 					&   (35.549)         &    (0.513)         &   (35.433)         &    (0.514)         \\[1ex]
 					\hline \\[-1.4ex]
 					Cluster             &         401         &         401         &         401         &         401         \\
 					Observations        &   2,667,593         &   2,667,593         &   2,667,593         &   2,667,593         \\[1ex]
 					\hline\hline
 				\end{tabular}
 				\begin{tablenotes}
\begin{small}  \textit{Notes:} Difference-in-differences regression coefficients as specified in equation (1). The dependent variable is either the (monthly) contractual or the (monthly) paid working hours in levels or in logarithms, as indicated by column titles. Estimate of the constant not displayed. The $\boldsymbol{X}$-vector includes  the following covariates: age, dummies for gender, education, and industry. Bite gap is calculated based on wages in 2014 and equals to (real hourly wage - 8.5)/real hourly wage. Real hourly wage is censored from below  at 1 \euro. All regressions are weighted. Standard errors in parentheses are clustered at the county level. Asterisks indicate significance levels: *~$p<0.05$, **~$p<0.01$, ***~$p<0.001$. \\\textit{Data:} SES, 2010, 2014, and 2018, weighted analysis sample, establishments with at least 10 regular employees.
\end{small}
\end{tablenotes}
\end{threeparttable}
\end{table}


\begin{table}[ht!]\centering
\caption{Minimum wage effect on the mean and the variance of log hourly and monthly wages,
all plants}
\label{tab:robust_all_plants_wage}
\begin{threeparttable}
\begin{tabular}{lC{2.5cm}C{2.5cm}C{2.5cm}C{2.5cm}}	
	\hline\hline
 			&\multicolumn{2}{c}{log hourly wage}
			&\multicolumn{2}{c}{log monthly wage}\\
			\cmidrule(lr){2-3}\cmidrule(lr){4-5}
			&\multicolumn{1}{c}{mean}&\multicolumn{1}{c}{variance}&\multicolumn{1}{c}{mean}&\multicolumn{1}{c}{variance}\\[1ex]
			\hline \\ [-1.4ex]
   $Bite*Year_{2014}$       &	\multicolumn{4}{c}{Reference} \\[1ex]	
$Bite*Year_{2018}$  & 0.410\sym{***}      &     -0.631\sym{***}   & 0.350\sym{***}      & -0.786\sym{***}       \\
                    & (0.023)             & (0.024)               & (0.043)             &    (0.120)            \\[1ex]
$Year_{2018}$       & 0.021\sym{***}      & 0.032\sym{***}        &  0.021\sym{**}      & 0.021                 \\
                    & (0.004)             & (0.004)               &  (0.007)            & (0.017)               \\[1ex]
$Bite$              &-1.653\sym{***}      & 0.184\sym{***}        & -1.308\sym{***}     & -0.434\sym{*}         \\
                    &  (0.063)            &    (0.050)            & (0.161)             &  (0.181)              \\[1ex]
\hline \\[-1.4ex]
 					Cluster             &         401         &         401         &         401         &         401         \\
 					Observations        &   1,461,576         &   1,461,576         &  1,461,576          &   1,461,576       \\[1ex]
 					\hline\hline
 				\end{tabular}
 				\begin{tablenotes}
\begin{small}  \textit{Notes:} Difference-in-differences regression coefficients as specified in equation (1), but with no interaction with 2010 and no year dummy for 2010. The dependent variable is either the log hourly or the log monthly wage or the RIF of the variance of log hourly or log monthly wages, as indicated by column titles.
      Estimate of the constant not displayed.
      The $\boldsymbol{X}$-vector includes the following covariates: age, dummies for gender, education, and industry.  Standard errors in parentheses are clustered at the county level. Asterisks indicate significance levels: 
         *~$p<0.05$, **~$p<0.01$, ***~$p<0.001$.
      \\\textit{Data:} 2014 and 2018, weighted analysis sample, all establishments.
      \end{small}
\end{tablenotes}
\end{threeparttable}
\end{table}

\clearpage
\section{Minimum wage effect on establishment size composition}\label{app:size_selectivity}
\renewcommand*{\thefigure}{\thesection\arabic{figure}}
\renewcommand*{\thetable}{\thesection\arabic{table}}
\setcounter{figure}{0}
\setcounter{table}{0}

\begin{table}[ht!]\centering
\caption{Establishment-selection into less than 10 employees}
\begin{threeparttable}
\begin{tabular}{{L{3.5cm}C{3.3cm}C{3.3cm}}}
\hline\hline
&\multicolumn{2}{c}{Dep. var.: }   \\   
&\multicolumn{2}{c}{I[Establishment $<$10 regular employees]}      \\\cmidrule(lr){2-3}
                    &\multicolumn{1}{c}{OLS}&\multicolumn{1}{c}{Employee-weighted}\\
\hline\\[-1.4ex]
$Bite * Year_{2011}$ &   Reference  & Reference         \\[1ex]
$Bite * Year_{2012}$ &      0.001         &      0.005         \\
                    &    (0.002)         &    (0.003)         \\
$Bite * Year_{2013}$ &      0.002         &      0.006         \\
                    &    (0.002)         &    (0.005)         \\
$Bite * Year_{2014}$ &      0.000         &      0.005         \\
                    &    (0.003)         &    (0.006)         \\
$Bite * Year_{2015}$ &      0.001         &      0.006         \\
                    &    (0.004)         &    (0.007)         \\
$Bite * Year_{2016}$ &     -0.003         &      0.001         \\
                    &    (0.004)         &    (0.007)         \\
$Bite * Year_{2017}$ &     -0.005         &     -0.002         \\
                    &    (0.005)         &    (0.008)         \\
$Bite * Year_{2018}$ &     -0.004         &     -0.000         \\
                    &    (0.005)         &    (0.009)         \\[1ex]
\hline\\[-1.4ex]
Clusters            &         401         &         401         \\
Observations        &    24,142,842         &    24,142,842         \\[1ex]
\hline\hline
\end{tabular}
\begin{tablenotes}
\begin{small} \textit{Notes:} Treatment effect interactions from difference-in-differences estimations. Dependent variables indicate whether an establishment has less than 10 regular employees. Cluster-robust standard errors in parentheses (cluster=county). Asterisks indicate significance levels: *~$p<0.05$, **~$p<0.01$, ***~$p<0.001$. \\\textit{Data:} BHP, administrative population of all establishments in Germany.
\end{small}
\end{tablenotes}
\end{threeparttable}
\end{table}

\clearpage

\begin{landscape}
\section{RIF regressions of log working hours at each percentile of the hours distribution}
\label{app:rif_hours_distributions}
\renewcommand*{\thefigure}{\thesection\arabic{figure}}
\renewcommand*{\thetable}{\thesection\arabic{table}}
\setcounter{figure}{0}
\setcounter{table}{0}

\begin{table}[h!]\centering
\caption{RIF regressions of log contractual monthly working hours along the hours distribution}
\label{tab:hours_distribution}
\begin{threeparttable}
\begin{tabular}{l*{10}{c}}
\hline\hline
\multicolumn{1}{r}{Percentile}                   &\multicolumn{1}{c}{q1}&\multicolumn{1}{c}{q2}&\multicolumn{1}{c}{q3}&\multicolumn{1}{c}{q4}&\multicolumn{1}{c}{q5}&\multicolumn{1}{c}{q6}&\multicolumn{1}{c}{q7}&\multicolumn{1}{c}{q8}&\multicolumn{1}{c}{q9}&\multicolumn{1}{c}{q10}\\
\multicolumn{1}{r}{Monthly Hours}                   
&\multicolumn{1}{c}{9.6}
&\multicolumn{1}{c}{15}
&\multicolumn{1}{c}{19.5}
&\multicolumn{1}{c}{22}
&\multicolumn{1}{c}{26.1}
&\multicolumn{1}{c}{30}
&\multicolumn{1}{c}{33}
&\multicolumn{1}{c}{35}
&\multicolumn{1}{c}{39}
&\multicolumn{1}{c}{40}\\
\hline
$Bite*Year_{2010}$  &       2.330         &       0.568         &       0.405         &       0.207         &       0.064         &       0.069         &      -0.078         &      -0.054         &      -0.046         &      -0.016         \\
                    &     (2.030)         &     (0.674)         &     (0.507)         &     (0.271)         &     (0.269)         &     (0.238)         &     (0.251)         &     (0.162)         &     (0.151)         &     (0.131)         \\[1ex]
$Bite*Year_{2014}$  & \multicolumn{9}{c}{Reference} \\[1ex] 
$Bite*Year_{2018}$  &       3.717         &       1.388\sym{*}  &       1.110\sym{*}  &       0.618\sym{**} &       0.655\sym{**} &       0.578\sym{**} &       0.475\sym{*}  &       0.245         &       0.196         &       0.154         \\
                    &     (1.938)         &     (0.650)         &     (0.468)         &     (0.237)         &     (0.236)         &     (0.196)         &     (0.210)         &     (0.128)         &     (0.115)         &     (0.099)         \\[1ex]
$Year_{2010}$       &       0.199         &       0.105\sym{*}  &       0.071         &       0.022         &       0.043         &       0.030         &       0.031         &       0.008         &       0.003         &      -0.006         \\
                    &     (0.149)         &     (0.053)         &     (0.043)         &     (0.024)         &     (0.025)         &     (0.023)         &     (0.026)         &     (0.017)         &     (0.016)         &     (0.014)         \\[1ex]
$Year_{2018}$       &      -0.410\sym{**} &      -0.182\sym{***}&      -0.171\sym{***}&      -0.084\sym{***}&      -0.098\sym{***}&      -0.098\sym{***}&      -0.112\sym{***}&      -0.068\sym{***}&      -0.071\sym{***}&      -0.062\sym{***}\\
                    &     (0.157)         &     (0.055)         &     (0.041)         &     (0.021)         &     (0.022)         &     (0.020)         &     (0.021)         &     (0.014)         &     (0.012)         &     (0.010)         \\[1ex]
$Bite$              &      -3.841\sym{*}  &      -1.412\sym{*}  &      -1.017\sym{*}  &      -0.545\sym{*}  &      -0.419         &      -0.219         &       0.029         &       0.189         &       0.325         &       0.339\sym{*}  \\
                    &     (1.918)         &     (0.615)         &     (0.463)         &     (0.249)         &     (0.263)         &     (0.246)         &     (0.282)         &     (0.181)         &     (0.169)         &     (0.149)         \\
\hline
Cluster             &         401         &         401         &         401         &         401         &         401         &         401         &         401         &         401         &         401         &         401         \\
Observations        &   2,667,593         &   2,667,593         &   2,667,593         &   2,667,593         &   2,667,593         &   2,667,593         &   2,667,593         &   2,667,593         &   2,667,593         &   2,667,593         \\
\hline\hline
\end{tabular}
\begin{tablenotes}
\begin{small} \textit{Notes:} Treatment effect interactions from difference-in-differences estimations, as specified in equation (1). Dependent variable is the RIF of log (monthly) contractual  working hours, defined for various percentiles as indicated by column titles. The $\boldsymbol{X}$-vector includes  the following covariates: age, dummies for gender, education, and industry. All regressions are weighted. Standard errors in parentheses are clustered at the county level. Asterisks indicate significance levels:
 *~$p<0.05$, **~$p<0.01$, ***~$p<0.001$. For some adjacent percentiles, the coefficient estimates are identical because these percentiles are located at the same hours value (e.g., percentiles 31--32). \\\textit{Data:} SES, 2010, 2014, and 2018, weighted analysis sample, establishments with at least 10 regular employees. 
\end{small}
\end{tablenotes}
\raggedleft{Continued overleaf}
\end{threeparttable}
\end{table}
\end{landscape}

\setcounter{table}{0}

\begin{landscape}
\begin{table}[h!]\centering
\caption{Continued}
\begin{threeparttable}
\begin{tabular}{l*{10}{c}}
\hline\hline
\multicolumn{1}{r}{Percentile} 
                    &\multicolumn{1}{c}{q11}
                    &\multicolumn{1}{c}{q12}
                    &\multicolumn{1}{c}{q13}
                    &\multicolumn{1}{c}{q14}
                    &\multicolumn{1}{c}{q15}
                    &\multicolumn{1}{c}{q16}
                    &\multicolumn{1}{c}{q17}
                    &\multicolumn{1}{c}{q18}
                    &\multicolumn{1}{c}{q19}
                    &\multicolumn{1}{c}{q20}\\
\multicolumn{1}{r}{Monthly Hours}                   
&\multicolumn{1}{c}{43}
&\multicolumn{1}{c}{43.5}
&\multicolumn{1}{c}{45}
&\multicolumn{1}{c}{47.8}
&\multicolumn{1}{c}{50.1}
&\multicolumn{1}{c}{53}
&\multicolumn{1}{c}{60}
&\multicolumn{1}{c}{65.2}
&\multicolumn{1}{c}{73.7}
&\multicolumn{1}{c}{80}\\
\hline
$Bite*Year_{2010}$  &      -0.004         &      -0.003         &       0.047         &       0.061         &       0.033         &       0.115         &       0.230         &       0.076         &       0.249         &       0.016          \\
                    &     (0.080)         &     (0.075)         &     (0.108)         &     (0.177)         &     (0.142)         &     (0.180)         &     (0.299)         &     (0.159)         &     (0.352)         &     (0.146)          \\[1ex]
$Bite*Year_{2014}$  & \multicolumn{9}{c}{Reference} \\[1ex]
$Bite*Year_{2018}$  &       0.070         &       0.052         &       0.009         &      -0.189         &      -0.286\sym{*}  &      -0.379\sym{**} &      -0.479\sym{*}  &      -0.085         &       0.050         &       0.042          \\
                    &     (0.059)         &     (0.054)         &     (0.099)         &     (0.156)         &     (0.126)         &     (0.144)         &     (0.242)         &     (0.129)         &     (0.311)         &     (0.130)          \\[1ex]
$Year_{2010}$        &      -0.015         &      -0.013         &      -0.041\sym{**} &      -0.058\sym{**} &      -0.056\sym{**} &      -0.075\sym{***}&      -0.132\sym{***}&      -0.087\sym{***}&      -0.208\sym{***}&      -0.085\sym{***} \\
                    &     (0.009)         &     (0.009)         &     (0.013)         &     (0.021)         &     (0.017)         &     (0.022)         &     (0.036)         &     (0.020)         &     (0.043)         &     (0.018)          \\[1ex]
$Year_{2018}$       &      -0.038\sym{***}&      -0.034\sym{***}&      -0.034\sym{**} &      -0.063\sym{***}&      -0.061\sym{***}&      -0.046\sym{**} &      -0.005         &       0.010         &       0.046         &       0.023          \\
                    &     (0.006)         &     (0.006)         &     (0.010)         &     (0.017)         &     (0.013)         &     (0.015)         &     (0.027)         &     (0.015)         &     (0.036)         &     (0.015)          \\[1ex]
$Bite$              &        0.284\sym{**} &       0.291\sym{***}&       0.636\sym{***}&       1.200\sym{***}&       0.975\sym{***}&       1.261\sym{***}&       1.950\sym{***}&       0.948\sym{**} &       2.302\sym{**} &       1.114\sym{***} \\
                    &     (0.092)         &     (0.087)         &     (0.165)         &     (0.280)         &     (0.232)         &     (0.313)         &     (0.559)         &     (0.320)         &     (0.770)         &     (0.325)          \\

\hline
Cluster             &         401         &         401         &         401         &         401         &         401         &         401         &         401         &         401         &         401         &         401         \\
Observations        &   2,667,593         &   2,667,593         &   2,667,593         &   2,667,593         &   2,667,593         &   2,667,593         &   2,667,593         &   2,667,593         &   2,667,593         &   2,667,593         \\
\hline\hline
\end{tabular}
\begin{tablenotes}
\raggedleft{Continued overleaf} 		
\end{tablenotes}
\end{threeparttable}
\end{table}
\end{landscape}

\setcounter{table}{0}

\begin{landscape}
\begin{table}[h!]\centering
\caption{Continued}
\begin{threeparttable}
\begin{tabular}{l*{10}{c}}
\hline\hline
\multicolumn{1}{c}{Percentile}                &\multicolumn{1}{c}{q21}
                    &\multicolumn{1}{c}{q22}
                    &\multicolumn{1}{c}{q23}
                    &\multicolumn{1}{c}{q24}
                    &\multicolumn{1}{c}{q25}
                    &\multicolumn{1}{c}{q26}
                    &\multicolumn{1}{c}{q27}
                    &\multicolumn{1}{c}{q28}
                    &\multicolumn{1}{c}{q29}
                    &\multicolumn{1}{c}{q30}\\
\multicolumn{1}{r}{Monthly Hours}                   
&\multicolumn{1}{c}{83.6}
&\multicolumn{1}{c}{86.5}
&\multicolumn{1}{c}{86.9}
&\multicolumn{1}{c}{87}
&\multicolumn{1}{c}{97.5}
&\multicolumn{1}{c}{104.3}
&\multicolumn{1}{c}{108.6}
&\multicolumn{1}{c}{115.8}
&\multicolumn{1}{c}{123.3}
&\multicolumn{1}{c}{130}\\
\hline
$Bite*Year_{2010}$  &      0.003         &       0.054         &       0.061         &      -0.001         &       0.164         &       0.135         &       0.095        &       0.289         &       0.191         &       0.036           \\
                    &    (0.067)         &     (0.045)         &     (0.046)         &     (0.049)         &     (0.242)         &     (0.139)         &     (0.096)         &     (0.237)         &     (0.125)         &     (0.045)          \\[1ex]
$Bite*Year_{2014}$  &\multicolumn{9}{c}{Reference} \\[1ex]
$Bite*Year_{2018}$  &      0.044         &       0.031         &       0.031         &       0.035         &       0.237         &       0.157         &       0.116         &       0.419\sym{*}  &       0.252\sym{*}  &       0.057          \\
                    &    (0.060)         &     (0.041)         &     (0.042)         &     (0.045)         &     (0.207)         &     (0.115)         &     (0.080)         &     (0.197)         &     (0.105)         &     (0.039)          \\[1ex]
$Year_{2010}$        &     -0.045\sym{***}&      -0.044\sym{***}&      -0.047\sym{***}&      -0.041\sym{***}&      -0.214\sym{***}&      -0.124\sym{***}&      -0.086\sym{***}&      -0.192\sym{***}&      -0.100\sym{***}&      -0.036\sym{***} \\
                    &    (0.008)         &     (0.005)         &     (0.006)         &     (0.006)         &     (0.029)         &     (0.017)         &     (0.012)         &     (0.029)         &     (0.016)         &     (0.006)          \\[1ex]
$Year_{2018}$       &      0.011         &       0.008         &       0.008         &       0.009         &       0.055\sym{*}  &       0.027         &       0.010         &       0.003         &      -0.010         &      -0.005          \\
                    &    (0.007)         &     (0.005)         &     (0.005)         &     (0.005)         &     (0.025)         &     (0.014)         &     (0.010)         &     (0.024)         &     (0.012)         &     (0.005)          \\[1ex]
$Bite$              &      0.501\sym{***}&       0.438\sym{***}&       0.453\sym{***}&       0.529\sym{***}&       2.612\sym{***}&       1.594\sym{***}&       1.131\sym{***}&       2.821\sym{***}&       1.466\sym{***}&       0.535\sym{***}  \\
                    &    (0.150)         &     (0.111)         &     (0.115)         &     (0.118)         &     (0.582)         &     (0.338)         &     (0.234)         &     (0.564)         &     (0.298)         &     (0.109)          \\
 
\hline
Cluster             &         401         &         401         &         401         &         401         &         401         &         401         &         401         &         401         &         401         &         401         \\
Observations        &   2,667,593         &   2,667,593         &   2,667,593         &   2,667,593         &   2,667,593         &   2,667,593         &   2,667,593         &   2,667,593         &   2,667,593         &   2,667,593         \\
\hline\hline
\end{tabular}
\begin{tablenotes} 
\raggedleft{Continued overleaf}
\end{tablenotes}
\end{threeparttable}
\end{table}
\end{landscape}

\setcounter{table}{0}

\begin{landscape}
\begin{table}[h!]\centering
\caption{Continued}
\begin{threeparttable}
\begin{tabular}{l*{10}{c}}
\hline\hline
\multicolumn{1}{r}{Percentile}                    &\multicolumn{1}{c}{q31}
                    &\multicolumn{1}{c}{q32}
                    &\multicolumn{1}{c}{q33}
                    &\multicolumn{1}{c}{q34}
                    &\multicolumn{1}{c}{q35}
                    &\multicolumn{1}{c}{q36}
                    &\multicolumn{1}{c}{q37}
                    &\multicolumn{1}{c}{q38}
                    &\multicolumn{1}{c}{q39}
                    &\multicolumn{1}{c}{q40}\\
 \multicolumn{1}{r}{Monthly Hours}                   
&\multicolumn{1}{c}{130.4}
&\multicolumn{1}{c}{130.4}
&\multicolumn{1}{c}{134.7}
&\multicolumn{1}{c}{140}
&\multicolumn{1}{c}{147}
&\multicolumn{1}{c}{151}
&\multicolumn{1}{c}{152.1}
&\multicolumn{1}{c}{152.1}
&\multicolumn{1}{c}{152.1}
&\multicolumn{1}{c}{152.1}\\                   
\hline
$Bite*Year_{2010}$  &       0.046         &       0.046         &       0.079         &       0.089         &       0.062         &       0.010         &       0.007         &       0.007         &       0.007         &      -0.142\sym{***} \\
                    &     (0.046)         &     (0.046)         &     (0.080)         &     (0.112)         &     (0.043)         &     (0.024)         &     (0.023)         &     (0.023)         &     (0.023)         &     (0.034)          \\[1ex]
$Bite*Year_{2014}$  &\multicolumn{9}{c}{Reference} \\[1ex]
$Bite*Year_{2018}$  &       0.064         &       0.064         &       0.137         &       0.178         &       0.063         &       0.000         &      -0.007         &      -0.007         &      -0.007         &      -0.029          \\
                    &     (0.039)         &     (0.039)         &     (0.075)         &     (0.107)         &     (0.040)         &     (0.021)         &     (0.021)         &     (0.021)         &     (0.021)         &     (0.024)          \\[1ex]
$Year_{2010}$        &      -0.037\sym{***}&      -0.037\sym{***}&      -0.049\sym{***}&      -0.059\sym{***}&      -0.026\sym{***}&      -0.015\sym{***}&      -0.014\sym{***}&      -0.014\sym{***}&      -0.014\sym{***}&       0.034\sym{***} \\
                    &     (0.006)         &     (0.006)         &     (0.010)         &     (0.014)         &     (0.005)         &     (0.003)         &     (0.003)         &     (0.003)         &     (0.003)         &     (0.005)          \\[1ex]
$Year_{2018}$       &      -0.005         &      -0.005         &      -0.015         &      -0.029\sym{*}  &      -0.011\sym{*}  &      -0.006\sym{**} &      -0.006\sym{*}  &      -0.006\sym{*}  &      -0.006\sym{*}  &      -0.007\sym{**}  \\
                    &     (0.005)         &     (0.005)         &     (0.009)         &     (0.012)         &     (0.005)         &     (0.002)         &     (0.002)         &     (0.002)         &     (0.002)         &     (0.003)          \\[1ex]
$Bite$              &       0.491\sym{***}&       0.491\sym{***}&       0.498\sym{**} &       0.660\sym{**} &       0.243\sym{**} &       0.137\sym{**} &       0.125\sym{**} &       0.125\sym{**} &       0.125\sym{**} &       0.241\sym{***}  \\
                    &     (0.109)         &     (0.109)         &     (0.174)         &     (0.232)         &     (0.087)         &     (0.045)         &     (0.044)         &     (0.044)         &     (0.044)         &     (0.052)          \\
\hline
Cluster             &         401         &         401         &         401         &         401         &         401         &         401         &         401         &         401         &         401         &         401         \\
Observations        &   2,667,593         &   2,667,593         &   2,667,593         &   2,667,593         &   2,667,593         &   2,667,593         &   2,667,593         &   2,667,593         &   2,667,593         &   2,667,593         \\
\hline
\hline
\end{tabular}
\begin{tablenotes}
\raggedleft{Continued overleaf}
\end{tablenotes}
\end{threeparttable}
\end{table}
\end{landscape}

\begin{landscape}
\begin{table}[h!]\centering
\caption{RIF regressions of log contractual monthly working hours along the hours distribution}
\begin{threeparttable}
\begin{tabular}{l*{10}{c}}
\hline\hline
\multicolumn{1}{c}{Percentile}                    &\multicolumn{1}{c}{q41}
                    &\multicolumn{1}{c}{q42}
                    &\multicolumn{1}{c}{q43}
                    &\multicolumn{1}{c}{q44}
                    &\multicolumn{1}{c}{q45}
                    &\multicolumn{1}{c}{q46}
                    &\multicolumn{1}{c}{q47}
                    &\multicolumn{1}{c}{q48}
                    &\multicolumn{1}{c}{q49}
                    &\multicolumn{1}{c}{q50}\\
\multicolumn{1}{r}{Monthly Hours}                   
&\multicolumn{1}{c}{152.1}
&\multicolumn{1}{c}{152.3}
&\multicolumn{1}{c}{155.6}
&\multicolumn{1}{c}{157.5}
&\multicolumn{1}{c}{160}
&\multicolumn{1}{c}{160.8}
&\multicolumn{1}{c}{162.4}
&\multicolumn{1}{c}{162.9}
&\multicolumn{1}{c}{162.9}
&\multicolumn{1}{c}{163}\\
\hline
$Bite*Year_{2010}$  &      -0.142\sym{***}&       0.005         &       0.035         &       0.029         &       0.017         &       0.016         &       0.019         &       0.018         &       0.018         &       0.003          \\        
                    &     (0.034)         &     (0.028)         &     (0.031)         &     (0.029)         &     (0.022)         &     (0.020)         &     (0.016)         &     (0.014)         &     (0.014)         &     (0.015)          \\[1ex]   
$Bite*Year_{2014}$  &\multicolumn{9}{c}{Reference} \\[1ex]
$Bite*Year_{2018}$  &      -0.029         &      -0.036         &      -0.037         &      -0.040         &      -0.042\sym{*}  &      -0.044\sym{*}  &      -0.034\sym{*}  &      -0.033\sym{**} &      -0.033\sym{**} &      -0.039\sym{**}  \\        
                    &     (0.024)         &     (0.025)         &     (0.027)         &     (0.027)         &     (0.020)         &     (0.017)         &     (0.013)         &     (0.012)         &     (0.012)         &     (0.013)          \\[1ex]   
$Year_{2010}$        &       0.034\sym{***}&      -0.010\sym{**} &      -0.012\sym{**} &      -0.012\sym{***}&      -0.012\sym{***}&      -0.013\sym{***}&      -0.011\sym{***}&      -0.010\sym{***}&      -0.010\sym{***}&      -0.007\sym{***} \\    
                    &     (0.005)         &     (0.004)         &     (0.004)         &     (0.004)         &     (0.003)         &     (0.003)         &     (0.002)         &     (0.002)         &     (0.002)         &     (0.002)          \\[1ex]   
$Year_{2018}$       &      -0.007\sym{**} &      -0.007\sym{*}  &      -0.004         &      -0.005         &      -0.005\sym{*}  &      -0.005\sym{**} &      -0.004\sym{**} &      -0.003\sym{*}  &      -0.003\sym{*}  &      -0.002          \\        
                    &     (0.003)         &     (0.003)         &     (0.003)         &     (0.003)         &     (0.002)         &     (0.002)         &     (0.001)         &     (0.001)         &     (0.001)         &     (0.002)          \\[1ex]   
$Bite$              &       0.241\sym{***}&       0.228\sym{***}&       0.267\sym{***}&       0.277\sym{***}&       0.211\sym{***}&       0.189\sym{***}&       0.156\sym{***}&       0.147\sym{***}&       0.147\sym{***}&       0.193\sym{***}  \\       
                    &     (0.052)         &     (0.049)         &     (0.055)         &     (0.053)         &     (0.040)         &     (0.036)         &     (0.029)         &     (0.027)         &     (0.027)         &     (0.027)          \\ 
\hline
Cluster             &         401         &         401         &         401         &         401         &         401         &         401         &         401         &         401         &         401         &         401         \\
Observations        &   2,667,593         &   2,667,593         &   2,667,593         &   2,667,593         &   2,667,593         &   2,667,593         &   2,667,593         &   2,667,593         &   2,667,593         &   2,667,593         \\
\hline
\hline
\end{tabular}
\begin{tablenotes}
\raggedleft{Continued overleaf} 
\end{tablenotes}
\end{threeparttable}
\end{table}
\end{landscape}

\setcounter{table}{0}

\begin{landscape}
\begin{table}[h!]\centering
\caption{Continued}
\begin{threeparttable}
\begin{tabular}{l*{10}{c}}
\hline\hline
\multicolumn{1}{r}{Percentile}                &\multicolumn{1}{c}{q51}
                    &\multicolumn{1}{c}{q52}
                    &\multicolumn{1}{c}{q53}
                    &\multicolumn{1}{c}{q54}
                    &\multicolumn{1}{c}{q55}
                    &\multicolumn{1}{c}{q56}
                    &\multicolumn{1}{c}{q57}
                    &\multicolumn{1}{c}{q58}
                    &\multicolumn{1}{c}{q59}
                    &\multicolumn{1}{c}{q60}\\
\multicolumn{1}{r}{Monthly Hours}                   
&\multicolumn{1}{c}{164}
&\multicolumn{1}{c}{165.1}
&\multicolumn{1}{c}{165.1}
&\multicolumn{1}{c}{165.1}
&\multicolumn{1}{c}{165.3}
&\multicolumn{1}{c}{167}
&\multicolumn{1}{c}{167.3}
&\multicolumn{1}{c}{167.3}
&\multicolumn{1}{c}{167.3}
&\multicolumn{1}{c}{168}\\
\hline
$Bite*Year_{2010}$  &      -0.000         &       0.004         &       0.004         &       0.004         &       0.009         &       0.010         &       0.009         &       0.009         &       0.009         &       0.016          \\
                    &     (0.012)         &     (0.011)         &     (0.011)         &     (0.011)         &     (0.012)         &     (0.010)         &     (0.010)         &     (0.010)         &     (0.010)         &     (0.010)          \\[1ex]
$Bite*Year_{2014}$  &\multicolumn{9}{c}{Reference} \\[1ex]
$Bite*Year_{2018}$  &      -0.036\sym{**} &      -0.037\sym{***}&      -0.037\sym{***}&      -0.037\sym{***}&      -0.033\sym{**} &      -0.029\sym{**} &      -0.025\sym{**} &      -0.025\sym{**} &      -0.025\sym{**} &      -0.025\sym{*}   \\
                    &     (0.012)         &     (0.011)         &     (0.011)         &     (0.011)         &     (0.012)         &     (0.010)         &     (0.010)         &     (0.010)         &     (0.010)         &     (0.010)          \\[1ex]
$Year_{2010}$        &      -0.008\sym{***}&      -0.007\sym{***}&      -0.007\sym{***}&      -0.007\sym{***}&      -0.007\sym{***}&      -0.007\sym{***}&      -0.006\sym{***}&      -0.006\sym{***}&      -0.006\sym{***}&      -0.007\sym{***} \\
                    &     (0.002)         &     (0.001)         &     (0.001)         &     (0.001)         &     (0.001)         &     (0.001)         &     (0.001)         &     (0.001)         &     (0.001)         &     (0.001)          \\[1ex]
$Year_{2018}$       &      -0.003\sym{*}  &      -0.002         &      -0.002         &      -0.002         &      -0.002         &      -0.001         &      -0.001         &      -0.001         &      -0.001         &      -0.000          \\
                    &     (0.001)         &     (0.001)         &     (0.001)         &     (0.001)         &     (0.001)         &     (0.001)         &     (0.001)         &     (0.001)         &     (0.001)         &     (0.001)          \\[1ex]
$Bite$              &       0.175\sym{***}&       0.153\sym{***}&       0.153\sym{***}&       0.153\sym{***}&       0.125\sym{***}&       0.100\sym{***}&       0.101\sym{***}&       0.101\sym{***}&       0.101\sym{***}&       0.125\sym{***}  \\
                    &     (0.024)         &     (0.022)         &     (0.022)         &     (0.022)         &     (0.022)         &     (0.019)         &     (0.019)         &     (0.019)         &     (0.019)         &     (0.017)          \\
\hline
Cluster             &         401         &         401         &         401         &         401         &         401         &         401         &         401         &         401         &         401         &         401         \\
Observations        &   2,667,593         &   2,667,593         &   2,667,593         &   2,667,593         &   2,667,593         &   2,667,593         &   2,667,593         &   2,667,593         &   2,667,593         &   2,667,593         \\
\hline
\hline
\end{tabular}
\begin{tablenotes}
\raggedleft{Continued overleaf}  			
\end{tablenotes}
\end{threeparttable}
\end{table}
\end{landscape}

\setcounter{table}{0}

\begin{landscape}
\begin{table}[h!]\centering
\caption{Continued}
\begin{threeparttable}
\begin{tabular}{l*{10}{c}}
\hline\hline
\multicolumn{1}{r}{Percentile}                &\multicolumn{1}{c}{q61}
                    &\multicolumn{1}{c}{q62}
                    &\multicolumn{1}{c}{q63}
                    &\multicolumn{1}{c}{q64}
                    &\multicolumn{1}{c}{q65}
                    &\multicolumn{1}{c}{q66}
                    &\multicolumn{1}{c}{q67}
                    &\multicolumn{1}{c}{q68}
                    &\multicolumn{1}{c}{q69}
                    &\multicolumn{1}{c}{q70}\\
\multicolumn{1}{r}{Monthly Hours}                   
&\multicolumn{1}{c}{168}
&\multicolumn{1}{c}{168}
&\multicolumn{1}{c}{169}
&\multicolumn{1}{c}{169.5}
&\multicolumn{1}{c}{169.5}
&\multicolumn{1}{c}{169.5}
&\multicolumn{1}{c}{169.5}
&\multicolumn{1}{c}{169.5}
&\multicolumn{1}{c}{169.6}
&\multicolumn{1}{c}{171.6}\\
\hline
 $Bite*Year_{2010}$  &       0.016         &       0.016         &      -0.063\sym{***}&      -0.064\sym{***}&      -0.064\sym{***}&      -0.064\sym{***}&      -0.064\sym{***}&      -0.064\sym{***}&      -0.058\sym{***}&      -0.058\sym{***} \\
                    &     (0.010)         &     (0.010)         &     (0.013)         &     (0.013)         &     (0.013)         &     (0.013)         &     (0.013)         &     (0.013)         &     (0.013)         &     (0.012)          \\[1ex]
$Bite*Year_{2014}$    &\multicolumn{9}{c}{Reference} \\[1ex]
$Bite*Year_{2018}$  &      -0.025\sym{*}  &      -0.025\sym{*}  &      -0.112\sym{***}&      -0.116\sym{***}&      -0.116\sym{***}&      -0.116\sym{***}&      -0.116\sym{***}&      -0.116\sym{***}&      -0.116\sym{***}&      -0.111\sym{***} \\
                    &     (0.010)         &     (0.010)         &     (0.013)         &     (0.012)         &     (0.012)         &     (0.012)         &     (0.012)         &     (0.012)         &     (0.012)         &     (0.011)          \\[1ex]
$Year_{2010}$        &      -0.007\sym{***}&      -0.007\sym{***}&      -0.005\sym{***}&      -0.004\sym{***}&      -0.004\sym{***}&      -0.004\sym{***}&      -0.004\sym{***}&      -0.004\sym{***}&      -0.004\sym{**} &      -0.004\sym{***} \\
                    &     (0.001)         &     (0.001)         &     (0.001)         &     (0.001)         &     (0.001)         &     (0.001)         &     (0.001)         &     (0.001)         &     (0.001)         &     (0.001)          \\[1ex]
$Year_{2018}$       &      -0.000         &      -0.000         &       0.001         &       0.002         &       0.002         &       0.002         &       0.002         &       0.002         &       0.002\sym{*}  &       0.003\sym{*}   \\
                    &     (0.001)         &     (0.001)         &     (0.001)         &     (0.001)         &     (0.001)         &     (0.001)         &     (0.001)         &     (0.001)         &     (0.001)         &     (0.001)          \\[1ex]
$Bite$              &       0.125\sym{***}&       0.125\sym{***}&       0.117\sym{***}&       0.121\sym{***}&       0.121\sym{***}&       0.121\sym{***}&       0.121\sym{***}&       0.121\sym{***}&       0.154\sym{***}&       0.151\sym{***}  \\
                    &     (0.017)         &     (0.017)         &     (0.016)         &     (0.016)         &     (0.016)         &     (0.016)         &     (0.016)         &     (0.016)         &     (0.016)         &     (0.015)          \\
\hline
Cluster             &         401         &         401         &         401         &         401         &         401         &         401         &         401         &         401         &         401         &         401         \\
Observations        &   2,667,593         &   2,667,593         &   2,667,593         &   2,667,593         &   2,667,593         &   2,667,593         &   2,667,593         &   2,667,593         &   2,667,593         &   2,667,593         \\
\hline
\hline
\end{tabular}
\begin{tablenotes}
\raggedleft{Continued overleaf}  
\end{tablenotes}
\end{threeparttable}
\end{table}
\end{landscape}

\setcounter{table}{0}

\begin{landscape}
\begin{table}[h!]\centering
\caption{Continued}
\begin{threeparttable}
\begin{tabular}{l*{10}{c}}
\hline\hline
\multicolumn{1}{r}{Percentile}  &\multicolumn{1}{c}{q71}
                    &\multicolumn{1}{c}{q72}
                    &\multicolumn{1}{c}{q73}
                    &\multicolumn{1}{c}{q74}
                    &\multicolumn{1}{c}{q75}
                    &\multicolumn{1}{c}{q76}
                    &\multicolumn{1}{c}{\ldots}
                    &\multicolumn{1}{c}{\ldots}
                    &\multicolumn{1}{c}{q89}
                    &\multicolumn{1}{c}{q90}\\
\multicolumn{1}{r}{Monthly Hours}                   
&\multicolumn{1}{c}{173}
&\multicolumn{1}{c}{173}
&\multicolumn{1}{c}{173.3}
&\multicolumn{1}{c}{173.8}
&\multicolumn{1}{c}{173.8}
&\multicolumn{1}{c}{173.8}
&\multicolumn{1}{c}{\ldots}
&\multicolumn{1}{c}{\ldots}
&\multicolumn{1}{c}{173.8}
&\multicolumn{1}{c}{173.8}\\
\hline
$Bite*Year_{2010}$  &      -0.057\sym{***}&      -0.057\sym{***}&      -0.063\sym{***}&      -0.065\sym{***}&      -0.065\sym{***}&      -0.065\sym{***}&  & &      -0.065\sym{***}&      -0.065\sym{***} \\     
                    &     (0.012)         &     (0.012)         &     (0.013)         &     (0.014)         &     (0.014)         &     (0.014)         &  & &     (0.014)         &     (0.014)          \\[1ex]
$Bite*Year_{2014}$  &     \multicolumn{9}{c}{Reference} \\[1ex]      
$Bite*Year_{2018}$  &      -0.114\sym{***}&      -0.114\sym{***}&      -0.116\sym{***}&      -0.115\sym{***}&      -0.115\sym{***}&      -0.115\sym{***}&  & &      -0.115\sym{***}&      -0.115\sym{***} \\     
                    &     (0.012)         &     (0.012)         &     (0.012)         &     (0.013)         &     (0.013)         &     (0.013)         &  & &     (0.013)         &     (0.013)          \\[1ex]
$Year_{2010}$       &      -0.003\sym{**} &      -0.003\sym{**} &      -0.003\sym{*}  &      -0.003\sym{*}  &      -0.003\sym{*}  &      -0.003\sym{*}  &  & &      -0.003\sym{*}  &      -0.003\sym{*}   \\     
                    &     (0.001)         &     (0.001)         &     (0.001)         &     (0.001)         &     (0.001)         &     (0.001)         &  & &     (0.001)         &     (0.001)          \\[1ex]
$Year_{2018}$       &       0.004\sym{**} &       0.004\sym{**} &       0.003\sym{**} &       0.003         &       0.003         &       0.003         &  & &       0.003         &       0.003          \\     
                    &     (0.001)         &     (0.001)         &     (0.001)         &     (0.001)         &     (0.001)         &     (0.001)         &  & &     (0.001)         &     (0.001)          \\[1ex]
$Bite$              &       0.151\sym{***}&       0.151\sym{***}&       0.152\sym{***}&       0.150\sym{***}&       0.150\sym{***}&       0.150\sym{***}&  & &       0.150\sym{***}&       0.150\sym{***}  \\    
                    &     (0.015)         &     (0.015)         &     (0.016)         &     (0.016)         &     (0.016)         &     (0.016)         &  & &     (0.016)         &     (0.016)          \\     
\hline
Cluster             &         401         &         401         &         401         &         401         &         401         &         401         &                  &                  &         401         &         401         \\
Observations        &   2,667,593         &   2,667,593         &   2,667,593         &   2,667,593         &   2,667,593         &   2,667,593         &            &            &   2,667,593         &   2,667,593         \\
\hline
\hline
\end{tabular}
\begin{tablenotes}
\begin{small}
\textit{Notes:}  Percentiles 77--88 not reported. Coefficient estimates are identical for percentiles 74--90 because they are all located at the same hours value.     
\end{small}  
\end{tablenotes}
\raggedleft{Continued overleaf}
\end{threeparttable}
\end{table}
\end{landscape}

\setcounter{table}{0}

\begin{landscape}
\begin{table}[h!]\centering
\caption{Continued}
\begin{threeparttable}
\begin{tabular}{l*{9}{c}}
\hline\hline
\multicolumn{1}{r}{Percentile}                    &\multicolumn{1}{c}{q91}
                    &\multicolumn{1}{c}{q92}
                    &\multicolumn{1}{c}{q93}
                    &\multicolumn{1}{c}{q94}
                    &\multicolumn{1}{c}{q95}
                    &\multicolumn{1}{c}{q96}
                    &\multicolumn{1}{c}{q97}
                    &\multicolumn{1}{c}{q98}
                    &\multicolumn{1}{c}{q99}
                    \\
\multicolumn{1}{r}{Monthly Hours}                   
&\multicolumn{1}{c}{174}
&\multicolumn{1}{c}{174}
&\multicolumn{1}{c}{174}
&\multicolumn{1}{c}{174.7}
&\multicolumn{1}{c}{176}
&\multicolumn{1}{c}{178}
&\multicolumn{1}{c}{182.5}
&\multicolumn{1}{c}{187}
&\multicolumn{1}{c}{198.5}
\\
\hline
$Bite*Year_{2010}$  &    -0.097\sym{***}&      -0.097\sym{***}&      -0.097\sym{***}&      -0.100\sym{***}&      -0.118\sym{***}&      -0.013         &      -0.002         &       0.038         &       0.087         \\         
                    &   (0.012)         &     (0.012)         &     (0.012)         &     (0.012)         &     (0.014)         &     (0.008)         &     (0.021)         &     (0.048)         &     (0.074)         \\[1ex]    
$Bite*Year_{2014}$  &   \multicolumn{9}{c}{Reference} \\[1ex]                                                                       
$Bite*Year_{2018}$  &    -0.113\sym{***}&      -0.113\sym{***}&      -0.113\sym{***}&      -0.115\sym{***}&      -0.135\sym{***}&      -0.038\sym{***}&      -0.057\sym{*}  &      -0.071         &      -0.049         \\         
                    &   (0.012)         &     (0.012)         &     (0.012)         &     (0.011)         &     (0.013)         &     (0.008)         &     (0.024)         &     (0.052)         &     (0.075)         \\[1ex]    
$Year_{2010}$        &     0.004\sym{***}&       0.004\sym{***}&       0.004\sym{***}&       0.000         &       0.001         &      -0.002\sym{*}  &      -0.005\sym{*}  &      -0.014\sym{*}  &      -0.009         \\         
                    &   (0.001)         &     (0.001)         &     (0.001)         &     (0.001)         &     (0.001)         &     (0.001)         &     (0.003)         &     (0.006)         &     (0.009)         \\[1ex]    
$Year_{2018}$       &    -0.000         &      -0.000         &      -0.000         &       0.001         &       0.001         &      -0.001         &      -0.005         &      -0.014\sym{*}  &      -0.022\sym{*}  \\         
                    &   (0.001)         &     (0.001)         &     (0.001)         &     (0.001)         &     (0.001)         &     (0.001)         &     (0.003)         &     (0.006)         &     (0.009)         \\[1ex]    
$Bite$              &     0.112\sym{***}&       0.112\sym{***}&       0.112\sym{***}&       0.108\sym{***}&       0.126\sym{***}&       0.023\sym{**} &       0.014         &      -0.016         &      -0.070          \\        
                    &   (0.012)         &     (0.012)         &     (0.012)         &     (0.011)         &     (0.013)         &     (0.007)         &     (0.019)         &     (0.047)         &     (0.061)         \\         
\hline
Cluster             &         401         &         401         &         401         &         401         &         401         &         401         &         401         &         401         &         401          \\
Observations        &   2,667,593         &   2,667,593         &   2,667,593         &   2,667,593         &   2,667,593         &   2,667,593         &   2,667,593         &   2,667,593         &   2,667,593         \\
\hline
\hline
\end{tabular}
\end{threeparttable}
\end{table}
\end{landscape}

\clearpage
\section{Working hours distributions by type of employment, excluding small establishments} \label{app:hours_distributions}
\renewcommand*{\thefigure}{\thesection\arabic{figure}}
\renewcommand*{\thetable}{\thesection\arabic{table}}
\setcounter{figure}{0}
\setcounter{table}{0}

\begin{figure}[ht!]
	\captionabove{Histograms of total monthly working hours in establishments with at least 10 employees, 5-hour bins}
\label{fig:histogram_hours_10_plus}
	\centering
	\includegraphics[width=0.9\textwidth]{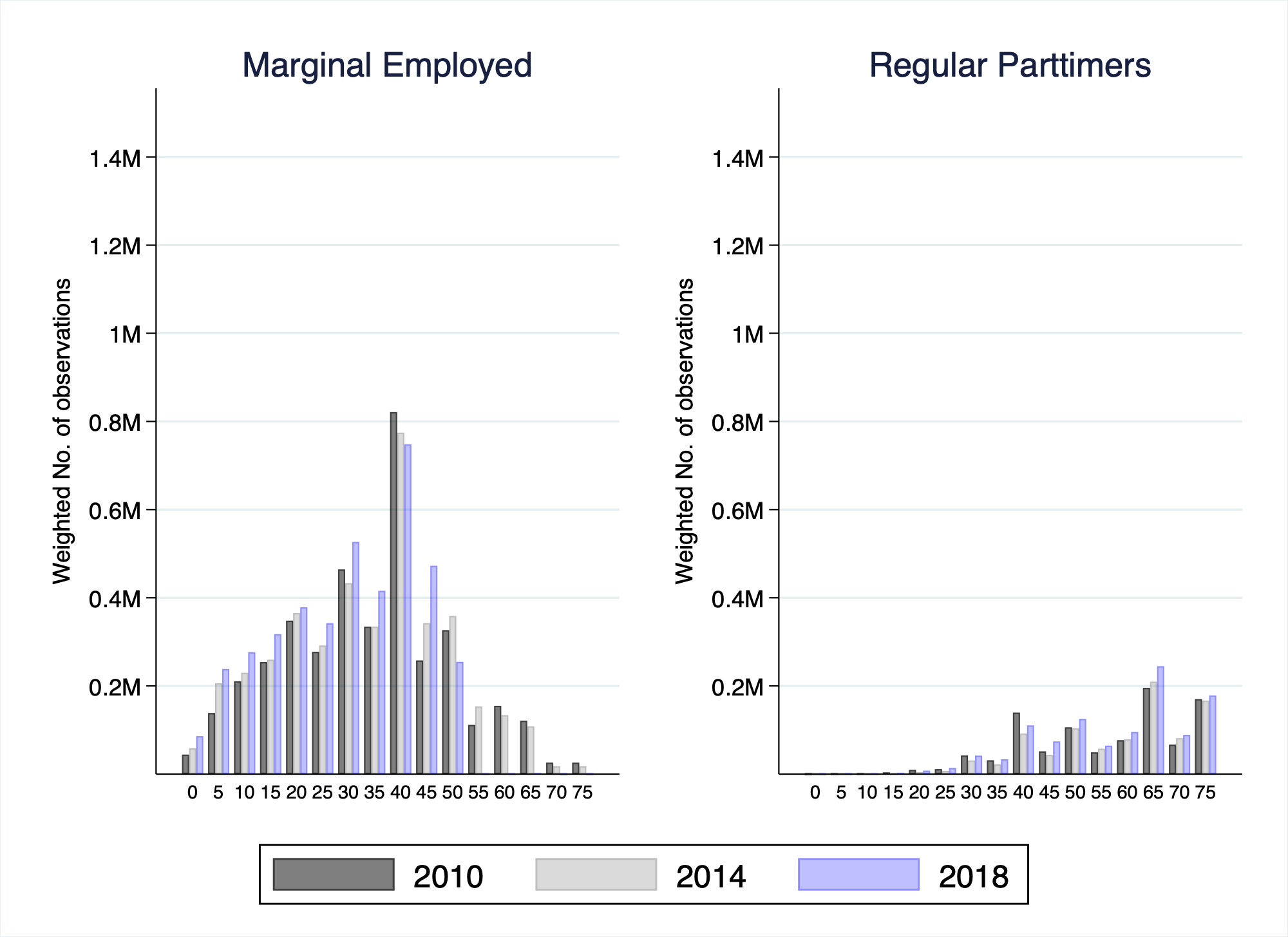}
\subcaption*{
\textit{Notes:} Histogram of monthly working hours of minijobs and part-time jobs as well as for regular full-time jobs, bin width is 5 hours. \\\textit{Data:} SES, 2014 and 2018, weighted analysis sample, establishments with at least 10 regular employees.\\
}
\end{figure}

\clearpage
\section{Inspection of employment trends}\label{app:employment_trends}
\renewcommand*{\thefigure}{\thesection\arabic{figure}}
\renewcommand*{\thetable}{\thesection\arabic{table}}
\setcounter{figure}{0}
\setcounter{table}{0}

\begin{table}[htbp]\centering
\caption{Minimum wage effect on log employment, not detrended}
\label{tab:employment_no_trend}
\begin{threeparttable}
				\begin{tabular}{{lcccccc}}
					\hline\hline
					&\multicolumn{2}{c}{all jobs}               &\multicolumn{2}{c}{regular employment}     &\multicolumn{2}{c}{minijobs}               \\\cmidrule(lr){2-3}\cmidrule(lr){4-5}\cmidrule(lr){6-7}
					&\multicolumn{1}{c}{10+ plants}&\multicolumn{1}{c}{all plants}&\multicolumn{1}{c}{10+ plants}&\multicolumn{1}{c}{all plants}&\multicolumn{1}{c}{10+ plants}&\multicolumn{1}{c}{all plants}\\[1ex]\hline \\ [-1.4ex]
$Bite*Year_{2010}$     &      0.225\sym{***}&      0.202\sym{***}&      0.123\sym{*}  &      0.148\sym{***}&      0.513\sym{***}&      0.291\sym{***}\\
                    &    (0.049)         &    (0.034)         &    (0.049)         &    (0.036)         &    (0.110)         &    (0.049)         \\[1ex]
$Bite*Year_{2014}$    & \multicolumn{6}{c}{Reference} \\[1ex]
$Bite*Year_{2018}$     &     -0.192\sym{***}&     -0.194\sym{***}&     -0.134\sym{***}&     -0.170\sym{***}&     -0.756\sym{***}&     -0.440\sym{***}\\
                    &    (0.039)         &    (0.027)         &    (0.033)         &    (0.026)         &    (0.221)         &    (0.090)         \\[1ex]
$Year_{2010}$           &     -0.099\sym{***}&     -0.091\sym{***}&     -0.102\sym{***}&     -0.101\sym{***}&     -0.080\sym{***}&     -0.058\sym{***}\\
                    &    (0.005)         &    (0.005)         &    (0.006)         &    (0.006)         &    (0.010)         &    (0.007)         \\[1ex]
$Year_{2014}$        & \multicolumn{6}{c}{Reference} \\[2ex]
$Year_{2018}$           &      0.095\sym{***}&      0.087\sym{***}&      0.111\sym{***}&      0.112\sym{***}&      0.058\sym{**} &      0.028\sym{**} \\
                    &    (0.005)         &    (0.004)         &    (0.005)         &    (0.004)         &    (0.018)         &    (0.010)         \\[1ex]
$Bite$                &     -5.140\sym{***}&     -4.966\sym{***}&     -5.229\sym{***}&     -4.937\sym{***}&     -4.656\sym{***}&     -5.058\sym{***}\\
                    &    (1.395)         &    (1.247)         &    (1.463)         &    (1.313)         &    (1.205)         &    (1.048)         \\[1ex]
$Constant$            &     12.014\sym{***}&     12.358\sym{***}&     11.827\sym{***}&     12.109\sym{***}&     10.065\sym{***}&     10.717\sym{***}\\
                    &    (0.222)         &    (0.230)         &    (0.236)         &    (0.247)         &    (0.170)         &    (0.180)         \\[1ex]
\hline \\[-1.4ex]
Clusters            &         401         &         401         &         401         &         401         &         401         &         401         \\
Observations        &        1203         &        1203         &        1203         &        1203         &        1203         &        1203         \\[1ex]
\hline\hline
\end{tabular}
\begin{tablenotes}
\begin{small}  \textit{Notes:} Difference-in-differences regression coefficients from a specification similar to equation~(\ref{eq:did_regional}), but without the bite-specific trend (Trend$\ast$Bite). The regressions are run at the county-level and weighted by county-level employment.
The dependent variable is logarithmic employment of either all jobs, regular employment, or minijobs, as indicated by column titles. 
Standard errors in parentheses are clustered at the county level.
Asterisks indicate significance levels: 
   *~$p<0.05$, **~$p<0.01$, ***~$p<0.001$.
 \\ \textit{Data:} BHP of 2010, 2014, and 2018; full population aggregated to counties. 
\end{small}
\end{tablenotes}
\end{threeparttable}
\end{table}

Table \ref{tab:employment_no_trend} presents the findings of the minimum wage effect on log employment at the county level, without accounting for pre-determined bite-specific trends. In all columns, the coefficient of $Bite*Year_{2010}$ indicates a positive time trend before 2014, suggesting that regions with higher bite experienced greater decreases in employment (between 2010 and 2014) even before the minimum wage implementation. This trend is particularly pronounced in the case of minijobs.

Furthermore, the coefficients of $Bite*Year_{2018}$ reveal a continuation of this negative trend until 2018, as the two effect sizes are similar in absolute terms (i.e., the coefficients of $Bite*Year_{2010}$ and $Bite*Year_{2018}$). Yet, for minijobs the coefficients of $Bite*Year_{2018}$ are larger (in absolute terms) than those of $Bite*Year_{2010}$, suggesting a reduction in minijob employment after the minimum wage implementation.

Based on the occurrence of pre-treatment bite-specific trends reported above, we detrend the employment regressions presented in the main text (Section \ref{sec:employment}).


\clearpage
\section{Transition analysis for establishments with at least 10 regular employees}\label{app:transitions_10+}
\renewcommand*{\thefigure}{\thesection\arabic{figure}}
\renewcommand*{\thetable}{\thesection\arabic{table}}
\setcounter{figure}{0}
\setcounter{table}{0}

\begin{table}[htbp]\centering
\caption{Minimum wage effect on forward-looking transitions out of minijobs, establishments with at least 10 regular employees}
\begin{threeparttable}
\begin{tabular}{lC{3.7cm}C{3.7cm}C{3.7cm}}
\hline \hline
                    &\multicolumn{1}{c}{out of minijob}&\multicolumn{1}{c}{minijob to regular job}&\multicolumn{1}{c}{minijob to non-empl.}\\[1ex]
\hline \\[-1.4ex]
$Bite * Year_{2012} $ & Reference & Reference & Reference \\[1ex]
$Bite * Year_{2013} $ &      -0.019         &      -0.008         &      -0.011         \\
                    &     (0.010)         &     (0.005)         &     (0.009)         \\
$Bite * Year_{2014} $ &       0.227\sym{***}&       0.093\sym{***}&       0.133\sym{**} \\
                    &     (0.042)         &     (0.010)         &     (0.042)         \\[1ex]
\hline \\[-1.4ex]
Clusters            &         401         &         401         &         401         \\
Observations        &     8,847,758         &     8,847,758         &     8,847,758         \\[1ex]
\hline\hline
\end{tabular}
\begin{tablenotes}
\begin{small} \textit{Notes:} Treatment effect interactions from difference-in-differences estimations, as specified in equation~(\ref{eq:minijob_trans}). Dependent variable captures forward-looking transitions out of minijobs, as indicated by column titles.  
Standard errors in parentheses are clustered at the county level. Asterisks indicate significance levels:    *~$p<0.05$, **~$p<0.01$, ***~$p<0.001$. \\\textit{Data:} IEB, 2012-2015, population of all minijobbers in establishment with at least 10 regular employees, multiple jobs excluded.
\end{small}
\end{tablenotes}
\end{threeparttable}
\end{table}

\begin{table}[htbp]\centering
\caption{Minimum wage effect on backward-looking transitions in minijobs, establishments with at least 10 regular employees}
\begin{threeparttable}
\begin{tabular}{lC{3.7cm}C{3.7cm}C{3.7cm}}
\hline \hline
                    &\multicolumn{1}{c}{in minijob}&\multicolumn{1}{c}{regular job to minijob}&\multicolumn{1}{c}{non-empl. to minijob}\\[1ex]
\hline \\[-1.4ex]
$Bite * Year_{2013}$ & Reference & Reference & Reference \\[1ex]
$Bite * Year_{2014}$ &      -0.053\sym{**} &      -0.022\sym{***}&      -0.028         \\
                    &     (0.018)         &     (0.004)         &     (0.018)         \\
$Bite * Year_{2015}$  &       0.037         &       0.062\sym{***}&      -0.021         \\
                    &     (0.024)         &     (0.008)         &     (0.025)         \\[1ex]
\hline \\[-1.4ex]
Clusters            &         401         &         401         &         401         \\
Observations        &     8,729,539         &     8,729,539         &     8,729,539         \\[1ex]
\hline\hline
\end{tabular}
\begin{tablenotes}
\begin{small} \textit{Notes:}    Treatment effect interactions from difference-in-differences estimations, as specified in equation~(\ref{eq:minijob_trans}). Dependent 
variable captures backward-looking transitions in minijobs, as indicated by column titles.  
Standard errors in parentheses are clustered at the county level.  Asterisks indicate significance levels: *~$p<0.05$, **~$p<0.01$, ***~$p<0.001$. \\\textit{Data:} IEB, 2012-2015, population of all minijobbers in establishment with at least 10 regular employees, multiple jobs excluded.
\end{small}
\end{tablenotes}
\end{threeparttable}
\end{table}


\clearpage
\section{Transitions in minijobs}\label{app:in_minijobs}
\renewcommand*{\thefigure}{\thesection\arabic{figure}}
\renewcommand*{\thetable}{\thesection\arabic{table}}
\setcounter{figure}{0}
\setcounter{table}{0}

\begin{table}[htbp]\centering
\caption{Minimum wage effect on backward-looking transitions in minijobs}
\label{tab:in_minijobs}
\begin{threeparttable}
\begin{tabular}{lC{3.7cm}C{3.7cm}C{3.7cm}}
\hline\hline
                    &\multicolumn{1}{c}{in minijob}&\multicolumn{1}{c}{regular job to minijob}&\multicolumn{1}{c}{non-empl. to minijob}\\[1ex]
\hline \\[-1.4ex]
$Bite * Year_{2013}$ & Reference & Reference & Reference \\[1ex]
$Bite * Year_{2014}$ &      -0.057\sym{***}&      -0.029\sym{***}&      -0.025\sym{*}  \\
                    &     (0.012)         &     (0.003)         &     (0.012)         \\
$Bite * Year_{2015}$ &       0.006         &       0.041\sym{***}&      -0.031         \\
                    &     (0.017)         &     (0.006)         &     (0.017)         \\[1ex]
\hline \\[-1.4ex]
Clusters           &     401  &     401  &     401    \\
Observations        &    13,456,194         &    13,456,194         &    13,456,194         \\[1ex]
\hline\hline
\end{tabular}
\begin{tablenotes}
\begin{small}  \textit{Notes:} 
Treatment effect interactions from difference-in-differences estimations, as specified in equation~(\ref{eq:minijob_trans}).
Dependent variable captures backward-looking transitions in minijobs, as indicated by column titles.  
Standard errors in parentheses are clustered at the county level. Asterisks indicate significance levels:    *~$p<0.05$, **~$p<0.01$, ***~$p<0.001$. \\\textit{Data:} IEB, 2012-2015, population of all minijobbers, multiple jobs excluded.
\end{small}
\end{tablenotes}
\end{threeparttable}
\end{table}

\begin{table}[ht!]\centering
\caption{Minimum wage effect on demotions (from regular to minijob) within and across establishments}
\label{tab:transition_demotion_within}
\begin{threeparttable}
\begin{tabular}{lC{4.2cm}C{4.2cm}}
\hline\hline
 & \multicolumn{2}{c}{regular job to minijob}\\
                    &\multicolumn{1}{c}{within establishments}&\multicolumn{1}{c}{across establishments}\\[1ex]
\hline \\[-1.4ex]
$Bite * Year_{2013}$ & Reference  & Reference     \\[1ex]
$Bite * Year_{2014}$ &      -0.012\sym{***}&      -0.017\sym{***}\\
                    &     (0.002)         &     (0.002)         \\
$Bite * Year_{2015}$ &       0.048\sym{***}&      -0.008\sym{**} \\
                    &     (0.005)         &     (0.003)         \\[1ex]
\hline \\[-1.4ex]
Clusters         &    401      &         401       \\
Observations     & 13,456,194         &    13,456,194      \\[1ex]
\hline\hline
\end{tabular}
\begin{tablenotes}
\begin{small} \textit{Notes:} 
   Treatment effect interactions from difference-in-differences estimations, as specified in equation~(\ref{eq:minijob_trans}).
Dependent variable captures backward-looking transitions in minijobs, as indicated by column titles.  
Standard errors in parentheses are clustered at the county level. Asterisks indicate significance levels:    *~$p<0.05$, **~$p<0.01$, ***~$p<0.001$. \\\textit{Data:} IEB, 2012-2015, population of all minijobbers, multiple jobs excluded.
\end{small}
\end{tablenotes}
\end{threeparttable}
\end{table}

\clearpage
\section{Transitions by initial hours}\label{app:additional}
\renewcommand*{\thefigure}{\thesection\arabic{figure}}
\renewcommand*{\thetable}{\thesection\arabic{table}}
\setcounter{figure}{0}
\setcounter{table}{0}

\begin{table}[ht!]\centering
\caption{Minimum wage effect on forward-looking transitions out of minijobs by initial hours}
\label{tab:minijobs_initial_hours}
\begin{threeparttable}
\begin{tabular}{lC{3.7cm}C{3.7cm}C{3.7cm}}
\hline\hline
                    &\multicolumn{1}{c}{out of minijob}&\multicolumn{1}{c}{minijob to regular job}&\multicolumn{1}{c}{minijob to non-empl.}\\
\hline \\ [-1.4ex]
\multicolumn{4}{l}{A: Initially $\geq$ 55 hours}  \\[1ex]
\hline \\ [-1.4ex]
$Bite*Year_{2012}$ & Reference & Reference & Reference \\[1ex]
$Bite*Year_{2013}$ &  0.004    &     -0.012         &      0.014       \\
                    &    (0.014)         &    (0.010)         &    (0.012) \\
$Bite*Year_{2014}$  &  0.222\sym{***}&      0.161\sym{***}&      0.062\sym{***} \\
                    &    (0.019)         &    (0.014)         &    (0.017)                \\[1ex]
\hline \\ [-1.4ex]
Clusters            &         401         &         401         &         401  \\
Observations        &     2,022,023       &     2,022,023       &     2,022,023  \\[1ex]
\hline \\ [-1.4ex]
\multicolumn{4}{l}{B: Initially $<$ 55 hours}  \\[1ex]
\hline \\ [-1.4ex]
$Bite*Year_{2012}$ & Reference & Reference & Reference  \\[1ex]
$Bite*Year_{2013}$  &  -0.022\sym{**} &     -0.003         &     -0.018\sym{*}  \\
                    &    (0.008)      &    (0.004)         &    (0.007)         \\
$Bite*Year_{2014}$  &   0.148\sym{***}&      0.055\sym{***}&      0.093\sym{**} \\
         &    (0.032)         &    (0.008)         &    (0.031)         \\[1ex]
\hline \\ [-1.4ex]
Clusters            &         401         &         401         &         401  \\
Observations        &    11,657,151         &    11,657,151         &    11,657,151  \\[1ex]
\hline\hline
\end{tabular}
\begin{tablenotes}
\begin{small} \textit{Notes:} OLS regression coefficients. Dependent variables capture forward-looking transitions out of minijobs, as indicated by column titles. Cluster-robust standard errors in parentheses (cluster=county). Asterisks indicate significance levels: *~$p<0.05$, **~$p<0.01$, ***~$p<0.001$. \\\textit{Data:} IEB, population of all minijobbers, multiple jobs excluded.
\end{small}
\end{tablenotes}
\end{threeparttable}
\end{table}

\clearpage
\section{Transition analysis with covariates}\label{app:transitions_controls}
\renewcommand*{\thefigure}{\thesection\arabic{figure}}
\renewcommand*{\thetable}{\thesection\arabic{table}}
\setcounter{figure}{0}
\setcounter{table}{0}

\begin{table}[ht!]\centering
\caption{Minimum wage effect on forward-looking transitions out of minijobs, with covariates}
\label{tab:transition_mini_covars}
\begin{threeparttable}
\begin{tabular}{lC{3.7cm}C{3.7cm}C{3.7cm}}
\hline\hline
&\multicolumn{1}{c}{out of minijob}&\multicolumn{1}{c}{minijob to regular job}&\multicolumn{1}{c}{minijob to non-empl.}\\
\hline \\ [-1.4ex]
$Bite * Year_{2012} $ & Reference & Reference & Reference \\[1ex]
$Bite * Year_{2013} $ &      -0.008         &       0.000         &      -0.009         \\
                    &     (0.008)         &     (0.004)         &     (0.007)         \\
$Bite * Year_{2014} $ &       0.190\sym{***}&       0.089\sym{***}&       0.097\sym{***}\\
                    &     (0.025)         &     (0.008)         &     (0.025)         \\[1ex]
\hline \\ [-1.4ex]
Clusters   &    401         &         401    &         401      \\
Observations  & 13,678,421         &    13,678,421         &    13,678,421    \\[1ex]
\hline\hline
\end{tabular}
\begin{tablenotes}
\begin{small} \textit{Notes:} 
   Treatment effect interactions from difference-in-differences estimations, as specified in equation~(\ref{eq:minijob_trans}). Dependent variable captures forward-looking transitions out of minijobs, as indicated by column titles.  The $\boldsymbol{X}$-vector includes  the following covariates: age, dummies for gender, education, and industry. Standard errors in parentheses are clustered at the county level. Asterisks indicate significance levels:    *~$p<0.05$, **~$p<0.01$, ***~$p<0.001$.\\\textit{Data:} IEB, 2012-2015, population of all minijobbers, multiple jobs excluded. 
\end{small}
\end{tablenotes}
\end{threeparttable}
\end{table}

\begin{table}[ht!]\centering
\caption{Minimum wage effect on minijobber promotions within and across establishments, with covariates}
\label{tab:transition_mini_within_covars}
\begin{threeparttable}
\begin{tabular}{lC{4.2cm}C{4.2cm}}
\hline\hline
 & \multicolumn{2}{c}{minijob to regular job}\\
                    &\multicolumn{1}{c}{within establishments}&\multicolumn{1}{c}{across establishments}\\[1ex]
\hline \\[-1.4ex]
$Bite * Year_{2012}$ & Reference  & Reference     \\[1ex]
$Bite * Year_{2013}$ &      -0.001         &       0.001         \\
                    &     (0.003)         &     (0.003)         \\
$Bite * Year_{2014}$ &       0.081\sym{***}&       0.008\sym{*}  \\
                    &     (0.006)         &     (0.003)         \\[1ex]
\hline \\[-1.4ex]
Clusters   &    401         &         401       \\
Observations        &    13,678,421         &    13,678,421         \\[1ex]
\hline\hline
\end{tabular}
\begin{tablenotes}
\begin{small} \textit{Notes:} 
   Treatment effect interactions from difference-in-differences estimations, as specified in equation~(\ref{eq:minijob_trans}).
Dependent variable captures forward-looking transitions out of minijobs, as indicated by column titles.  The $\boldsymbol{X}$-vector includes  the following covariates: age, dummies for gender, education, and industry. Standard errors in parentheses are clustered at the county level. Asterisks indicate significance levels:    *~$p<0.05$, **~$p<0.01$, ***~$p<0.001$. \\\textit{Data:} IEB, 2012-2015, population of all minijobbers, multiple jobs excluded.
\end{small}
\end{tablenotes}
\end{threeparttable}
\end{table}

\begin{table}[ht!]\centering
\caption{Minimum wage effect on backward-looking transitions in minijobs, with covariates}
\begin{threeparttable}
\begin{tabular}{lC{3.7cm}C{3.7cm}C{3.7cm}}
\hline\hline
                    &\multicolumn{1}{c}{in minijob}&\multicolumn{1}{c}{regular job to minijob}&\multicolumn{1}{c}{non-empl. to minijob}\\[1ex]
\hline \\[-1.4ex]
$Bite * Year_{2013}$ & Reference & Reference & Reference \\[1ex]
$Bite * Year_{2014}$ &      -0.043\sym{***}&      -0.027\sym{***}&      -0.013         \\
                    &     (0.011)         &     (0.003)         &     (0.010)         \\
$Bite * Year_{2015}$ &       0.049\sym{***}&       0.041\sym{***}&       0.013         \\
                    &     (0.014)         &     (0.005)         &     (0.011)         \\[1ex]
\hline \\[-1.4ex]
Clusters           &     401  &     401  &     401    \\
Observations        &    13,455,389         &    13,455,389         &    13455389         \\[1ex]
\hline\hline
\end{tabular}
\begin{tablenotes}
\begin{small}  \textit{Notes:}     Treatment effect interactions from difference-in-differences estimations, as specified in equation~(\ref{eq:minijob_trans}). Dependent variable captures backward-looking transitions in minijobs, as indicated by column titles. The $\boldsymbol{X}$-vector includes  the following covariates: age, dummies for gender, education, and industry. Standard errors in parentheses are clustered at the county level. Asterisks indicate significance levels:    *~$p<0.05$, **~$p<0.01$, ***~$p<0.001$. \\\textit{Data:} IEB, 2012-2015, population of all minijobbers, multiple jobs excluded.
\end{small}
\end{tablenotes}
\end{threeparttable}
\end{table}

\begin{table}[ht!]\centering
\caption{Minimum wage effect on demotions (from regular to minijob) within and across establishments, with covariates}
\label{tab:transition_demotion_within_covars}
\begin{threeparttable}
\begin{tabular}{lC{4.2cm}C{4.2cm}}
\hline\hline
 & \multicolumn{2}{c}{regular job to minijob}\\
                    &\multicolumn{1}{c}{within establishments}&\multicolumn{1}{c}{across establishments}\\[1ex]
\hline \\[-1.4ex]
$Bite * Year_{2013}$ & Reference  & Reference     \\[1ex]
$Bite * Year_{2014}$ &      -0.012\sym{***}&      -0.015\sym{***}\\
                    &     (0.002)         &     (0.002)         \\
$Bite * Year_{2015}$ &       0.046\sym{***}&      -0.005\sym{*}  \\
                    &     (0.005)         &     (0.002)         \\[1ex]
\hline \\[-1.4ex]
Clusters         &    401      &         401       \\
Observations        &    13,455,389         &    13,455,389         \\[1ex]
\hline\hline
\end{tabular}
\begin{tablenotes}
\begin{small} \textit{Notes:}    Treatment effect interactions from difference-in-differences estimations, as specified in equation~(\ref{eq:minijob_trans}). Dependent variable captures backward-looking transitions in minijobs, as indicated by column titles.  The $\boldsymbol{X}$-vector includes the following covariates: age, dummies for gender, education, and industry. Standard errors in parentheses are clustered at the county level. Asterisks indicate significance levels:    *~$p<0.05$, **~$p<0.01$, ***~$p<0.001$. \\\textit{Data:} IEB, 2012-2015, population of all minijobbers, multiple jobs excluded.
\end{small}
\end{tablenotes}
\end{threeparttable}
\end{table}

\clearpage
\section{Establishment heterogeneities of minijob transitions}\label{app:transtions_establ_heterogeneities}
\renewcommand*{\thefigure}{\thesection\arabic{figure}}
\renewcommand*{\thetable}{\thesection\arabic{table}}
\setcounter{figure}{0}
\setcounter{table}{0}

In this appendix, we examine whether the transitions of minijobbers, i.e., upgrading of a minijob to a regular job or the transition of a minijobber to non-employment, differ by firm size, by the AKM wage effect, and by industry. The AKM-estimation jointly estimates employee and establishment fixed effects in a wage equation. For Germany, the AKM-estimation is thoroughly described and analyzed in \cite{Card2013}. The establishments' AKM effect is typically interpreted as the employer wage premium or the employer quality net of employee quality. 

\begin{figure}[ht!]
	\captionabove{Establishment-level effect heterogeneities for transitions out of minijobs}\label{fig:heterogeneities_establishments}
	\centering
\begin{subfigure}[a]{0.7\textwidth}	
\centering
\caption{Transitions in regular jobs}
\includegraphics[width=\textwidth]{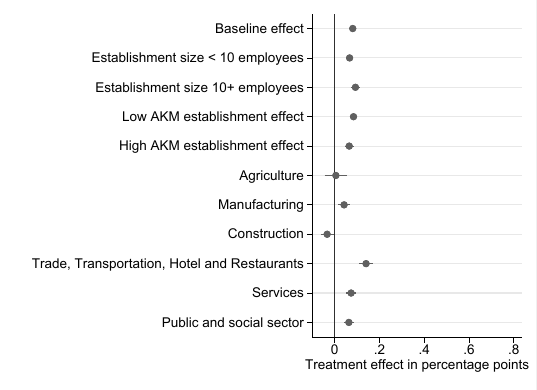}
\end{subfigure}
     \hfill
\begin{subfigure}[b]{0.7\textwidth}	
\centering
\caption{Transitions in non-employment}
\includegraphics[width=\textwidth]{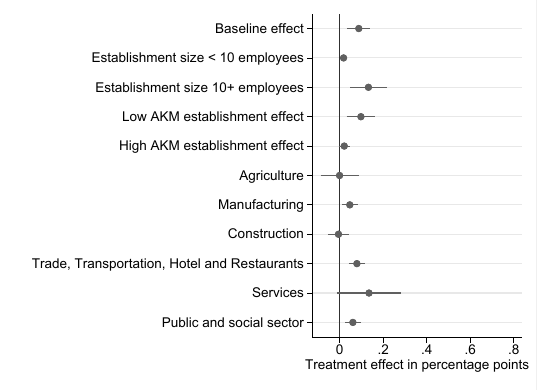}
\end{subfigure}
\subcaption*{
\textit{Notes:} Transition regressions in regular jobs and non-employment respectively, as specified in the second and third column of Table \ref{tab:transition_mini}. \\\textit{Data:} IEB, 2012-2015, population of all minijobbers, multiple jobs excluded.
}
\end{figure}

Regarding the upgrading of minijobs (shown in the upper part of Figure \ref{fig:heterogeneities_establishments}), we observe no heterogeneities in establishment size or the establishments' AKM effect. The upgrading is particularly pronounced in the industry aggregate which includes Trade, Transportation, Hotel, and Restaurants. It is smaller in agriculture and in construction. 

Regarding transitions in non-employment, we observe a stronger effect for establishments with at least 10 employees, establishments with AKM-effects below the median, and in services, although the effect for services is imprecisely estimated.

\end{appendix}

\end{document}